\documentclass[graybox, nosecnum]{svmult}

\usepackage{mathptmx}       
\usepackage{helvet}         
\usepackage{courier}        
\usepackage{type1cm}        
\usepackage{makeidx}         
\usepackage{graphicx}        
\usepackage{multicol}        
\usepackage[bottom]{footmisc}
\usepackage{hyperref}        
\usepackage{soul}            

\usepackage{amsmath}
\usepackage{amssymb}	
\usepackage{url}

\usepackage{enumitem}

\hypersetup{colorlinks=true,urlcolor=blue}
\usepackage[square,numbers]{natbib}

\newcommand {\comment}[1]{}
\newcommand{\red}[1]{\textcolor{black}{#1}}

\newcommand {\ergs}{{\rm erg\ \rm s^{-1}}}
\newcommand {\lum}{{\rm erg\ \rm s$^{-1}$}}
\newcommand {\be}{\begin {equation}}
\newcommand {\ee}{\end {equation}}
\newcommand {\beq}{\begin {eqnarray}}
\newcommand {\eeq}{\end {eqnarray}}

\def\apj{ApJ}
\def\apss{Astroph.Sp.Sci.}
\def\aap{A\&A}
\def\mnras{MNRAS}
\def\apjs{ApJS}

\def\nat{Nature}
\def\apjl{ApJLett}
\def\aapr{A\&ARv}
\def\apspr{Astrophys.\ Space Phys.\ Res.}

\def\ssr{Space Sci.\ Rev.}
\def\prd{Phys.\ Rev.\ D}
\def\sovast{Soviet\ Ast.}
\def\pasj{PASJ}
\def\nar{New Astron. Rev.}

\makeindex          

\begin{document}
\title*{Accreting strongly magnetised neutron stars: X-ray Pulsars}
\author{
Alexander Mushtukov, 
Sergey Tsygankov
}
\institute{Alexander Mushtukov \at 
Astrophysics, Department of Physics, University of Oxford, OX1 3RH, UK \\
Leiden Observatory, NL-2300RA Leiden, The Netherlands,\\
\email{alexander.mushtukov@physics.ox.ac.uk}
\and Sergey Tsygankov \at Department of Physics and Astronomy,  FI-20014 University of Turku, Finland \\
\email{sergey.tsygankov@utu.fi}
}

\maketitle

\abstract{
X-ray pulsars (XRPs) are accreting strongly magnetised neutron stars (NSs) in binary systems with, as a rule, massive optical companions. Very reach phenomenology and high observed flux put them into the focus of observational and theoretical studies since first X-ray instruments were launched into space. The main attracting characteristic of NSs in this kind of systems is the magnetic field strength at their surface, about or even higher than $10^{12}\,{\rm G}$, that is about six orders of magnitude stronger than what is attainable in terrestrial laboratories. Although accreting XRPs were discovered about 50 years ago, the details of the physical mechanisms responsible for their properties are still under debate.
Here we review recent progress in observational and theoretical investigations of XRPs as a unique laboratory for studies of fundamental physics (plasma physics, QED and radiative processes) under extreme conditions of ultra-strong magnetic field, high temperature, and enormous mass density.
}

\section{Keywords} 

X-ray pulsars,
neutron stars, 
accretion,
accretion discs, 
strong magnetic fields,
radiative transfer,
neutrinos,
X-rays.

\section{\red{1\,} Introduction}

\begin{figure}
\centering 
\includegraphics[width=12.cm, angle =0]{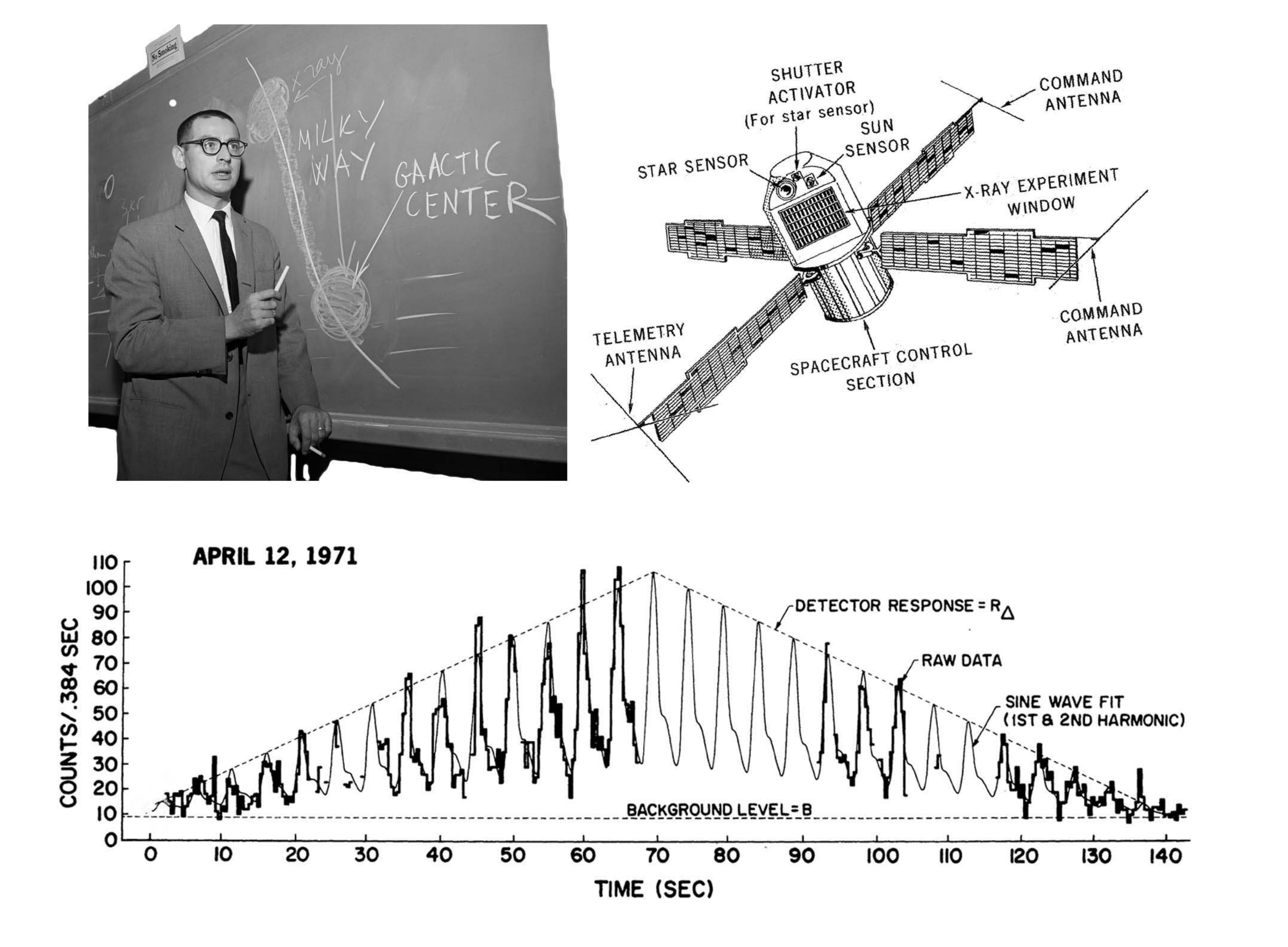} 
\caption{
2002 Noble Prize for Physics Riccardo Giacconi, the X-ray observatory \textit{UHURU} and X-ray pulsations detected from Cen X-3 \citep{1971ApJ...167L..67G}.
}
\label{pic:Giacconi}
\end{figure}

First coherent pulsations of the X-ray flux from the cosmic source Cen X-3 was discovered back in the early 70s by the first X-ray space observatory \textit{UHURU} (Fig.\,\ref{pic:Giacconi}, 
\citep{1971ApJ...167L..67G}
) and almost immediately recognized as a result of accretion of matter onto a strongly magnetized NS \citep{1973ApJ...184..271L,1973ApJ...179..585D}. The strong magnetic field (B-field) of the NS in XRP (typically $\gtrsim 10^{12}\,{\rm G}$ at the surface) affects both the global geometry of accretion flow in the system and physical processes in close proximity to the NS surface. As a result, the primary observational features of XRPs (temporal variability at different scales and spectral properties) bear information about interaction of matter and radiation with extremely strong magnetic fields.

\begin{figure}
\centering 
\includegraphics[width=10.cm, angle =0]{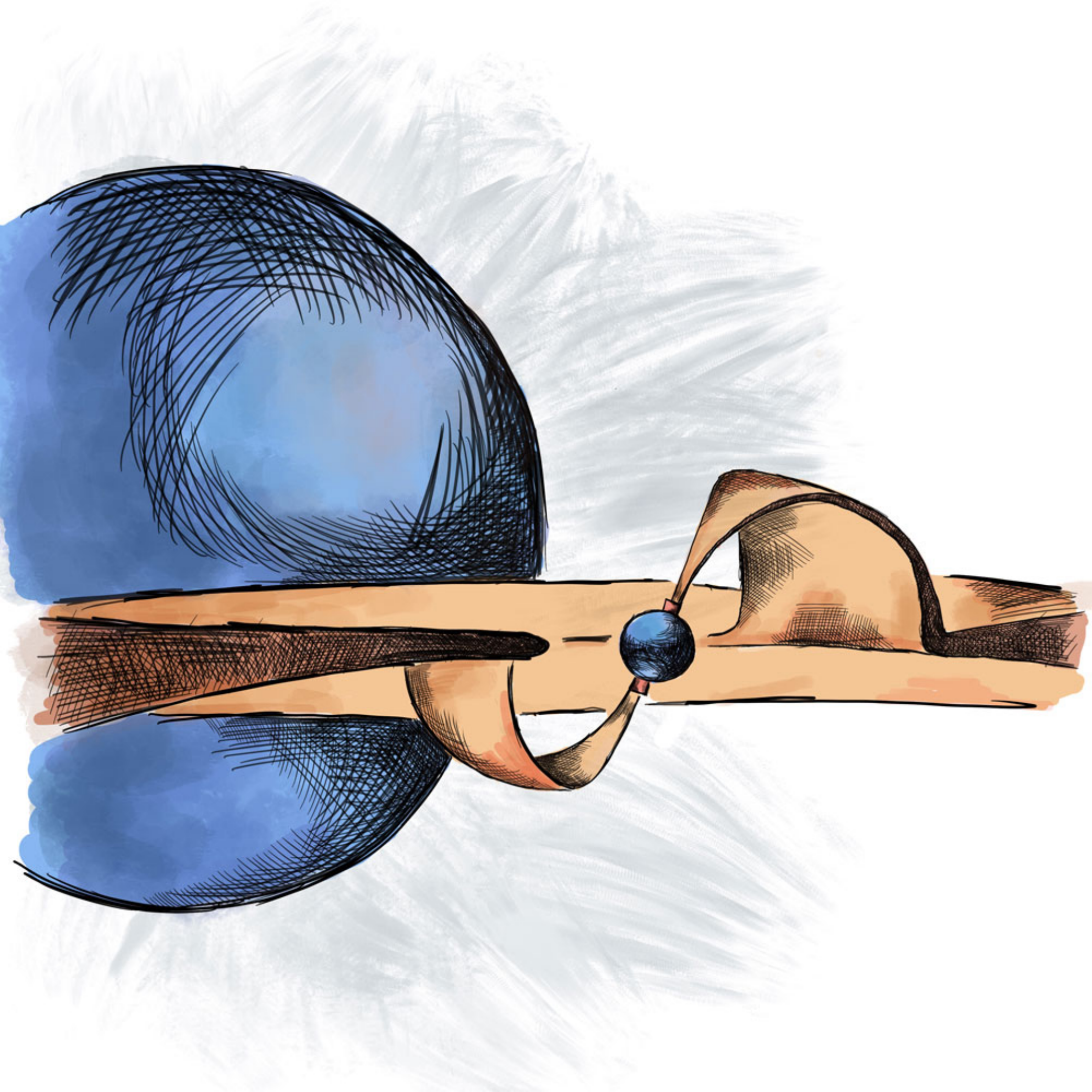} 
\caption{
\red{Schematic illustration of accretion geometry in X-ray pulsars.
Companion star provides material for the accretion process.
Accreting material forms an accretion disc or accretes via the stellar wind.
At a certain distance from a central object called the magnetospheric radius, the accretion flow experiences strong pressure of NS magnetic field and cannot move across magnetic field lines anymore.
Due to the instabilities at the magnetospheric radius, the flow settles magnetic field lines and follows them towards the NS surface.
Finally, accretion flow reaches NS in small regions located close to the magnetic poles of a star, where it loses its energy emitting it predominantly in the form of X-rays.
From \citep{2022ApJ...941L..14T}.}
}
\label{pic:XRPs}
\end{figure}

The matter in the form of plasma from the optical companion can be accreted by a compact object via the stellar wind, accretion disc or their combination. In the case of XRPs, at a certain distance from a NS, the flow cannot move towards the compact object without being disturbed by the magnetic field and is stopped forming the boundary at the NS magnetosphere \red{(see Fig.\,\ref{pic:XRPs})}. 
The material penetrates into the magnetosphere due to various instability mechanisms, and then follows the magnetic field lines towards small regions ($\sim 10^{10}\,{\rm cm^2}$) at the stellar surface close around the magnetic poles.
Here the kinetic energy of accreting matter is released predominantly in the form of X-rays. Due to the misalignment of the NS magnetic and rotation axes, the emission detected by a distant observer exhibits pulsations with the NS period of rotation.
Pulsation periods in XRPs lie in the range from $\sim 0.1\,{\rm s}$ to thousands of seconds and show complex evolution with time \citep{1997ApJS..113..367B}. 

About one hundred XRPs in our Galaxy and the Magellanic Clouds are known up to date \citep{2015A&ARv..23....2W}. 
Conventionally they can be divided into persistent and transient sources, with the latter ones covering more than five orders of magnitude in luminosity during outbursts \citep{2020MNRAS.491.1857D}. 
The apparent X-ray luminosity of XRPs may vary from $10^{32}$ up to $10^{41}\,\ergs$, with the brightest sources belonging to the recently discovered class of pulsating ultraluminous X-ray sources 
{(see \citep{2014Natur.514..202B,2017Sci...355..817I} and \citep{2021AstBu..76....6F} for review).}
Accreting XRPs can be studied observationally not only in X-rays but also in the radio band 
\citep{2018MNRAS.473L.141V,2018Natur.562..233V}.
It is also expected that the brightest members of this family can be equally bright in neutrinos being the so-called ``neutrino pulsars" \citep{2018MNRAS.476.2867M}. 
In the nearest future, we will have a chance to measure X-ray polarisation in XRPs thanks to the upcoming dedicated X-ray polarimeters.  

The basic phenomenological interpretation of the main observational features of XRPs has remained essentially unchanged since their discovery. At the same time a number of exciting details and complications related to questions of stellar evolution, accretion flow dynamics, and radiation physics have emerged thanks to the unprecedented quality of the data collected by modern X-ray instruments. Below we review a recent progress in both observational and theoretical studies of XRPs in a broad range of physical parameters.

\section{\red{2\,} Magnetic field: the reason for the XRPs uniqueness}


NSs are born as a result of a core collapse in massive supergiant O or B stars with initial mass greater than $8-10$ solar masses during the supernova explosion.
The earliest and simplest predictions for the NS magnetic field were based on the assumption that the magnetic flux of a progenitor star is conserved 
\citep{1964SPhD....9..329G}.
Indeed, the typical radius of the NS progenitor is $R_{\rm OB}\sim 10^{11}\,{\rm cm}$, while the NS radius is $R\sim 10^6\,{\rm cm}$.
If magnetic flux is conserved during supernova explosion, the field is amplified by a factor of $(R_{\rm OB}/R)^2\sim 10^{10}$.
These simple calculations predict magnetic field strengths on the NS surface of the order of $\sim 10^{12}\,{\rm G}$, in good agreement 
with estimates obtained from the energy of the cyclotron resonant scattering spectral features in XRPs \citep{2019A&A...622A..61S} and spin periods and spin period derivatives in radio pulsars 
\citep{1968Natur.218..731G}.
However, since the NS is formed from the progenitor's core comprising only $\sim15$\% of its total original mass, and it is not correct to use the whole stellar radius to estimate the magnetic field amplification due to the flux conservation \citep{2008AIPC..983..391S}.
More realistic scenarios for the formation of extremely strong magnetic field in NSs involve dynamo amplification of magnetic fields at the proto-NS stage \citep{1993ApJ...408..194T}.
Overall, it is expected that NSs are born with a range of magnetic field strength covering a few orders of magnitude from $~10^{11}$ and up to $10^{15}\,{\rm G}$.

The extremely strong magnetic field of NSs in XRPs is one of the key features causing a keen interest in this class of objects. Indeed, the typical strength of the magnetic fields we are dealing in the everyday life is less than $1\,{\rm G}$.
The field strength in the active regions of normal stars can be of the order of $10^3\,{\rm G}$.
The strongest magnetic field available in terrestrial laboratories is about $10^6\,{\rm G}$ \citep{2019StrongestMagnet} which can be generated for a few seconds only and used by experimental physicists to probe plasma physics under extreme conditions.
Unfortunately, the magnetic field of a few million gauss is the maximum of our experimental abilities. 
A magnetic field of strength
\be
B>\frac{m_{\rm e}^2 e^3 c}{\hslash^{3}}
=2.35\times10^9\,{\rm G}
\ee
destroys the basic laws of ``school chemistry". 
In this field, the Lorentzian force of interaction between electrons and the external magnetic field becomes stronger than the electric force of interaction between electrons and atomic nuclei. 
Thus, the magnetic field becomes strong enough to change the structure of the energy levels in atoms and their chemical properties 
\citep{2001RvMP...73..629L}.

A magnetic field exceeding $10^{11}\,{\rm G}$ affects the wave functions of charged elementary particles, influences the properties of the magnetized vacuum and of photon propagation in the medium itself.
In particular, the electron motion becomes quantized in the direction perpendicular to the magnetic field with electrons occupying the so-called Landau energy levels. 
The characteristic scale of the extreme magnetic field is given by the critical field value:
\be 
B_Q\equiv \frac{m_{\rm e}^2 c^3}{e \hslash}\simeq
4.414\times 10^{13}\,{\rm G}.
\ee 
At this field strength, the transverse motion of electrons becomes relativistic, and the linear scale of electron de Broglie wave becomes equal to the scale of electron's gyroradius.
In such a strong magnetic field, one can expect processes, which are impossible in a weak field limit: photon splitting \citep{1994PhRvD..50.1125M} and one-photon pair production and annihilation \citep{1983ApJ...273..761D}. 

Equating the magnetic energy ($\sim R^3 B^2/6$) and the gravitational binding energy ($\sim 0.6 GM^2/R$) we can estimate the maximal ``virial" value of the magnetic field for a NS \citep{2000ApJ...537..351B}: 
\be 
B_{\rm max}\approx 10^{18}\,
\left(\frac{M}{1.4 M}\right)
\left(\frac{R}{10\,{\rm km}}\right)^{-2}\,\,{\rm G}.
\ee 
A stronger magnetic field would induce a dynamical instability of a hydrostatic configuration and cannot be sustained in a considerable fraction of the star 
\citep{1953ApJ...118..116C}.

The initial structure of the NS magnetic field can be different depending on the details of magnetic field formation during the dynamo process.
Further evolution of the NS magnetic field is regulated by many factors, including Omhic decay, Hall effect and accretion process (see \citep{2021Univ....7..351I} for review). 
These processes influence both magnetic field strength and structure. 
Typically, the large-scale magnetic field in XRPs is assumed to be dominated by the dipole component because of its slow weakening with distance ($\propto r^{-3}$).
In spherical coordinates $(r,\theta,\varphi)$ the dipole magnetic field ${\bf B}$ is given by
\be\label{eq:dipole_field}
B^{\rm (d)}_r=\frac{2\mu_{\rm d}\cos\theta_B}{r^3},\quad\quad
B^{\rm (d)}_\theta=\frac{\mu_{\rm d}\sin\theta_B}{r^3},
\ee 
where $\mu_{\rm d}$ is a dipole magnetic moment, and $\theta_B$ is the colatitude related to the dipole magnetic axis.
In the case of pure dipole magnetic field, the dipole magnetic moment 
$\mu_{\rm d}\equiv \frac{B_0}{2}\, r^3$, where $B_0$
is the field strength at the magnetic pole, i.e. at $\theta_B=0$.
At small distance from a NS surface, higher multipoles, like the quadrupole component 
\be\label{eq:quadrupole_field}
B^{\rm (q)}_r=\frac{\mu_{\rm q}(3\cos^2\theta_B^{\rm (q)}-1)}{r^4},\quad\quad
B^{\rm (q)}_\theta=\frac{\mu_{\rm q}\sin2\theta_B^{\rm (q)}}{r^4},
\ee 
where $\mu_{\rm q}$ is the quadrupole magnetic moment and $\theta_B^{\rm (q)}$ is the colatitude related to the quadrupole magnetic axis, can come into play \citep{2007MNRAS.374..436L}.
In the case of pure quadrupole field, the quadrupole magnetic moment is given by 
$$\mu_{\rm q}\equiv \frac{B_0}{2}\,r^4,$$ 
where $B_0$ is the field strength $\theta_B^{\rm (q)}=0$.
Some XRPs already provide evidence of possibly strong non-dipole components of their magnetic field \citep{2017Sci...355..817I,2017A&A...605A..39T,2022MNRAS.515..571M}.

\section{\red{3\,} Observational appearance of X-ray pulsars}

XRPs are among the most bright sources on the X-ray sky that allowed many observatories to collect rich datasets covering several decades. 
In this section, we will discuss the observational appearance of XRPs at different time scales and energy bands.
This is a complicated task considering the richness of phenomenology available nowadays.
We will consider only the basic features here with the choice of topics being inevitably subjective.
We will start with a discussion of the coherent flux pulsations, the definitive feature of the XRPs.
It will be shown that the analysis of pulsations and their temporal evolution can tell us a lot about a system and the accretion process in it. 
Then we will talk about the luminosity of XRPs, that is directly related to the mass accretion rate in a system, and its variability.
Estimates of the mass accretion rate, in turn, helps to identify the place of XRPs among other classes of accreting compact objects, and to figure out which physical processes come into play.
We will see that different sub-classes of XRPs exhibit various types of luminosity variation patterns opening a door into a world of unstable processes in XRPs developing on different time scales and involving various aspects of physics.
Even stochastic flickering of XRPs can tell us a lot about the geometry of accretion flows and physical conditions there.
The spectral analysis of XRPs emission gives us an opportunity to go further in the understanding of accretion onto magnetised NSs. 
Spectral features bear fingerprints of elementary processes shaping radiation and matter interaction in the emitting regions of a NS.
XRPs are binary systems, and the properties of the NS companions make it possible to estimate the age of a system and understand the place of XRPs in the global evolution of binaries.


\subsection{\red{3.1\,} Coherent pulsations: the definitive feature of XRPs}

\begin{figure}
\centering 
\includegraphics[width=0.7\columnwidth]{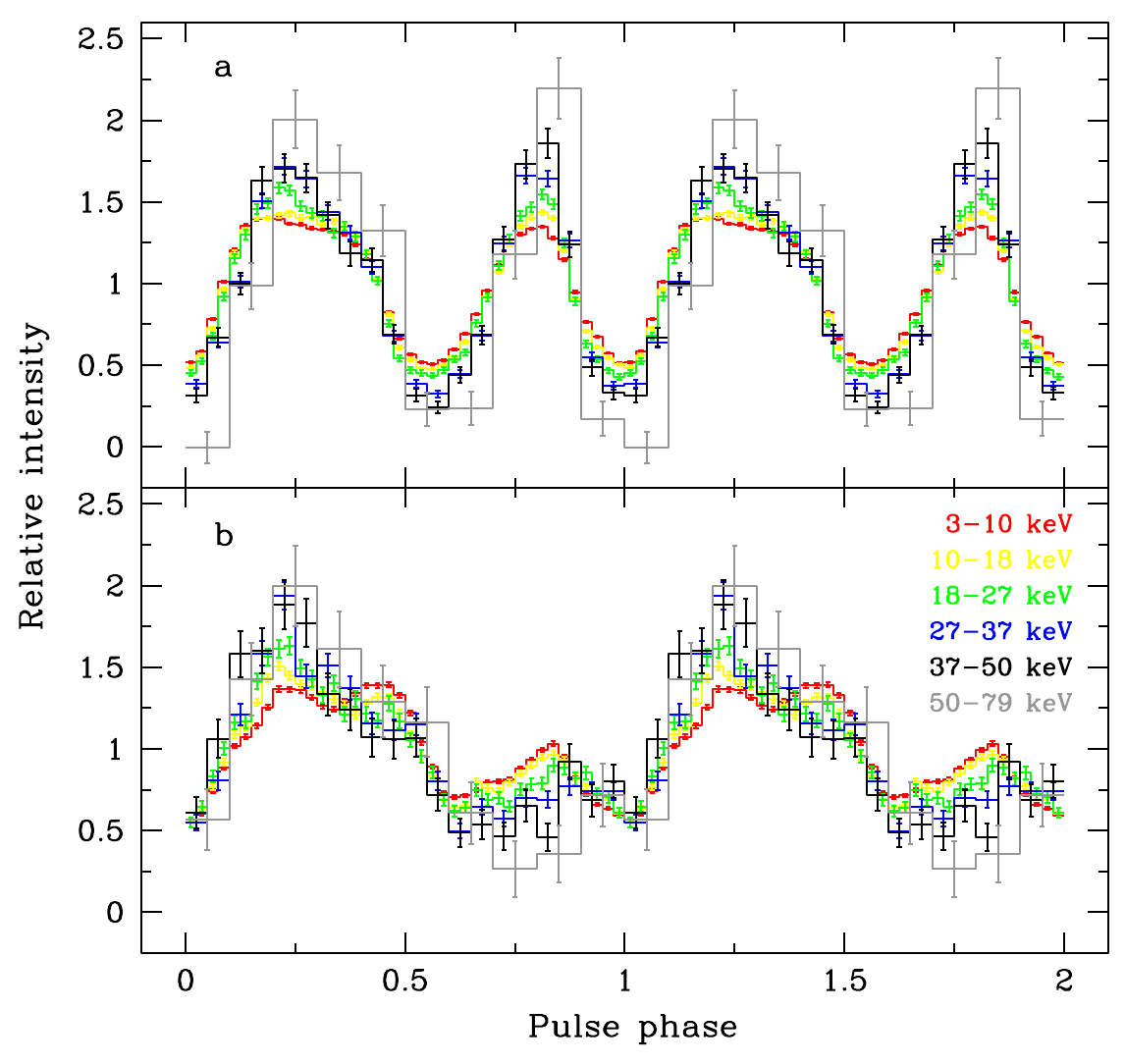} 
\caption{
Dependence of the pulse profile of SMC~X-3 on energy in two {\it NuSTAR} observations with different X-ray luminosities: (panel a) $L_{\rm bol}=10.2\times10^{38}$ \lum, (panel b) $L_{\rm bol}=1.9\times10^{38}$ \lum. 
Different energy bands are shown with different colours (shown in the bottom panel). 
The profiles are normalized by the mean flux in each energy band and plotted twice for clarity.
One can see that the pulse profiles tent to be dependent both on the energy band and accretion luminosity, which reflects the dependence of the beam pattern on energy and mass accretion rate. 
From \citep{2017A&A...605A..39T}.
}
\label{pic:pprof_smcx3}
\end{figure}

\begin{figure}
\vbox{\centering
\includegraphics[width=0.7\columnwidth]
{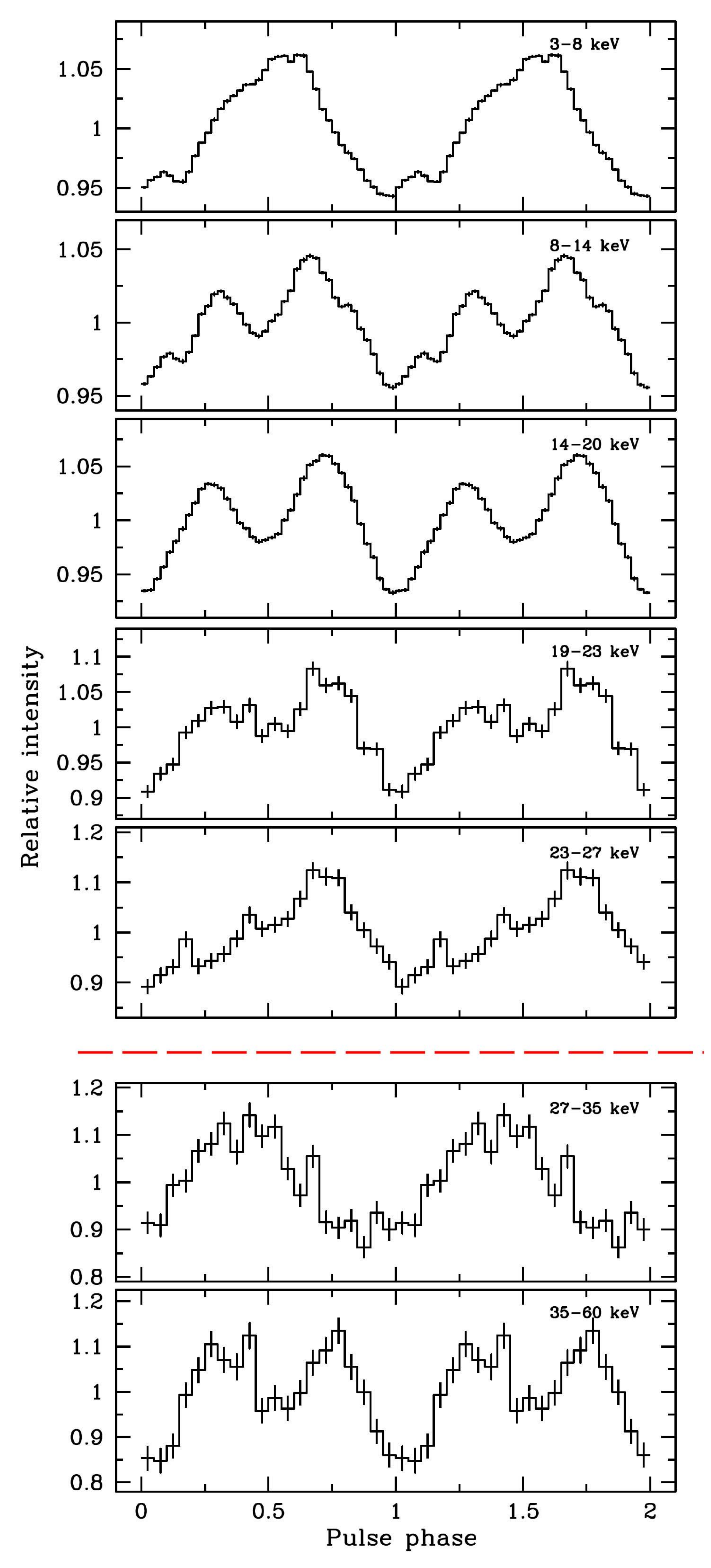}

\caption{Pulse profiles of V\,0332+53 at different energies obtained by the {\it RXTE} observatory during the brightening phase of the 2004--2005 outburst at the luminosity $L\!\simeq\!16\times10^{37}$ \lum. The cyclotron line in the source spectrum during this observation is
$E_\mathrm{cycl}\simeq\!27.2$~keV (shown with the red dashed line). From \citep{2010MNRAS.401.1628T}.
}
\label{pic:pprof_v0332}}
\end{figure}

\begin{figure}
\centering 
\includegraphics[width=0.85\columnwidth]{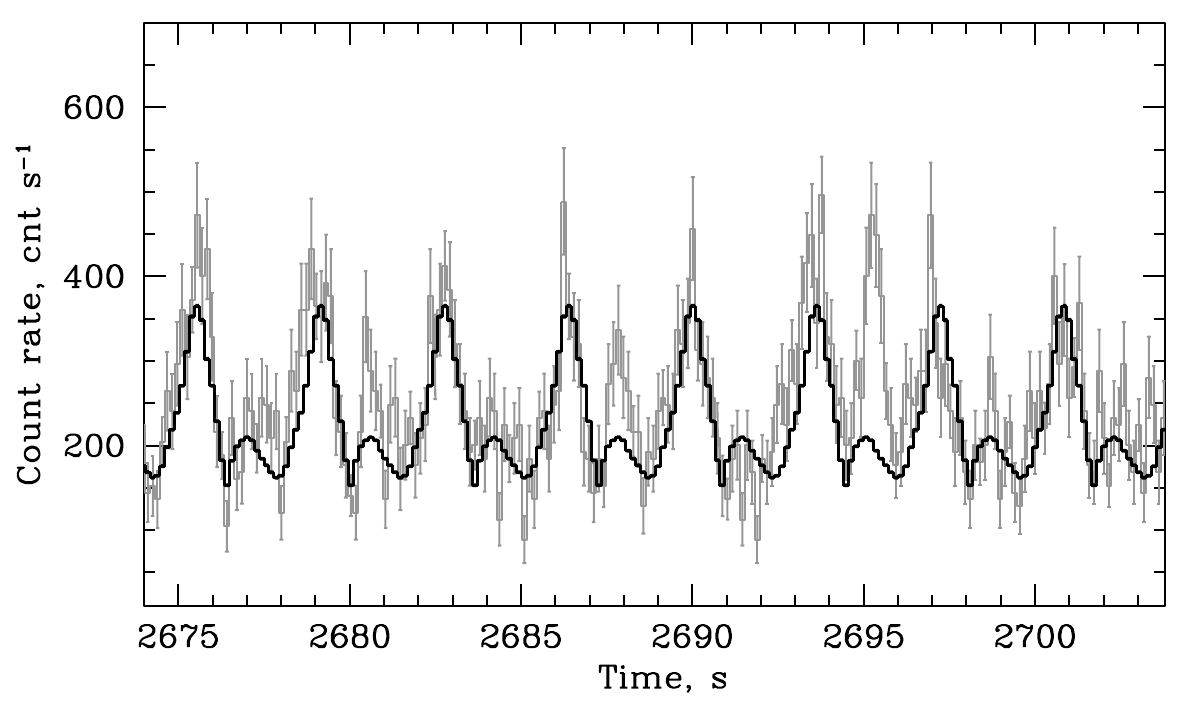} 
\caption{Light curve of 4U~0115+63 in the 17-20.3 keV energy band obtained with the {\it RXTE}/PCA instrument at a luminosity of $\sim6.6\times10^{37}$ \lum\ (grey curve).  The solid black line indicates the average pulse profile shape in this energy band for the entire observing session. 
One can clearly see the pulse-to-pulse variability of the pulse profile shape with second peak being much more variable than the main one.
From \citep{2007AstL...33..368T}.
}
\label{pic:pprof_4u0115}
\end{figure}

Coherent pulsations of the flux observed from XRPs, naturally explained by a misalignment between the rotation and magnetic axes of a NS, are characterized by phase lightcurves (aka pulse profiles) of very different shapes (see examples in  Fig.\,\ref{pic:pprof_smcx3},\ref{pic:pprof_v0332},\ref{pic:pprof_4u0115}). 
In contrast to radio pulsars, typical pulsations of XRPs have a large duty cycle ($\gtrsim 50\%$), and emission never drops to zero intensity at the pulse minimum. The convenient way to characterized pulsations amplitude in a given energy band is the so-called pulsed fraction (PF), which can be calculated as: \be 
PF\equiv \frac{F_{\rm max}-F_{\rm min}}{F_{\rm max}+F_{\rm min}}, 
\ee 
where $F_{\rm min}$ and $F_{\rm max}$ are minimum and maximum X-ray flux detected within the pulse period, respectively. 
The pulsed fraction of XRPs may depend on the source luminosity and energy band, and can be as high as $100\,\%$ at high energies.

The pulse profile shape in XRPs may have complex morphology and, similarly to the pulsed fraction, vary with the energy band (see Fig.\,\ref{pic:pprof_smcx3},\ref{pic:pprof_v0332}) and accretion luminosity (compare the upper and lower panels in Fig.\,\ref{pic:pprof_smcx3}). 
Variations of pulse profiles with the energy band can be caused by the dependence of the beam pattern on photon energy and/or local absorption due to not spherically symmetric distribution of matter in the system. 
Changes in the pulse profiles with accretion luminosity indicate the dependence of the geometry of the emission forming region and, therefore, of the beam pattern on the mass accretion rate (see basic model types of beam patterns in Fig.\,\ref{pic:beam_paterns}, \citep{1973A&A....25..233G,2015MNRAS.448.2175L}). 
Remarkably, the pulse profiles can also demonstrate a strong short-term variability on a pulse period timescale (see Fig.\,\ref{pic:pprof_4u0115}).
In practice, the average observed X-ray energy flux over the pulse profile is used to estimate the accretion luminosity of XRP. 

The relative stability of the NS rotation rate can be used to determine the orbital parameters of the binary system. Indeed, the observed pulse period in XRPs is affected by the Doppler effect due to the NS orbital motion.
Thanks to that, periodic variations of the pulse period are widely used to establish orbital parameters in a binary system. From another side, strong variations of pulse period due to the orbital motion 
may preclude from the discovery of pulsations from some sources, e.g. as it was in the case of pulsating ULXs \citep{2017Sci...355..817I}. 
The long-term variations of the pulse period (see Fig.\,\ref{pic:pdot_obs}) are related to the physics of the accretion process, in particular angular momentum exchange between accretion flow and NS.
Simultaneous measurements of the spin-up/-down trends and the corresponding luminosity are typically used to probe the NS magnetic field and physical features of the accretion flow.

In addition to the long-term steady evolution of the NS spin period, sudden changes of rotation frequency called ``glitches" (in the case of frequency increase), and ``anti-glitches" (in the case of frequency decrease) were detected in XRPs and pulsating ULXs (see \citep{2020ApJ...891...44B} and references there).
Although the details of physics standing behind glitches are not very well known, it is thought that the sudden changes of spin frequency can be related to a transfer of angular momentum between superfluid and non-superfluid components of a NS (see discussion and references in \citep{2019ApJ...879..130R}). 
It is remarkable that glitches are known to be typical events for isolated radio pulsars, whereas anti-glitches seems to be a specific feature of the accretion process.
Indeed, the phenomenon of anti-glitch can be caused by prolonged rapid spin-up of the NS due to the accretion process. 
In this case, the sudden transfer of angular momentum from NS outer crust to superfluid inner crust (and possibly core) leads to the drop of the pulsation frequency \citep{2015A&A...578A..52D}.

\subsection{\red{3.2\,} How bright are they?}

The actual luminosity of an XRP can be estimated from the observed X-ray energy flux if one knows the distance to the source, beaming factor and spectral distribution in the broad energy band.
For normal XRPs, the distances remain the largest uncertainty, being often known with errors as large as 50 per cent. At the same time, thanks to the {\it Gaia} observatory, distance to a substantial number of the Galactic XRPs is known with an accuracy of about 10 per cent.
For the bright XRPs, unknown beaming factor due to the geometry of accretion flow results in additional contribution to the uncertainty \citep{2017MNRAS.468L..59K,2021MNRAS.501.2424M}.

\begin{figure}
\centering 
\includegraphics[width=11.5cm, angle =0]{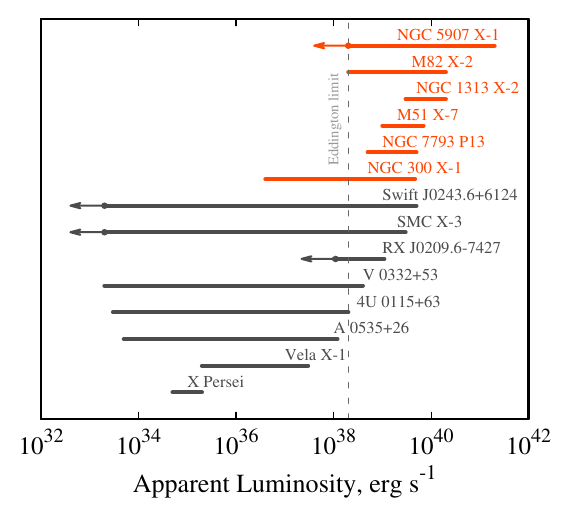} 
\caption{
Apparent luminosity and luminosity range observed in some X-ray transients powered by accretion onto magnetised NSs.
Grey segments correspond to ``normal" X-ray transients, while the red ones correspond to ULX pulsars. 
Dashed line represents the Eddington luminosity for a NS.
}
\label{pic:luminosity_range}
\end{figure}

The apparent luminosity of XRPs covers many orders of magnitude from $\sim 10^{32}\,\ergs$ up to $\sim 10^{41}\,\ergs$ (see Fig.\,\ref{pic:luminosity_range} for several representative examples).
At the low end of this range, we see either accreting systems at very low mass accretion rates  \citep{2017MNRAS.470..126T}, or XRPs in the propeller state \citep{1975A&A....39..185I,2006ApJ...646..304U}, where accretion is stopped by a strong magnetic field and X-rays emission is produced by cooling of the NS surface heated during the episodes of intensive accretion \citep{2016A&A...593A..16T,2016MNRAS.463L..46W}.  
The upper limit in the luminosity range is provided by the recently discovered XRPs in ultra-luminous X-ray sources (ULXs, \citep{2014Natur.514..202B,2017Sci...355..817I}), where the apparent luminosity can exceed the Eddington luminosity
\be 
L_{\rm Edd}=\frac{4\pi GM m_{\rm p}c}{\sigma_{\rm T}}\approx 1.26\times 10^{38}
\left(\frac{M}{M_\odot}\right)\,\ergs,
\ee 
where $M$ is a mass of a NS, $m_{\rm p}$ is a mass of a proton, and $\sigma_{\rm T}=6.65\times 10^{-25}\,{\rm cm^2}$ is the Thomson scattering cross section, by a factor of hundreds.

\begin{figure}
\centering 
\includegraphics[width=12.cm, angle =0]{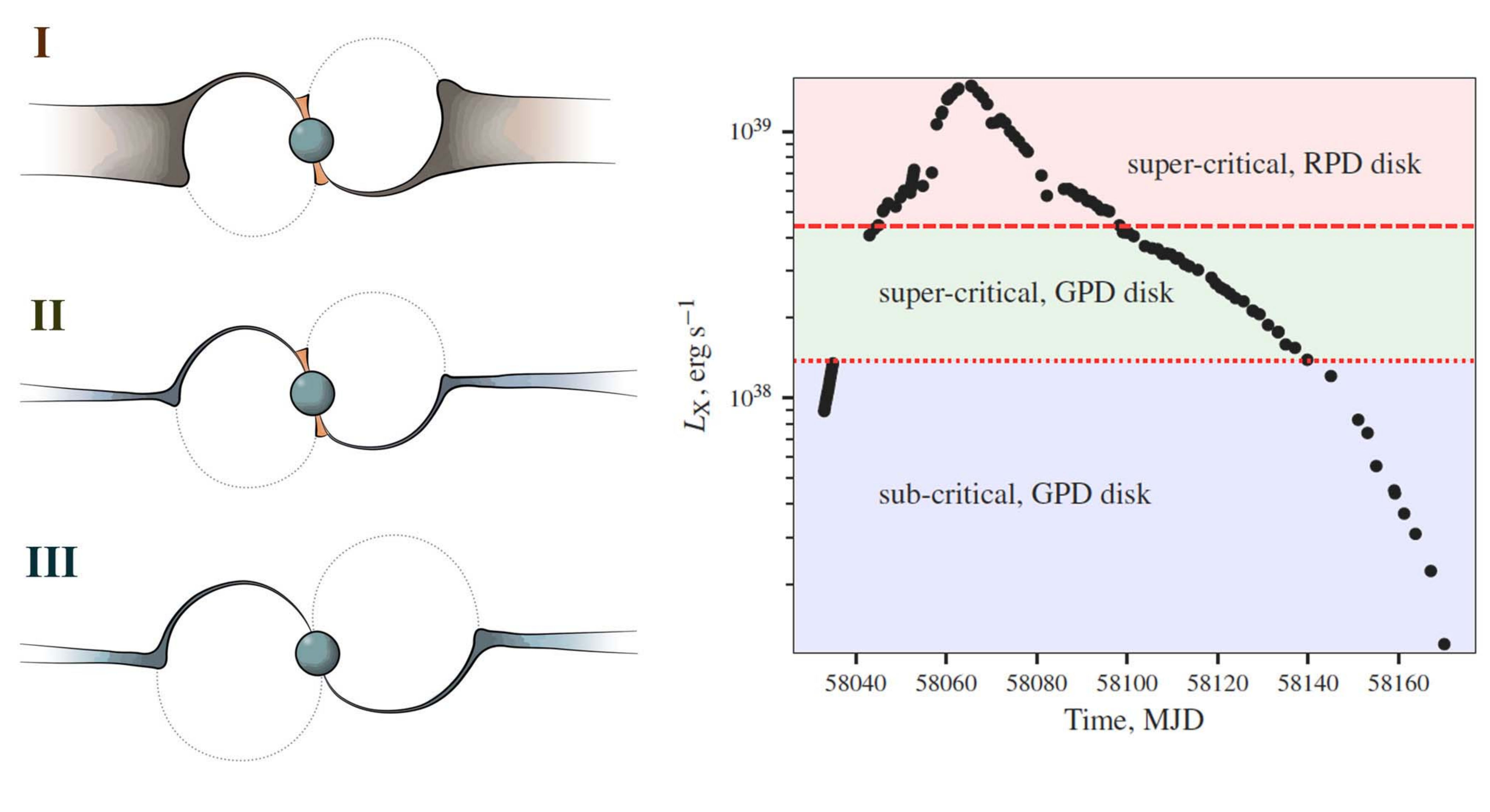} 
\caption{
Variations of accretion disc geometry and geometry of the emitting region expected at different mass accretion rates {\it(left)} and the observed light curve during the giant outburst in XRP Swift~J0243.6+6124 {\it(right)}.
At relatively low mass accretion rates, accretion disc is expected to be gas pressure dominated and geometrically thin, accretion process results in hot spots located close to the NS magnetic poles (III).
Increase of the mass accretion rate leads to increase of X-ray luminosity. The radiative force starts to affect dymanics of accretion flow in the vicinity of the NS surface and leads to appearance of radiation dominated shock and accretion columns above stellar surface (II). 
Further increase of the mass accretion rate turns accretion disc into geometrically thick radiation pressure dominated mode (I). 
At extremely high mass accretion rates the disc can be even advection dominated, when the photons are confined in the flow and participate in accretion process. 
From \citep{2020MNRAS.491.1857D}.
}
\label{pic:light_curve_Swift}
\end{figure}

\begin{figure}
\centering 
\includegraphics[width=12.cm, angle =0]{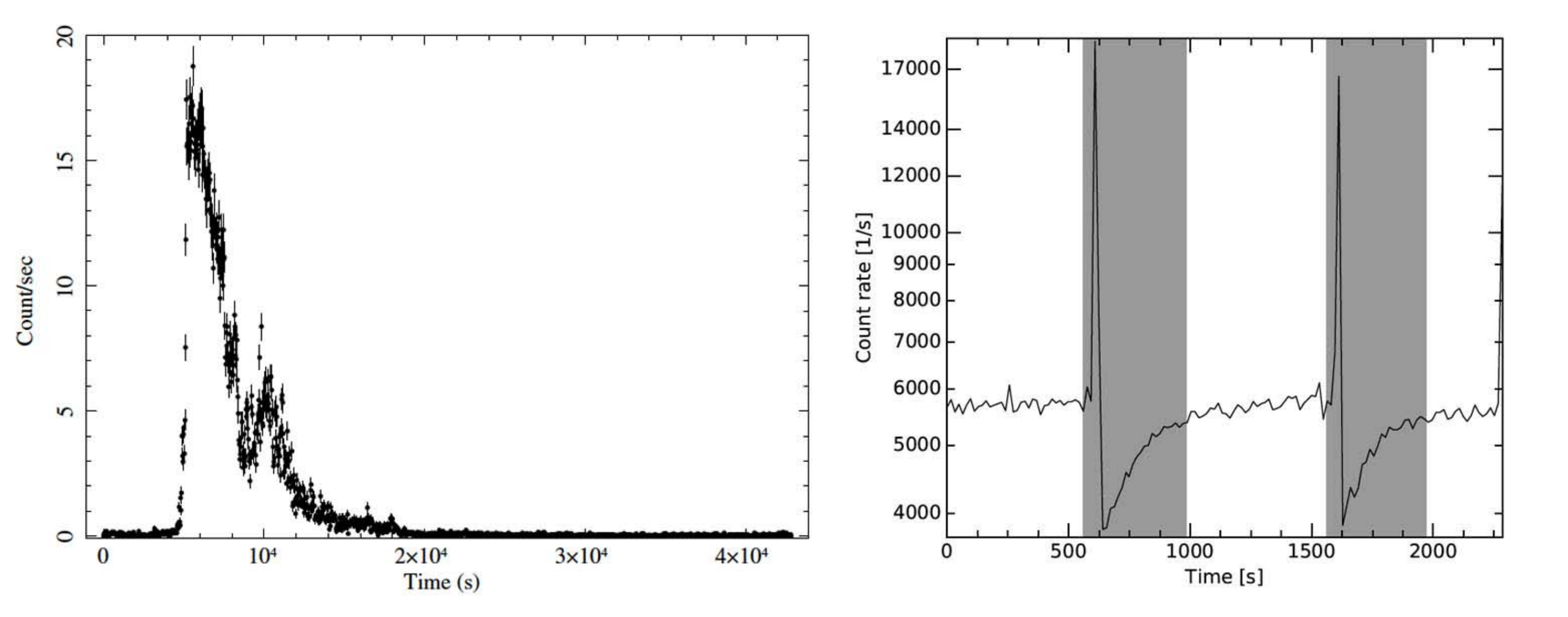} 
\caption{
Light curve of Supergiant Fast X-ray Transient IGR~J18410$-$0535 ({\it left}, from \citep{2011A&A...531A.130B}) and
short-term brightenings observed in bursting XRP GRO~J1744$-$28 at the peak of its outburst ({\it right}, from \citep{2019A&A...626A.106M}).
These kinds of short-term transient activity in XRPs can be caused by the development of various instabilities in the accretion flow. 
In particular, the activity observed in the SFXT can be a result of sudden matter penetration through the centrifugal barrier of a NS in a propeller state or stellar wind clumpiness. 
The fast (tens of seconds) brightening in GRO~J1744$-$28 can be explained by the development of instability in the radiation-pressure-dominated disc when the inner disc parts suddenly fall onto the NS (in this scenario, the deeps following the brightenings correspond to accretion disc recovery).
}
\label{pic:light_curve_J1744}
\end{figure}

Vast majority of known XRPs are transient sources and demonstrate dramatic variations of X-ray luminosity on time scales from weeks and months during the outbursts (see Fig.\,\ref{pic:light_curve_Swift},\,\ref{pic:propeller_LC} and \ref{pic:cold_disc_LC}) down to seconds (see Fig.\,\ref{pic:light_curve_J1744} {\it right}, \citep{1996ApJ...462L..39L}).
The transient nature can be caused by variations of mass accretion rate from the companion in a binary system, processes in accretion flow in between NS and its companion, and processes happening in close proximity to the NS.

One of the most numerous class of the transient XRPs is systems with Be optical companions (BeXRBs). Such systems are characterised by variability of two types. Type I outbursts, related to the enhanced mass accretion rate near the periastron passage, have short duration (about 10-20\% of the orbital period) and relatively low peak luminosity ($L\lesssim10^{37}\,\ergs$). 
On the contrary, type II (giant) outbursts, caused by the formation of the circumstellar disc around the companion, are rare events. They last for several orbital periods, during which NSs luminosity may exceed the Eddington limit. Example of a giant outburst from BeXRB Swift~J0243.6+6124 discovered in 2017 is presented in Fig.~\ref{pic:light_curve_Swift}.

Another sub-class of strongly variable high-mass X-ray binaries (HMXBs) are Super-giant Fast X-ray Transients (SFXTs), discovered with the {\it INTEGRAL} observatory \citep{2005A&A...444..221S}. In these binaries, strongly magnetised NS accretes matter via wind from the OB super-giant companion. However, in contrast to the classical wind-fed supergiant HMXBs, which exhibit themselves as persistent and relatively bright sources with luminosity around 10${^{36}}\,$\lum, SFXTs have very low quiescence luminosity and produce short and bright flares with typical duration of a few hours. This class of objects is not homogeneous in terms of the observed properties. To explain the variety of variability patterns, several models were proposed, including clumpy stellar wind \citep{2007A&A...476..335W}, propeller mechanism \citep{2007AstL...33..149G} and quasi-spherical settling accretion \citep{2014MNRAS.442.2325S}. Example of a powerful flare from SFXT IGR~J18410$-$0535 is shown in Fig.~\ref{pic:light_curve_J1744} {\it (left)}. 

{Apart from the discussed types of variability a unique behaviour is demonstrated by GRO~J1744$-$28, Galactic XRP in low-mass X-ray binary. 
Namely, in the peak of very rare and bright outbursts it exhibits extremely bright flares with duration of only several ($\sim 10$) seconds. 
This variability patterns appears due to transition of the inner accretion disc into the radiation-dominated regime when the Lightman-Eardley instability can be developed with subsequent decrease of the inner disc mass, reflected in the flux drop right after the burst (see Fig.~\ref{pic:light_curve_J1744}\,{\it right}, \citep{1996ApJ...466L..31C})}.

The brightest X-ray pulsars belong to the recently discovered class of pulsating ULXs.
ULXs are off-nuclear, extragalactic X-ray sources with isotropic luminosities exceeding the Eddington limit for a stellar-mass black hole, i.e. $L\gtrsim 10^{39}\,\ergs$ (see \citep{2021AstBu..76....6F} for a review).
ULXs were discovered back in the 80s by the \textit{Einstein} space observatory \citep{1983adsx.conf..117L} and considered for decades as accreting black holes with intermediate \citep{1999ApJ...519...89C} or stellar masses
\citep{2007MNRAS.377.1187P}.
Nowadays, there are several hundred detected ULXs \citep{2019MNRAS.483.5554E} and intermediate-mass black holes are almost excluded from theoretical models in the absolute majority of cases \citep{2004NuPhS.132..369G,2007MNRAS.377.1187P}. 
Discovery of coherent pulsations from some ULXs
{(see \citep{2014Natur.514..202B,2017Sci...355..817I} and references in \citep{2021AstBu..76....6F})}
shows that some of them are powered by accretion onto strongly magnetised NSs. 
There are 6 pulsating ULXs known up to date (see Fig.\,\ref{pic:luminosity_range} and Tab.\,\ref{tab:ULXs}).
It is remarkable that there are only $\sim 15$ ULXs out of $\sim 300$ known to provide the statistics sufficient for detection of pulsations \citep{2020ApJ...895...60R}, and $\sim 25$ per cent of them are proven to be accreting NSs. 
Therefore, one can speculate that a significant fraction of ULXs is represented by X-ray binaries hosting NSs.
This guess is in agreement with the recent result of population synthesis models \citep{2020AstL...46..658K} and the analysis based on the observed high mass X-ray binaries luminosity function \citep{2015MNRAS.454.2539M}.

It is still debated by the scientific community how the apparent luminosity of pulsating ULXs is related to the actual one. 
It is possible that X-ray radiation from these sources is highly beamed due to strong outflows from this kind of systems, and the apparent luminosity exceeds the actual luminosity by a large factor \citep{2017MNRAS.468L..59K,2019MNRAS.485.3588K,2020MNRAS.494.3611K}.
At the same time, there are strong evidence that actual and apparent luminosity are close one to another \citep{2021MNRAS.501.2424M,2020AstL...46..658K,2019MNRAS.488.5225V}.

\begin{table}
\centering
\caption{
The basic properties of detected ULX pulsars: 
their maximal apparent luminosity, spin period, spin period derivative, a pulsed fraction (PF), orbital period, and the estimation of companion mass in a binary.
Two ULX pulsars show significant changes in their pulse periods on the time scale of ten years. 
For these objects, we represent parameters corresponding to different epochs.
}
\tabcolsep=0.3cm
  \label{tab:ULXs}
  \begin{tabular}{l c c c c c c}
    \hline\hline
    \vspace{0.1cm}
  Name & $L_{X}$ (max) & $P$   & $\dot{P}$ & PF & $P_{\rm orb}$ & $M_{2}$ \\
       & $[\ergs]$.    & $[{\rm s}]$& ${\rm [10^{-10}\,s\,s^{-1}]}$ & \% & $[{\rm d}]$ & $[M_\odot]$ 
        \vspace{0.1cm} \\
    \hline 
    M82~X-2 & $1.8\times 10^{40}$ & $1.37$ & {$\sim 2$} & $>20$ & $2.52$ & $>5.2$ \\
    NGC~7793~P13 & $5\times 10^{39}$ & $0.42$ & {$\sim 0.35$} & $\sim20$ & $64$ & $18-23$ \\
    NGC~5907~X-1 & $2\times10^{41}$ & $1.42$ & 115 & $\sim15$ & $5.3$ & ? \\
                   &                  & $1.13$ & 47 & $\sim15$ &       &   \\
    NGC~300~X-1 & $4.7\times10^{39}$ & $125$ & $1.4\times 10^5$ & ?        & ? & - \\
                  &                    & $31.5$ & $5.5\times 10^3$ & $\sim90$ &  &  \\
                  &                    & $20$   & $1.7\times 10^3$ & $\sim90$ &  &  \\
    M51~X-7 & $7\times10^{39}$ & $2.8$ & $1.6-9.4$ & $5-20$ & $\sim2$ & ? \\
    NGC~1313~X-2 & $2\times 10^{40}$ & $1.5$ & ? & $5-6.5$ & ? & $<12$ \\    
    \hline\hline
  \end{tabular}
\end{table}



\subsection{\red{3.3\,} Apperiodic variability or flickering XRPs}

\begin{figure}
\centering 
\includegraphics[width=12.cm, angle =0]{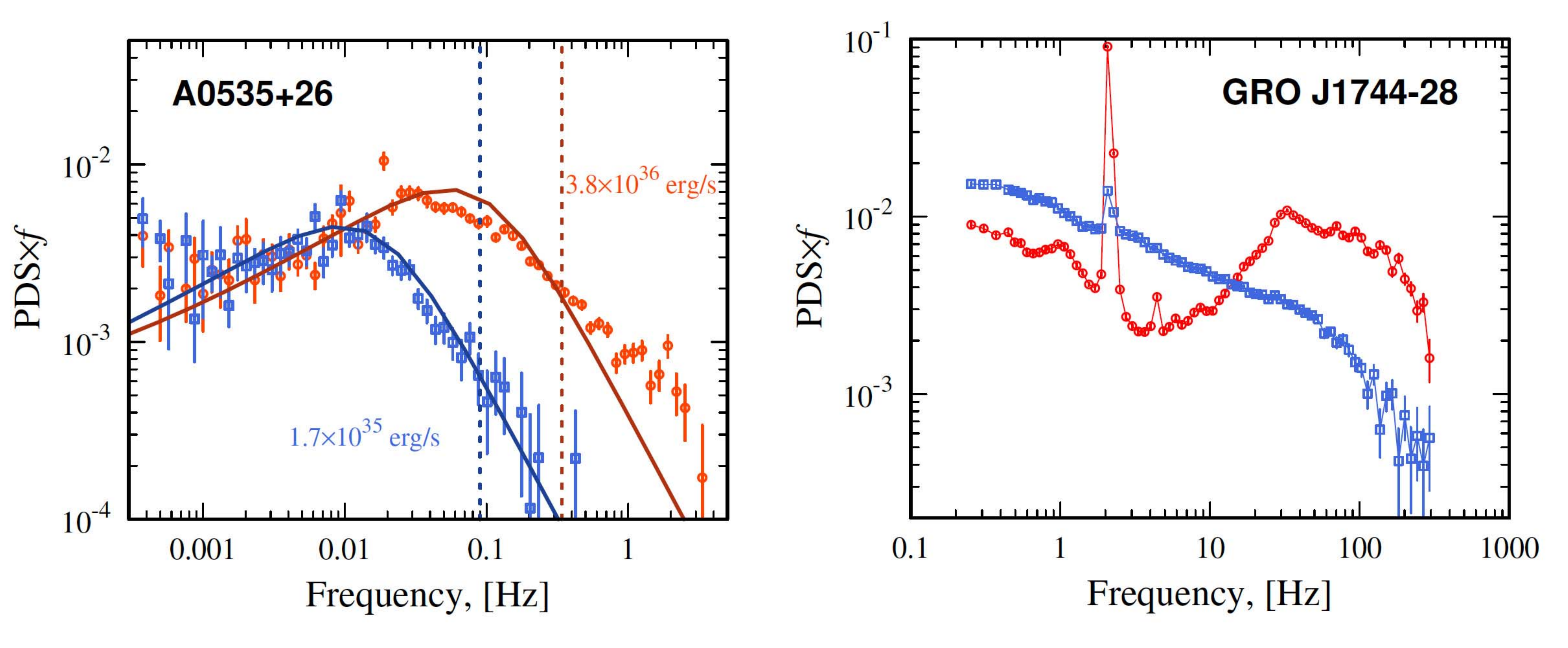}
\caption{
{\it Left:} The PDS of the X-ray flux fluctuations in two luminosity states of the X-ray transient A~0535+26 during its outburst:
$L_1\simeq 1.7\times 10^{35}\,\ergs$ (blue squares) and $L_2 \simeq 3.8\times 10^{36}\,\ergs$ (red circles).
The power spectra exhibit a break, whose frequency varies with the accretion luminosity: the higher the luminosity, the higher the break frequency. 
Blue and red solid lines represent the theoretical PDS expected from propagating fluctuations of the mass accretion rate in the disc \citep{2019MNRAS.486.4061M}.
Blue and red dashed lines represent Keplerian frequencies at the inner disc radii, at  $L_1\simeq 1.7\times 10^{35}\,\ergs$ and $L_2 \simeq 3.8\times 10^{36}\,\ergs$, respectively.
{\it Right:} The PDS of X-ray flux fluctuations in two states of the X-ray transient GRO~J1744$-$28: during the giant outburst at $L_1=8\times 10^{37}\,\ergs$ (red circles) and during the mini-outburst at $L_2=9\times 10^{37}\,\ergs$ (blue squares).
It is remarkable that the PDS shows dramatic changes and at a certain conditions, a strong high frequency aperiodic variability arises at $f>10\,{\rm Hz}$.
The nature of strong additional variability at high frequency is still under debates. 
Different authors relate it to various phenomena from the photon bubble instability in accretion column \citep{1996ApJ...469L.119K} to the rapidly propagating mass accretion rate fluctuation in the inner radiation pressure dominated parts of a disc \citep{2019A&A...626A.106M}.
The plot represents data published in \citep{2019A&A...626A.106M}.
}
\label{pic:PDS_2}
\end{figure}

\begin{figure}
\centering 
\includegraphics[width=8.cm, angle =0]{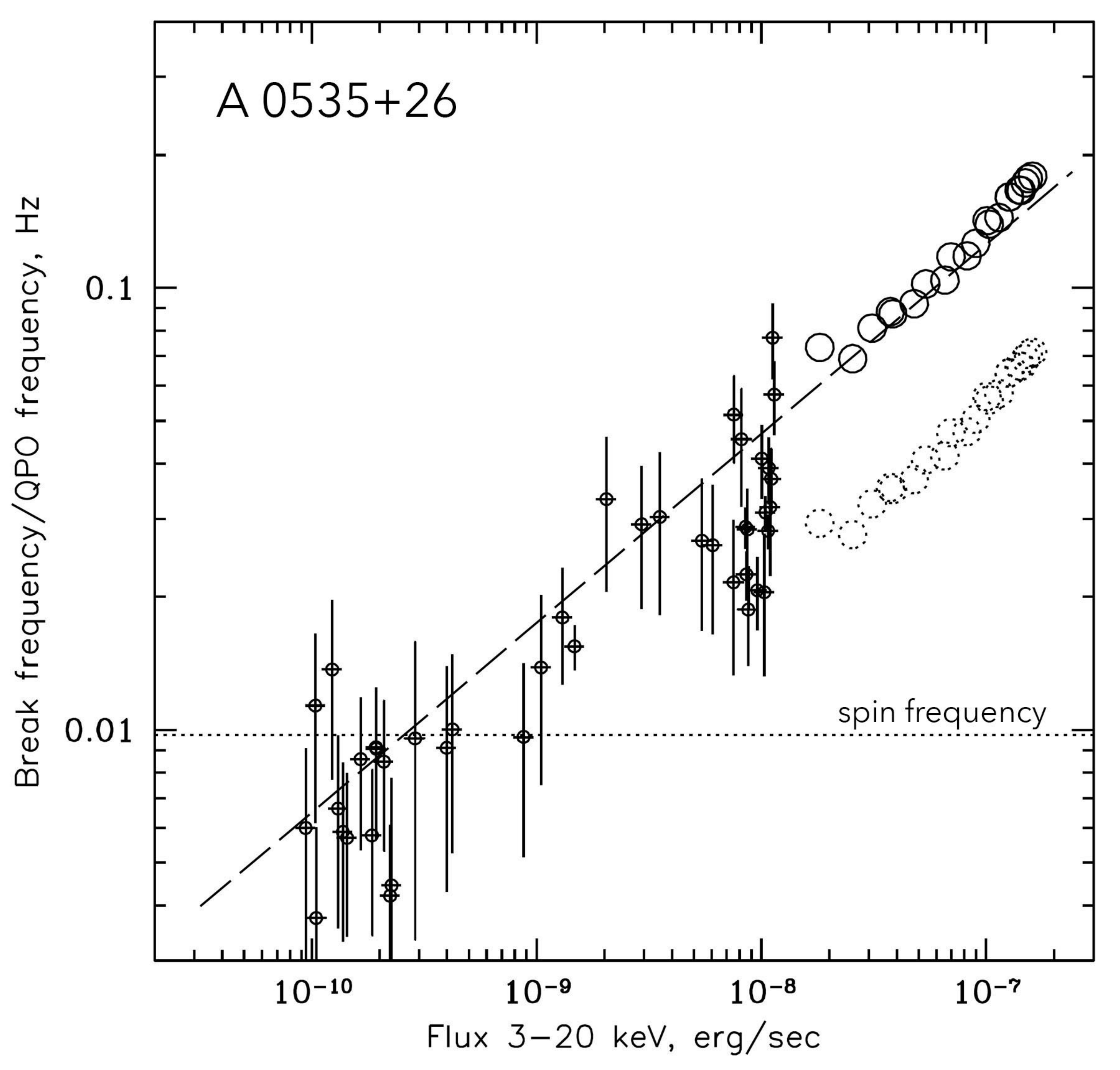}
\caption{
Dependence of the break frequency in the PDS of A~0535+26 on the observed flux (circles with error bars). 
The dependence of the QPO frequency observed in A~0535+26 is shown by open dotted circles. 
The solid circles show the results of the QPO frequency multiplied by $2.5$ to match the break frequency trend. 
The dashed line shows frequency dependence $f\propto L^{3/7}$ (similar to the dependence of the expected Keplerian frequency at the inner disc radius). 
The dotted line shows the observed spin frequency of the pulsar.
\red{The plot represents data published in \citep{2009A&A...507.1211R}.}
}
\label{pic:PDS_Mike}
\end{figure}

XRPs show strong aperiodic variability of X-ray flux over a very broad frequency range similar (modulo mass scaling) to what is detected in accreting black holes (BHs, see, e.g. \citep{2000A&A...363.1013R}) and active galactic nuclei (AGN, see, e.g. \citep{2004MNRAS.348..783M}).
Strong aperiodic variability can be considered as a typical feature of the accretion process and sometimes used as an argument confirming or ruling out accretion in X-ray sources \citep{2020A&A...643A.173D}.
At a short time scale, aperiodic variability in XRPs sometimes extends down to milliseconds. 
In addition to the peak related to the coherent pulsations, the observed power density spectrum (PDS) typically includes a broad continuum component, which can be approximated by a broken (or double broken) power-law, and narrow features that are classified as quasi-periodic oscillations (QPOs). 
Both components in accreting compact objects, including XRPs, are known to evolve with the observed luminosity (see  \citep{2009A&A...507.1211R} and Fig.\,\ref{pic:PDS_2}).
A typical feature of PDS in XRP is a high-frequency break ($f\gtrsim 10^{-2}\,{\rm Hz}$).
The break frequency can depend on the accretion luminosity.
In some sources, the break frequency depends on luminosity as $f\propto L^{3/7}$, which is similar to the expected dependence of the Keplerian frequency at the inner disc radius on the luminosity (see Fig.\,\ref{pic:PDS_Mike}).

Fluctuations of the X-ray flux are largely caused by fluctuations of the mass accretion rate onto the NS surface.
The latter carries information about geometry and physical condition of the accretion flow: wind/disc accretion, inner and outer radii of accretion disc, gas or radiation pressure dominated accretion flow, development of instabilities in the accretion flow, etc.
Fluctuations of the flux, however, do not exactly replicate fluctuations of the mass accretion rate due to several reasons: 
\begin{itemize}[noitemsep,topsep=0pt]
{\setlength{\itemsep}{0pt}
\item superposition between fluctuating mass accretion rate and NS rotation can strongly distort the PDS \citep{1997ApJ...476..267L};
\item fluctuations of the X-ray flux can be affected by variations of the beam pattern formed in the emitting regions \citep{1996ApJ...469L.119K};
\item fluctuations of the X-ray flux can be disturbed by the accretion flow in between the inner disc radius and NS surface \citep{2019MNRAS.484..687M}. 
}
\end{itemize}

\subsection{\red{3.4\,} Energy spectrum}

The effective temperature in XRPs can be estimated from the accretion luminosity $L$ and the expected area $S$ of the emitting region at the NS surface:
\footnote{
We define $Q=Q_{x}10^{x}$ in cgs units if not mentioned otherwise.
}
\be\label{eq:T_eff}
T_{\rm eff}=\left(\frac{L}{\sigma_{\rm SB}S}\right)^{1/4}
\simeq 5.6\,L_{37}^{1/4}S_{10}^{-1/4}\,\,{\rm keV},
\ee 
where $\sigma_{\rm SB}$ is the Stefan–Boltzmann constant.
The estimation (\ref{eq:T_eff}) gives $T_{\rm eff}\sim 1\,{\rm keV}$ for $L_{37}=1$ even if one takes the total area to a NS.
It explains why the most of radiation in XRPs is emitted in the X-ray band. 
However, one has to keep in mind that the energy spectra of XRPs can not be approximated with a simple blackbody. 

\begin{figure}
\centering 
\includegraphics[width=8.cm, angle =0]{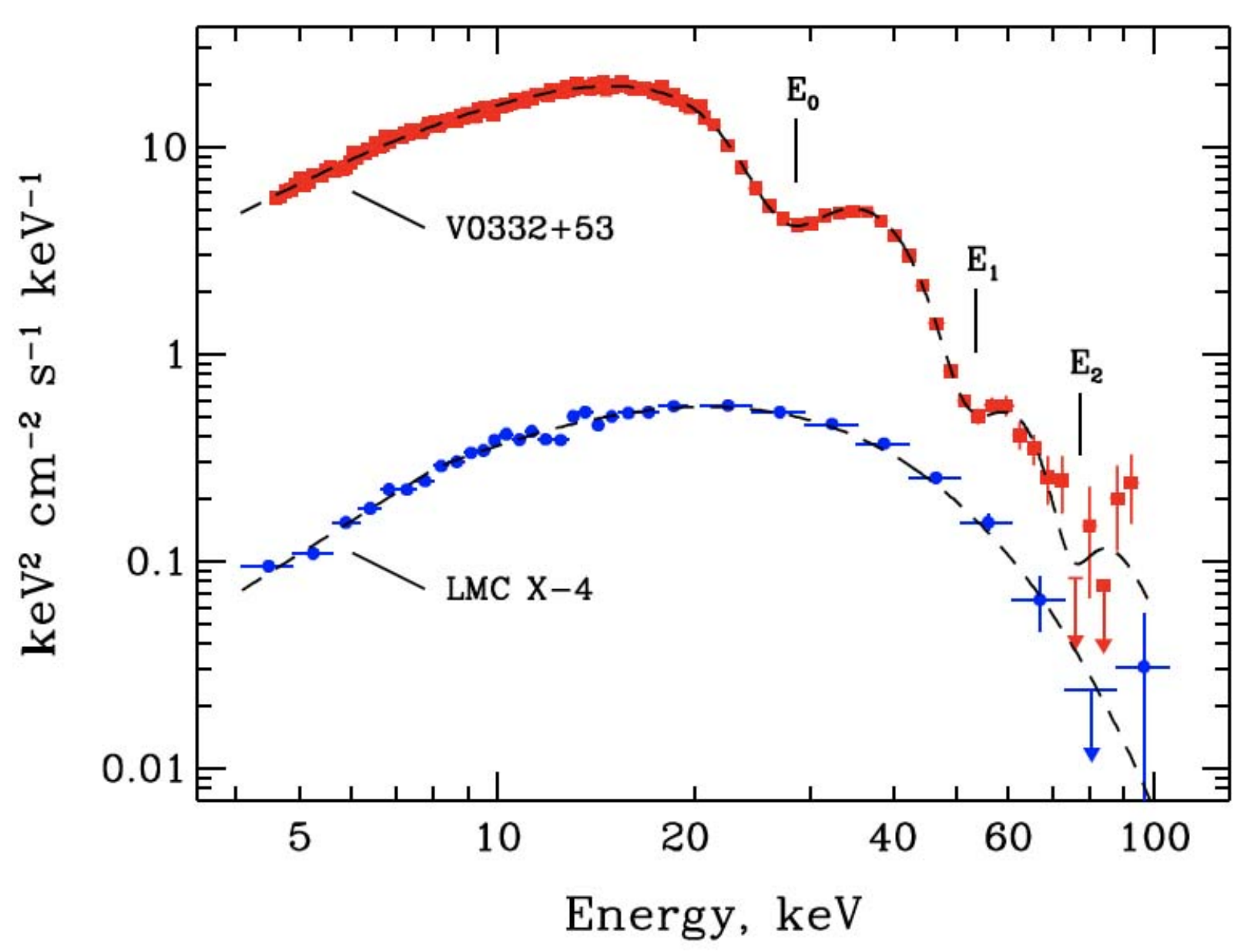} 
\caption{
Energy spectra of two XRPs: V~0332+53 (red squares) with three harmonics of the cyclotron absorption line and LMC X-4 (blue circles) without cyclotron features in the spectrum.
From \citep{2015A&ARv..23....2W}.
}
\label{pic:XRP_sp}
\end{figure}

\begin{figure}
\centering 
\includegraphics[width=12.cm, angle =0]{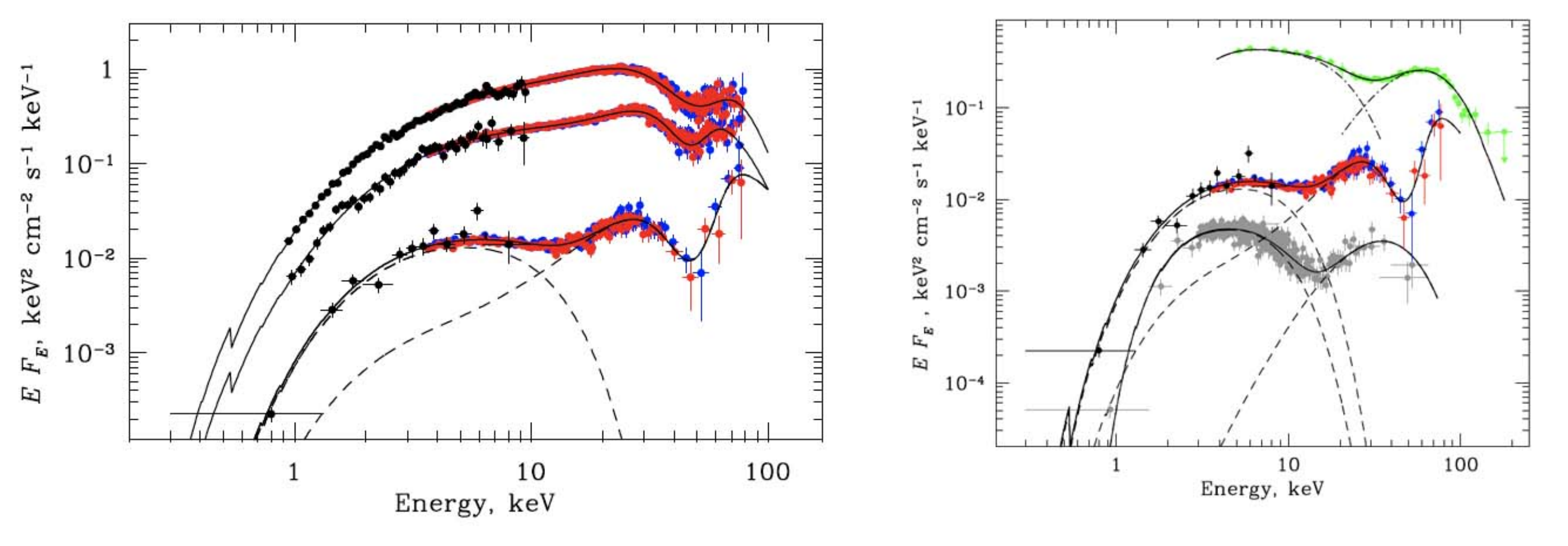} 
\caption{
Variations of the X-ray energy spectrum of A~0535+262 with accretion luminosity {\it(left)} and spectra of three XRPs at low mass accretion rates {\it(right)}: GX~304$-$1 (grey points), A~0535+262 (blue and red points), and X~Persei (green points)
\citep{2019MNRAS.483L.144T,2019MNRAS.487L..30T,2012A&A...540L...1D}.
It appears that the broadband energy spectrum of some XRPs experiences dramatic changes at low mass accretion rates, which indicates changes in the physical structure of the emitting regions at the NS surface.
Note, however, that such spectral changes are not a general feature of XRPs, and several sources do not show dramatic spectral changes at low mass accretion rates (see, for example, Fig.\,\ref{pic:sp_MAXI_J0903}). 
}
\label{pic:low_state_sp}
\end{figure}

\begin{figure}
\centering 
\includegraphics[width=8.cm, angle =0]{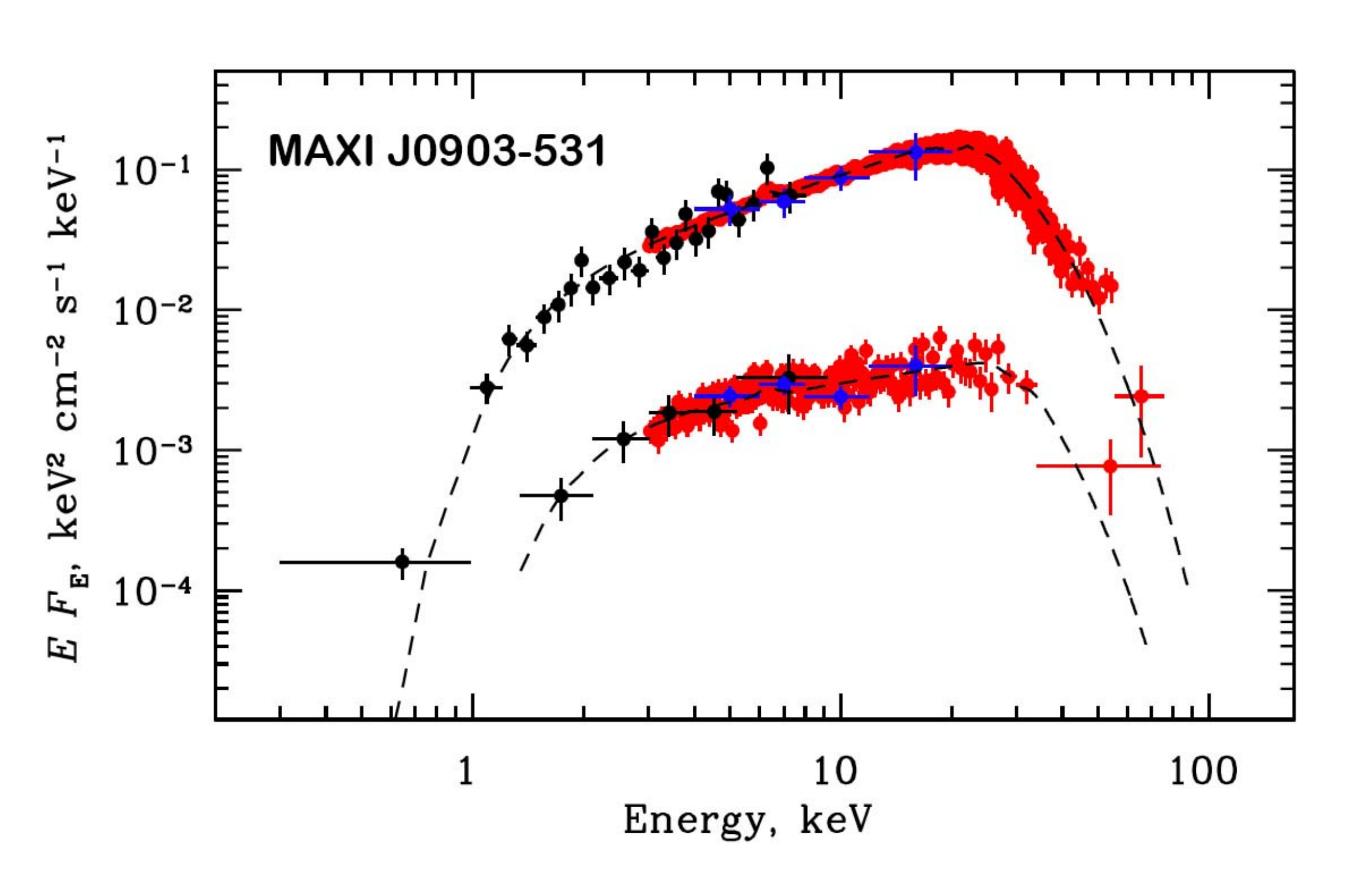}
\caption{
Unfolded energy spectra of transient XRP MAXI~J0903$-$531 at different states with luminosity varying by a factor of $\sim 30$.
Remarkably, the energy spectrum conserves its shape over the wide range of accretion luminosity.
From \citep{2022A&A...661A..45T}.
}
\label{pic:sp_MAXI_J0903}
\end{figure}

Spectra of all bright XRPs {($L\gtrsim 10^{35}\,\ergs$)} are quite similar to each other and can be roughly described with a cutoff power-law continuum 
\citep{1989PASJ...41....1N}.
In a few dozens of sources, on top of continuum spectra, one can detect cyclotron absorption scattering feature (for the details see a review  \citep{2019A&A...622A..61S} and Tab.\,\ref{table:cycl}). 
In some sources, the scattering feature can be accompanied by several (up to five) higher harmonics (see Fig.\,\ref{pic:XRP_sp}).
Because the accretion flow reaches the NS surface with a velocity comparable to the speed of light and temperature near the NS surface is about a few keV (or higher), bulk and thermal Comptonization play a key role in the formation of non-thermal X-ray emission and define the observed shape of its spectrum \citep{2007ApJ...654..435B}.  
Recently, it has been discovered that the spectra of some XRPs experience dramatic spectral changes when the mass accretion rate drops below $\sim 10^{35}\,\ergs$:
instead of a classical single-component shape, the spectra show two distinct components with the one peaked at a few keV, and the other peaked at a few tens of keV (see Fig.\,\ref{pic:low_state_sp}, 
\citep{2019MNRAS.483L.144T,2019MNRAS.487L..30T,2021ApJ...912...17L}).
In some cases, the cyclotron absorption scattering feature is detected on top of the high-energy component \citep{2019MNRAS.487L..30T}. 
However, these spectral changes at low luminosity are not a general trend for all XRPs, and some sources conserve the classical spectral shape down to very low mass accretion rates (see Fig.\,\ref{pic:sp_MAXI_J0903}). \red{At the same time, if an XRP demonstrates a two-component spectrum, it can be misinterpreted as a classical pulsar spectrum with CRSF. This can be a reason why some sources with known CRSF don't show higher harmonics of the line. This effect potentially can significantly reduce number of XRPs with known cyclotron features.}

\begin{table}
\centering
\caption{
Cyclotron line sources discovered after \citep{2019A&A...622A..61S}. 
}
\tabcolsep=0.15cm
\label{table:cycl}
\begin{tabular}{l l c l c l l l}
 \hline\hline
  \vspace{0.1cm}
System                   & Type  & P$_{\rm spin}$ & P$_{\rm orb}$  & Ecl. & E$_{\rm cyc}$     & Instr. of            & Reference  \\
                &                     &  (s)    &  (days)        &      & (keV)         & detection            &             \\
 \hline     
\red{Swift~J0243.6+6124}               & Be trans.        & \red{9.8}           & \red{28}     &  no      & \red{120-146}               & \textit{\red{Insight-HXMT}}              & \citep{2022ApJ...933L...3K}  \\ 
\red{Cen X-3}               & HMXB        & \red{4.8}           & \red{2.09}     &  yes      & \red{28,\,47$^a$}               & \textit{\red{Insight-HXMT}}              & \citep{2023MNRAS.519.5402Y}  \\ 
Swift~J1626.6$-$5156         & Be trans.        & 15.3           & 132.9   &  no      & 5,9,13,17               & \textit{NuSTAR}              & \citep{2021ApJ...915L..27M}  \\  
 GRO~J1750$-$27               & Be trans.        & 4.45           & 29.8     &  no      & 43               & \textit{NuSTAR}              & \citep{2022ATel15241....1M}  \\  
\red{Swift~J1808.4$-$1754}               & Be trans.        & \red{910}           & \red{$-$}     &  no      & \red{21}               & \textit{\red{NuSTAR}}              & \citep{2022MNRAS.514.2707S}  \\ 
XTE~J1858+034                & SyB              & 220           & 81?,380?     &  no      & 48                     & \textit{NuSTAR}              & \citep{2021ApJ...909..154T,2021ApJ...909..153M}  \\  
 GRO~J2058+42                 & Be trans.        & 195           & 110     &  no      & 10,20,30               & \textit{NuSTAR}              & \citep{2019ApJ...883L..11M}  \\  
 \hline\hline
\end{tabular}
\red{$^a$ Second harmonic. The fundamental one at $\sim28$~keV was known before.}
\end{table}

\begin{figure}
\centering 
\includegraphics[width=12.cm, angle =0]{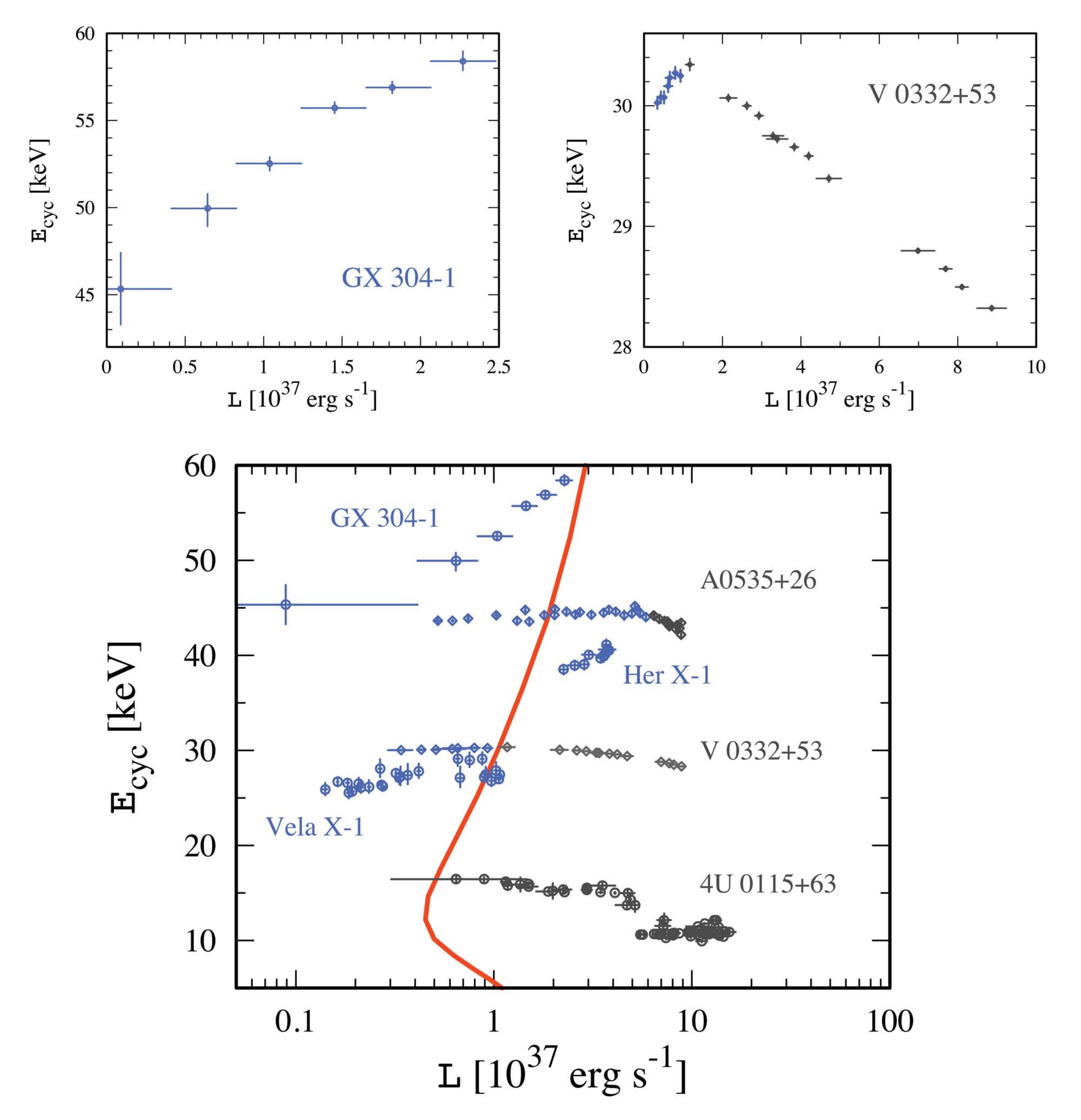} 
\caption{
{\it Top left}: Positive correlation of cyclotron line centroid energy with accretion luminosity in XRP GX~304$-$1. The plot reproduces results reported in  \citep{2017MNRAS.466.2752R}.\\
{\it Top right}: Negative correlation of the cyclotron energy with the luminosity in V~0332+53. 
The plot reproduces results reported in \citep{2017MNRAS.466.2143D}.
Note that XRPs V~0332+53 and A~0535+26 show both positive and negative correlations, which can be considered as an evidence of the emitting region geometry dependence on the mass accretion rate, i.e. transition through the critical luminosity. \\
{\it Bottom}: Cyclotron energy variations with the accretion luminosity observed in a set of variable XRPs. 
Blue circles and diamonds represent the data for the subcritical XRPs (i.e., demonstrating the positive correlation), black circles and diamonds show pulsars with supercritical behaviour (i.e., the negative correlation).
For Vela~X-1, the energy of the first harmonic divided by two is used.
We used the data points used in \citep{2015MNRAS.447.1847M} for 4U~0115+63, Vela~X-1, Her~X-1, the data reported for GX~304$-$1 in \citep{2017MNRAS.466.2752R}, for V~0332+53 in \citep{2017MNRAS.466.2143D}, and for XRP A~0535+26 in \citep{2021ApJ...917L..38K}.
Red solid line represents predictions for the critical luminosity calculated according to \citep{2015MNRAS.447.1847M}.
}
\label{pic:cyc_lines_variations}
\end{figure}

Cyclotron scattering features appear in spectra of XRPs due to the resonant Compton scattering in a strong magnetic field 
\citep{1979PhRvD..19.2868H,1986ApJ...309..362D} 
in a line forming region, which is located in close proximity to the NS surface.
Cyclotron lines were predicted in the 1970s \citep{1974A&A....36..379G} and discovered shortly after that \citep{1978ApJ...219L.105T}.
The fundamental cyclotron line (corresponding to the electron transition between the first excited and the ground Landau levels) appears at 
\be\label{eq:Ecyc_vs_B}
E_{\rm cyc}\simeq \frac{\hslash e B}{m_{\rm e} c}
\approx 11.6\,\left(\frac{B}{10^{12}\,{\rm G}}\right)\,{\rm keV}.
\ee 
Because of a simple relation between the cyclotron energy and magnetic field strength, the detection of the cyclotron scattering feature in the source spectrum can be used as a direct probe of the NS magnetic field strength. 
Since the discovery of the cyclotron lines in XRPs \citep{1978ApJ...219L.105T}, it has been found that the centroid energy of the cyclotron features could vary with the accretion luminosity, pulsation phase and on long (years) time scales.
In particular, the long-timescale reduction and evolution of the line centroid energy were found in Her X-1 \citep{2017A&A...606L..13S,2019JHEAp..23...29X}.

The variations of cyclotron line energy with accretion luminosity has been detected in a number of XRPs including
V~0332+53 \citep{2006MNRAS.371...19T,2010MNRAS.401.1628T}, 
Her~X-1 
\citep{2007A&A...465L..25S,2014A&A...572A.119S},
A~0535+26 \citep{2007A&A...465L..21C}, 
Vela~X-1 \citep{2014ApJ...780..133F,2014MNRAS.440.1114W}, 
GX~304$-$1 \citep{2012A&A...542L..28K,2017MNRAS.466.2752R}, 
Cep~X-4 \citep{2017A&A...601A.126V}, 
4U~0115+63 
\citep{2004ApJ...610..390M,2007AstL...33..368T},
GRO~J1008$-$57 \citep{2021ApJ...919...33C}, 
and 2S~1553-542 \citep{2022ApJ...927..194M}.
Moreover, sources with relatively low mass accretion rates show a positive correlation between the line centroid energy and luminosity, while sources with relatively high mass accretion rates show a negative correlation (see Fig.\,\ref{pic:cyc_lines_variations}). 
In two sources - V0332+53 \citep{2017MNRAS.466.2143D} and A~0535+26 \citep{2021ApJ...917L..38K} - both correlations were observed with the critical luminosity dividing positive and negative dependencies robustly measured.
In the same two sources, it was also shown that the relation between the line centroid energy and accretion luminosity is not the same during the raising and fading parts of the outburst, i.e. the same source can have different line centroid energies at the same apparent luminosity \citep{2016MNRAS.460L..99C,2021ApJ...917L..38K}.

Cyclotron lines were shown to be variable with the pulse phase in several XRPs 
\citep{2015MNRAS.448.2175L,2017MNRAS.466..593L}.
Recently, it has also been found that the cyclotron line can be pulse-phase-transient and appear only in a narrow range of the pulse phases 
\citep{2019ApJ...883L..11M}. 
The variability of the cyclotron line energy in the spectra of XRPs is considered to be related to the geometry of accretion flow in close proximity to the NS surface.
The geometry of the emitting region, in turn, is related to the mass accretion rate and magnetic field strength and structure \citep{1976MNRAS.175..395B,2015MNRAS.454.2539M}.

\smallskip

In addition to the main continuum components discussed above, some XRPs demonstrate the appearance of the soft emission component (aka soft excess) and fluorescent iron line in their spectra. 
The former is often modelled as a blackbody with a temperature of about 0.1 keV. 
It was shown by \citep{2004ApJ...614..881H} that reprocessing of hard X-rays from the NS by the inner region of the accretion disc is the most probable process that can explain the soft excess in the brightest pulsars (with $L>10^{38}$~\lum). 
In less bright XRPs, the soft excess may be explained by diffuse gas or thermal emission from the NS surface.

The fluorescent K$\alpha$ iron line is another emission component frequently detected in the XRPs spectra and can be utilized to study the spatial distribution and ionization state of the cold matter around the X-ray sources 
(see, e.g., \cite{1974A&A....31..249B,1985SSRv...40..317I, 2010LNP...794...17G,2019ApJS..243...29A}).
In accreting XRPs, the fluorescent emission may be produced at any point from the surface of the massive donor star itself or stellar wind/accretion disc, down to the Alfv\'en surface and accretion stream/column (see, e.g., \cite{1985SSRv...40..317I}).

Some XRPs exhibit variability of the equivalent width of the iron line with the rotational phase of the neutron star. 
For instance, the pulsating iron line was detected in LMC X-4 \citep{2017AstL...43..175S}, 
Cen~X-3 \citep{1993ApJ...408..656D}, 
GX~301$-$2 \citep{2018MNRAS.480.4746L, 2020MNRAS.491.4802Z}, 
Her~X-1 \citep{1994ApJ...437..449C}, 
and 4U~1538$-$522 \citep{2014ApJ...792...14H}.  
Most recently variability of iron line equivalent width and iron K-edge at $\sim$7.1 keV with NS spin and orbital period were discovered in the transient XRP V~0332+53 \citep{2010arXiv1002.1898T,2021MNRAS.506.2156B}.

\smallskip\smallskip

X-ray energy spectra of pulsating ULXs are softer than the spectra of normal XRPs. 
They do not show a significant difference with the spectra of ULXs, where pulsations were not detected (\citep{2018ApJ...856..128W} and \citep{2021AstBu..76....6F} for review).
Recently, the discovery of a cyclotron scattering feature at $4.5\,{\rm keV}$ in the spectrum of ULX-8 in galaxy M51 \citep{2018NatAs...2..312B} (note, that pulsations were not detected in this particular ULX so far)
and potential cyclotron scattering feature around $13\,{\rm keV}$ in pulsating ULX-1 in the galaxy NGC300 \citep{2018ApJ...857L...3W} were reported.

\smallskip\smallskip

Due to complexity of the problem of the emission formation in XRPs, no self-consistent physical model able to describe the observed spectra from these objects in a broad range of mass accretion rates was proposed yet.
Therefore, in the vast majority of the observational studies the easily-parameterised phenomenological models are used in order to characterise spectral shape and to obtain some physical parameters in the emission regions of a NS. 
The list of the most commonly applied models available in the spectral fitting package {\sc xspec} \citep{Arnaud1996} is presented in Table~\ref{tab:models}. 
Detailed description of these models can be found in the {\sc xspec} manual.\footnote{\url{https://heasarc.gsfc.nasa.gov/xanadu/xspec/manual/Models.html}} \red{It worth mentioning however, that physical interpretation of the best-fit parameters obtained from such phenomenological models should be taken with great caution. Typical example of such over-interpretation is a discussion of the emission region properties obtained from the physical black body component, added to compensate residuals in the fit with absolutely nonphysical power law component.}

\begin{table}
\centering
\caption{
List of the phenomenological models utilised for the approximation of different components of emission from XRPs.
}
\tabcolsep=0.3cm
  \label{tab:models}
  \begin{tabular}{p{3.5cm} p{2cm} p{4.5cm}}
    \hline\hline
    \vspace{0.05cm}&\vspace{0.05cm}&\vspace{0.05cm}\\
  Model & {\sc xspec} notation  & Description  \\
   \vspace{0.01cm} \\
    \hline 
Power law with high energy exponential cutoff & {\sc cutoffpl}  & Additive model for the continuum emission from XRPs. \\

A high energy cutoff & {\sc highecut}  & Multiplicative model for the continuum emission. Is used in combination with a power law component. \\

A blackbody spectrum & {\sc bbodyrad}  & Additive blackbody component with normalization proportional to the surface area. Is used to account for the soft excess. \\

Cyclotron absorption line & {\sc cyclabs}  & Multiplicative model for the cyclotron absorption line component.\\

Gaussian absorption line & {\sc gabs}  & Another version of a multiplicative model for the cyclotron absorption line component.\\

Gaussian line profile & {\sc gau}  & Additive model component in the form of gaussian line profile. Is used, e.g. to approximate iron fluorescent line emission. \\

A photoelectric absorption & {\sc tbabs, phabs}  & Multiplicative model used to account for the photoelectric absorption. \\

    \hline\hline
  \end{tabular}
\end{table}

\subsection{\red{3.5\, Polarization properties of XRPs}}

\red{
Polarization can be considered as the most direct way to probe geometrical configuration of highly-magnetized NSs: the inclination of their rotation axis, the angle between the rotation axis and magnetic field axis, possible asymmetries of dipole magnetic field configuration, presence of a non-dipole component in the magnetic field structure, etc. }

\red{
Until recently, emission of XRPs was expected to be strongly polarized (up to 80\%) with specific behaviour of the polarization degree over the pulse phase expected for different accretion geometries (see, e.g., \citep{1988ApJ...324.1056M,2021MNRAS.501..109C}). 
The reason for that is a strong dependence of cross section of processes of interaction between radiation and matter (Compton scattering, free-free magnetic absorption/emission and cyclotron scattering/absorption, in the first place) on photon energy and momentum direction in respect to magnetic field (see \citep{2006RPPh...69.2631H} for review).
}

\begin{figure}
\centering 
\includegraphics[width=6.cm, angle =0]{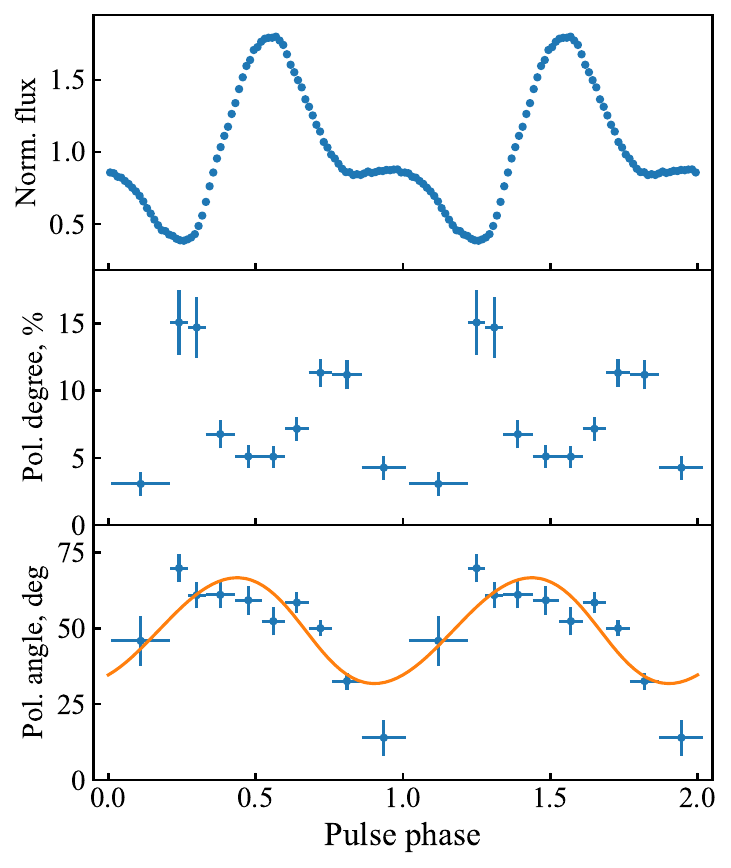} 
\caption{
\red{{\it Top}: Dependence of the normalized flux of Cen X-3 in the 2--8 keV energy band on the pulse phase. {\it Middle}: Dependence of the polarization degree on phase from the spectro-polarimetric analysis. {\it Bottom}: Dependence of the polarization angle on the pulse phase. The orange line corresponds to the best-fit rotating vector model (RVM). Adopted from \citep{2022ApJ...941L..14T}.}
}
\label{pic:cenx3_pol}
\end{figure}

\red{
Unfortunately, sensitive enough polarimeters able to operate in the X-ray band were not available for astronomers until recently. 
This situation has changed with the launch of the {\it Imaging X-ray Polarimeter Explorer} \citep[{\it IXPE},][]{Weisskopf2022} on 2021 December 9. 
Already first observations of XRPs performed with this instrument led to the completely unexpected results. 
Namely, it was found that even bright XRPs (with luminosities exceeding $10^{37}$~\lum) show polarization degree (PD) well below 20\% even in the phase-resolved data (Fig.\,\ref{pic:cenx3_pol} and \citep{2022NatAs...6.1433D,2022ApJ...940...70M,2022ApJ...941L..14T}). 
To some extent, this result can be explained by the structure of the atmosphere of a NS, where the upper layers are expected to the hotter than the underling ones due to the accretion process.
Nevertheless, the problem about the reasons for the low PD remains open and awaits a solution.
Consideration of low polarisation degree requires analyses of additional to the intrinsic polarization from the hot spot scenarios and mechanisms influencing X-ray polarisation. 
In particular, polarisation on the level of percents can be due to 
(i) X-ray reflection from the atmosphere of a NS, 
(ii) reflection/reprocessing of photons by the accretion flow covering magnetosphere of a NS, 
(iii) X-ray reflection from accretion disc in a system, 
(iv) scattering by the stellar wind, and
(v) reflection of X-ray by the companion star (see discussion and references in \citep{2022ApJ...941L..14T}). 
}

\red{
Parameters of NS rotation in XRPs can be obtained from the variations of the polarisation angle during the pulse period on the base of the rotating vector model (RVM, see e.g. \citep{1969ApL.....3..225R,2020A&A...641A.166P}), which is a standard method in determination of NS rotation geometry in radio astronomy for years and applicable for decoding the data on X-ray polarisation.
RVM, however, assumes the dipole configuration of NS magnetic field in XRP, which is not necessarily a case and possibility on non-dipole field structure was already discussed in literature \citep{2017Sci...355..817I,2017A&A...605A..39T,2022MNRAS.515..571M}.
Nonetheless, applying RVM to the data on a few XRPs obtained by {\it IXPE}, it was possible to get the position angles of NS spin axis, NS inclinations and a magnetic obliquity (the angle between the spin and magnetic field axis, see details in \citep{2022NatAs...6.1433D,2022ApJ...941L..14T}).
}

\subsection{\red{3.6\,} Optical companions in XRPs}

\begin{figure}
\centering 
\includegraphics[width=7.cm, angle =0]{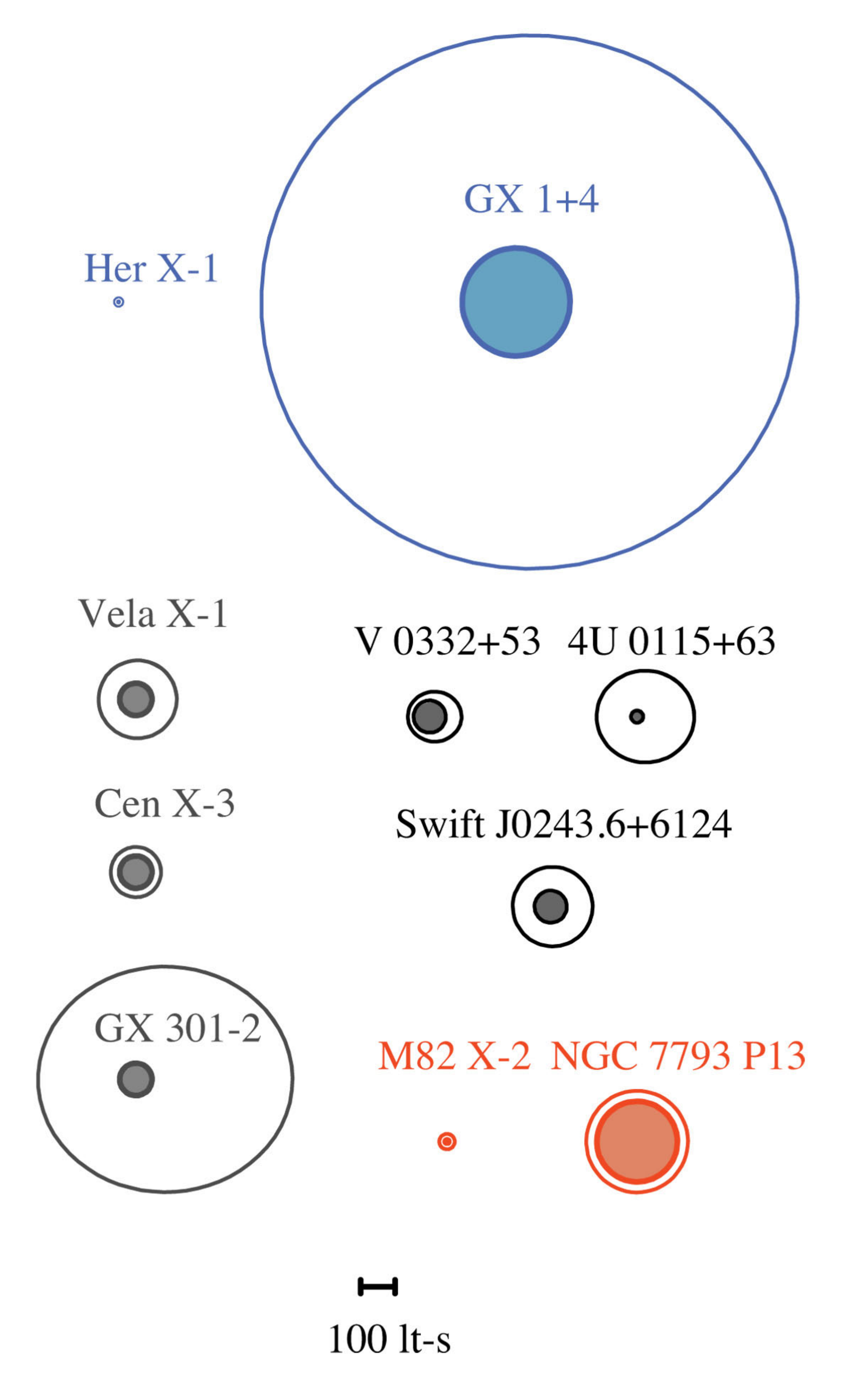} 
\caption{
The schematic view of the geometry of some XRPs, including LMXBs (Her X-1 and GX 1+4), 
wind-fed sources (Vela X-1, Cen X-3 and GX 301$-$2), 
BeXRBs (V~0332+53, 4U~0115+63 and Swift~J0243.6+6124)
and 
ULX pulsars (M82~X-2 and NGC~7793~P13).
The black section illustrates the linear scale corresponding to 100 light seconds. 
The geometry of XRPs is reproduced from the orbital parameters reported by \textit{Fermi}/GBM Accreting Pulsars Program (\url{http://gammaray.nsstc.nasa.gov/gbm/science/pulsars/}) and review \citep{2015A&ARv..23....2W}.
Note that due to the unknown orbital inclination of the binary systems, the represented sizes of the orbits can be underestimated.
}
\label{pic:orbit}
\end{figure}

Observational appearances and properties of XRPs, like orbital variability and parameters of a binary system, its age, NS magnetic field strength and mechanism of the mass transfer, are related to the type of a companion star and orbital separation in a system.
Schematic view on geometry (including companion star size, orbital separation and eccentricity) of some XRPs of different types is represented in Fig.\,\ref{pic:orbit}.
Accreting highly magnetised NSs are hosted both in HMXBs, where companion stars are young and have masses $\gtrsim 8\,M_\odot$ (e.g. V~0332+53, 4U~0115+63, Vela~X-1), 
and in low-mass X-ray binaries (LMXBs), where the companions are older and have masses $\lesssim 2\,M_\odot$ (e.g. GRO~J1744$-$28, Her X-1, GX~1+4). 

In the HMXBs, it is possible to distinguish two sub-classes of companions: 
massive early-type stars (Cen~X-3, Vela~X-1, etc.),
and Be-stars (BeXRBs, e.g. V~0332+53, A0535+26, GRO~J1008$-$57, etc.). 
In contrast to white dwarfs or black holes binaries, the contribution of the accretion disc to the total optical emission of a HMXB is negligibly small.
In the non-Be HMXBs, the companions are typically a massive giant with mass $\gtrsim 20\,M_\odot$ and orbital periods $P_{\rm orb}\sim 1.5 - 10\,{\rm days}$. 
Such companions are near to filling their Roche lobe, and the mass loss occurs either via atmospheric Roche lobe overflow or via the stellar wind.
These systems show eclipses, and the orbits are generally circular.
BeXRBs host a Be star of mass $\sim 10-20\,M_\odot$ lying deep inside its Roche lobe and demonstrating emission lines in its spectrum, originating from the circumstellar disc arising as a result of a rapid rotation of the star. 
Orbits in Be systems have long periods ($P_{\rm orb}\gtrsim 15\,{\rm days}$) and large eccentricity, resulting in strong flux variability in BeXRBs.
Several BeXRBs belong to the small class of persistent sources and are characterised by circular orbits and relatively low luminosity (e.g. X~Persei, RX~J0440.9+4431). 
The detailed review of BeXRBs properties can be found in \citep{2011Ap&SS.332....1R}.
There are few HMXB XRPs associated to supernova remnants \citep{2013ApJ...779..171H,2019MNRAS.490.5494M}, which indicate extreme youth ($\lesssim {\rm few}\times 10^4\,{\rm yrs}$) of NSs in these particular binaries. 

XRPs in the LMXBs are very rare, that is related to the typical ages of LMXBs ($\gtrsim 10^9\,{\rm years}$) and the fact that strong magnetic field tends to decay with time due to Omhic processes, Hall evolution and accretion onto the NS surface.
There are two types of LMXBs {hosting XRPs:
(i) {\it Type I} LMXBs (age $\lesssim 10^9\,{\rm years}$), where companions are represented by moderate-age main-sequence star or an evolved companion and orbital periods $P_{\rm orb}\simeq 0.2 - 10\,{\rm days}$ (example: Her~X-1).
(ii) {\it Type II} LMXBs are represented by older systems (age $\gtrsim 5\times 10^9\,{\rm years}$) with companions represented by low-luminous main-sequence or dwarf star and orbital periods of the order of a few hours (e.g. 4U~1626$-$67). }

Optical companions in pulsating ULXs are poorly studied because ULXs, in general, are extragalactic sources, and their multiwavelength counterparts are often faint.
It was reported about blue supergiant as a donor star pulsating ULX NGC~7793~P13 \citep{2011AN....332..367M} and about red supergiant in NGC~300 X-1 \citep{2019ApJ...883L..34H}.
In other pulsating ULXs, companions are unknown. 
However, it remains possible to estimate companion mass on the base of measured orbital periods in systems. 

\section{\red{4\,} Physics and geometry of accretion in XRPs}

XRPs are powered by the accretion of matter ejected from the companion star and accelerated by the gravitational field of a NS.
The velocity of accreting material in the vicinity of the NS surface is close to the free-fall velocity $v_{\rm ff}$ and can be estimated as
\be 
v\approx v_{\rm ff}\approx c\sqrt{\frac{R_{\rm Sh}}{R}}\simeq 0.54\,c \sqrt{\frac{m}{R_6}},
\ee 
where $c$ is a speed of light, $R_{\rm Sh}=2GM/c^2\simeq 2.95\times 10^5\,m\,\,{\rm cm}$ is the Schwarzschild radius, $R$ is the radius of a NS, and $m\equiv M/M_\odot$ is the NS mass in units of solar masses.
The kinetic energy of matter accreting onto the NS surface with free-fall velocity is comparable to the rest mass energy
\be\label{eq:E_kin}
E_{\rm kin}\approx 
m_{\rm acc}c^2\left[\frac{1}{\sqrt{1-(v_{\rm ff}/c)^2}}-1\right]
\sim\, 0.2\,m_{\rm acc}c^2.
\ee 
As soon as matter reaches the surface of a NS, its kinetic energy is released and emitted mostly in the X-ray energy band.
The total luminosity is related to the mass accretion rate as 
\be\label{eq:L_tot}
L_{\rm tot}=\frac{GM\dot{M}}{R}
\approx
1.33\times 10^{37}\,
\left(\frac{\dot{M}}{10^{17}\,{\rm g\,s^{-1}}}\right)
\frac{m}{R_6}\,\,\ergs.
\ee 
If the mass accretion rate per unit area of the NS surface 
\be
\dot{m}\equiv \frac{\dot{M}}{S}\gtrsim {3\times }10^{ 4}\,{\rm g\,s^{-1}\,cm^{-2}},
\ee
the temperature and pressure of matter funnelled by a strong magnetic field to the relatively small polar areas of the NS are sufficient for stable thermonuclear burning, precluding the appearance of thermonuclear bursts in XRPs \citep{1997ApJ...477..897B}. 
But still, the energy release due to thermonuclear burning is negligible compared to the one from the conversion of kinetic energy: nuclear fusion  yields only
\be
E_{\rm nuc}\approx 0.007\,m_{\rm acc} c^2,
\ee
which is $\sim 20$ times smaller than the energy release due to the accretion onto a NS (\ref{eq:E_kin}).

The phenomenon of XRP requires directional motion of accreting material toward small regions onto the NS surface.
The key factor defining accretion flow geometry in XRPs is the strong magnetic field of a NS.
The accreting material in XRPs is represented by highly ionised plasma, which becomes subject to the Lorentz force in the magnetic field of a NS.
At large distances from the compact object, the accretion flow is unaffected by the NS magnetic field, and one can use well-known solutions describing disc or spherical accretion processes. 
However, in the vicinity of a NS, the magnetic field becomes strong enough to shape entirely the geometry of the flow directing it towards NS magnetic poles.

Thus, we already have the basic physical picture of XRP powered by the accretion of matter onto strongly magnetised NS.
Now we will go into more detail and closely consider the processes developing on different spatial scales.
We will start with large scales, comparable to the size of the entire binary system and see how the material can be captured from the companion star.
Then we will discuss the geometry of accretion flow forming within the Roche lobe of a NS, the region where the gravity of the compact object dominates. 
We will figure out how and where the strong magnetic field of a NS becomes important and what kind of observational phenomenon arises because of that.
Finally, we will see what happens at the polar regions of a NS, where material lands and loses its kinetic energy.

\subsection{\red{4.1\,} Mass transfer in the binary system}

The geometry of accretion inside the Roche lobe of a NS is determined by the characteristic distance at which matter becomes gravitationally captured by the compact object and by the mean specific angular momentum carried by the matter at this distance. 
Thus, the geometry is largely determined by the way how the companion star losses its mass.
We would mention three basic mass-loss mechanisms typical to XRPs: \\
\indent (i) Roche lobe overflow {(e.g. Her~X-1, GRO~J1744-28)}; \\ 
\indent (ii) Capture of matter from the decretion disc in Be-system {(e.g. V~0332+53, 4U~0115+63)}; \\
\indent (iii) Mass loss due to stellar wind {(e.g. Vela~X-1, Cen~X-3)}.

\smallskip\smallskip\smallskip

Both geometrical size of the companion star and orbital separation in a binary system evolve due to mass exchange and physical processes in stars \citep{1982ApJ...254..616R}.
Roche lobe overflow starts when the companion star during the evolution of a binary becomes large enough and starts to lose its mass through the inner Lagrangian point.
The rate of mass transfer through the inner Lagrangian point is determined by the stellar evolution of the companion and variations of the distance between companions in a system due to the mass exchange and energy losses due to gravitational waves.
The specific angular momentum of the matter captured by the NS can be estimated as
\be\label{eq:ang_mom_binary}
l_0\sim \eta_0 a^2 \Omega_{\rm orb},
\ee 
where $a$ is orbital separation, $\Omega_{\rm orb}$ is the angular velocity in a system, and $\eta_0$ is a dimensionless parameter of the order of unity.
The specific angular momentum given by (\ref{eq:ang_mom_binary}) is much larger than that in the Keplerian rotation near the NSs.
As a result, the Roche lobe overflow in a system results in the formation of an accretion disc around the compact object. 

\smallskip\smallskip 

Capture of matter from the decretion disc in BeXRBs is a much more complicated mechanism, whose analysis requires detailed numerical simulations.
BeXRBs contain a Be star in a relatively wide orbit (typical orbital periods are tens or hundreds days) of significant eccentricity ($e\gtrsim 0.1$) with compact object (often it is a NS, see e.g. \citep{2011Ap&SS.332....1R}, but BHs are also represented in BeXRBs, see e.g. \citep{2014ApJ...786L..11M}). 
The orbital angular momentum of BeXRB is typically misaligned to the spin of the Be star, which is likely a result of kick experienced during the supernova explosion 
\citep{2009MNRAS.397.1563M}.
Capture of material from the decretion disc results in formation of accretion disc of complicated dynamics around a compact objects \citep{2014ApJ...790L..34M}.

\smallskip\smallskip

The wind is the major source of accretion in a binary system if the companion star does not fill its Roche lobe.
The accretion fed by the stellar wind is particularly relevant for systems containing an early-type (O or B) star or red giant and a compact object in a close orbit.
The mass loss rate in early type stars can be as high as $10^{-6}-10^{-5}\,M_\odot\,{\rm yr}^{-1}\sim 6\times (10^{19}-10^{20})\,{\rm g\,s^{-1}}$ and the velocity of the wind is highly supersonic and typically estimated as $v_{\rm w}\sim 10^8\,{\rm cm\,s^{-1}}$.

\begin{figure}
\centering 
\includegraphics[width=7.cm, angle =0]{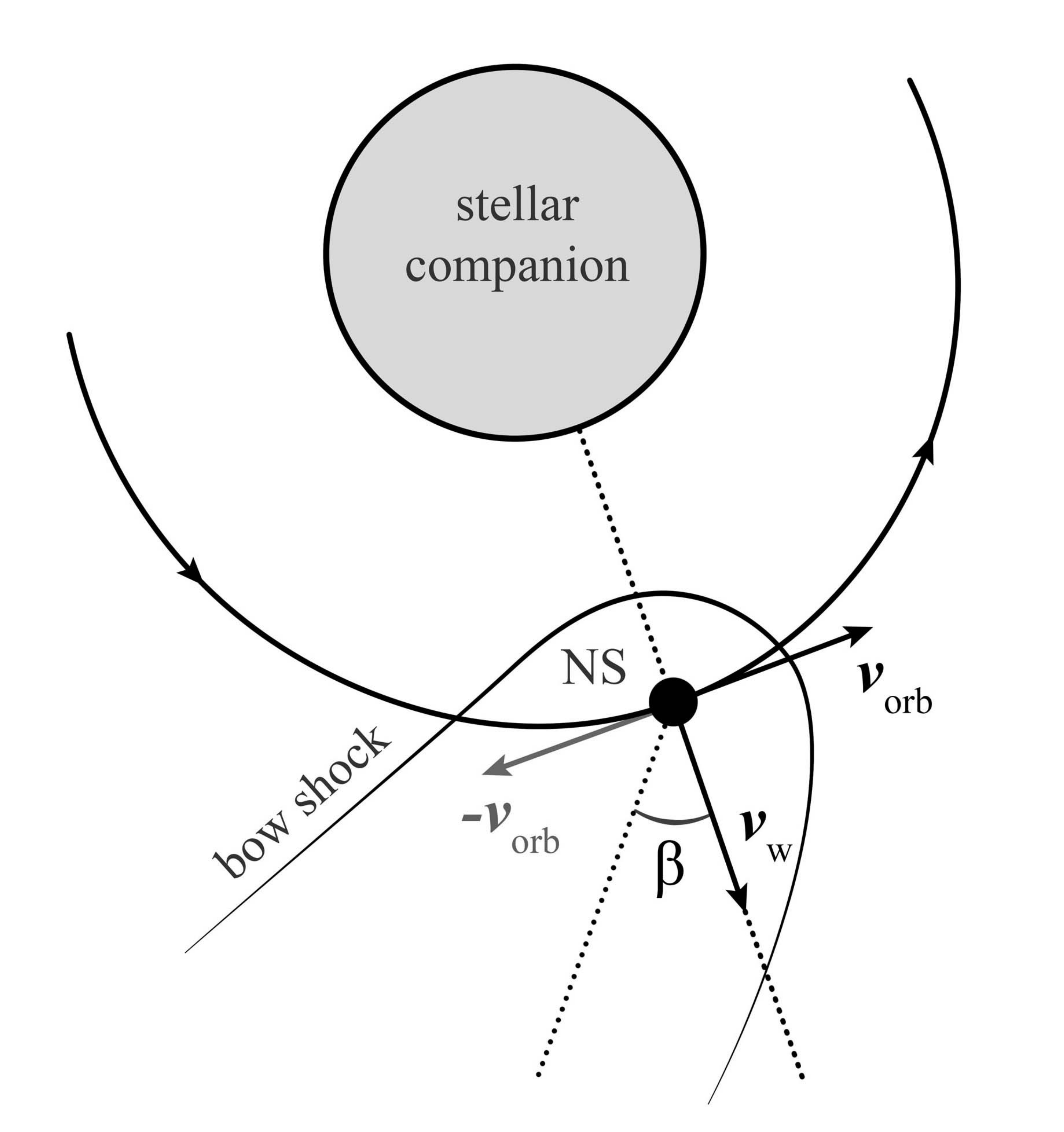} 
\caption{
XRP accreting from the stellar wind.
The wind is captured at the accretion radius and bow shock with a cone-shape cavity forms around a NS in the case of supersonic wind flow. 
}
\label{pic:bow_shock}
\end{figure}

The mass accretion rate from the wind is determined by the mass losses from the companion star, velocity of the stellar wind and velocity of a NS in respect to the companion. 
Equating the gravitational energy of wind material to its kinetic energy, one can estimate the typical radius of gravitational capture or so-called accretion radius:
\be\label{eq:R_acc}
R_{\rm acc}=\frac{2GM}{v^2_{\rm rel}}
\approx 2.6\times 10^{10}\,m v^{-2}_{\rm rel,8}\,\,{\rm cm},
\ee 
where $v_{\rm rel}$ is the relative velocity of a NS in respect to the wind.
Capture of the material from supersonic stellar wind results in formation of a bow shock with a cone-shape cavity around a NS (see Fig.\,\ref{pic:bow_shock}, \citep{1979SvA....23..201B,1996ApJ...459L..31W}).
The relative velocity of a stellar wind $v_{\rm rel}$ in a binary system is determined by the orbital velocity of a NS $v_{\rm orb}$ and the velocity of a stellar wind $v_{\rm w}$.
In the case of spherically symmetric mass loss from the companion and $v_{\rm w}\gg v_{\rm orb}$, the companion mass loss is
\be\label{eq:dotM_c}
\dot{M}_{\rm c}=4\pi r^2\rho v_{\rm w}
\ee 
and the mass accretion rate onto a compact object is
\be\label{eq:dotM_0}
\dot{M}_0 \approx \pi R_{\rm acc}^2 \rho v_{\rm w},
\ee 
where $r$ is the orbital separation between companions in a binary and $\rho$ is a mass density of a wind at the orbit of a NS.
Combining (\ref{eq:R_acc}), (\ref{eq:dotM_c}) and (\ref{eq:dotM_0}) we get an estimation of the mass accretion rate from the stellar wind:
\beq
\label{eq:dotM_wind_approx}
\dot{M}_0\approx\frac{\dot{M}_{\rm c}}{4}\left(\frac{R_{\rm acc}}{r}\right)^2
&\approx& 1.7\times 10^{-4}\,
\dot{M}_c\,\frac{m^2}{v^4_{\rm w,8}\,r_{12}^2}  \\
&\approx& 10^{16}\,
\left(\frac{\dot{M}_c}{10^{-6}\,M_\odot\,{\rm yr}^{-1}}\right)\,
\frac{m}{v^4_{\rm w,8}\,r_{12}^2}\,\,{\rm g\,s^{-1}}.
\nonumber
\eeq 
In the case of $v_{\rm w}\sim v_{\rm orb}$, one has to account for the influence of orbital motion on the relative velocity of the wind and the rough estimation given by (\ref{eq:dotM_wind_approx}) turns into:
\be
\dot{M}_0=\dot{M}_{\rm c}\,q^2(1+q)^2 \frac{\xi\tan^4\beta}{\pi(1+\tan^2\beta)^{3/2}},
\ee
where $q=M/M_{\rm c}$ is a mass ratio in a binary, parameter $\xi\sim 1$ and $\tan\beta=v_{\rm orb}/v_{\rm w}$ (see Fig.\,\ref{pic:bow_shock} and \citep{1973ApJ...179..585D,1992ans..book.....L} for details).

In very close binary systems, one has to account for the motion of a companion star around the mass centre, which leads to the wind collimation in the orbital plane.
Under this condition and in the case of relatively slow wind, the mass accretion rate onto the compact object from the stellar wind can be higher and even comparable to the mass accretion rates typical for ULX pulsars \citep{2019A&A...622L...3E}.

\subsection{\red{4.2\,} Accretion flow interacting with the NS magnetosphere}

\subsubsection{\red{4.2.1\,} Magnetospheric boundary}

Because the accreting material in XRPs is highly ionised and affected by the Lorentz force, the strong magnetic field of a NS shapes the geometry of the accretion flow in XPRs to a great extent.
In XRPs, the magnetic field of a NS is strong enough to disrupt the accretion flow at the magnetospheric boundary that is located at a large distance ($\sim 10^8\,{\rm cm}$) from the central object.
The distance where the accretion flow is disrupted by the field is called the magnetospheric radius $R_{\rm m}$.
Accreting material from the magnetospheric boundary is funnelled towards the polar cups of a NS, stopped at the magnetospheric boundary or ejected to infinity.
The physics of the interaction between the accretion flow and NS magnetic field is exceedingly complex (see \citep{2014EPJWC..6401001L} for review). Still, some useful estimations can be obtained even from a simplified physical picture. 

The magnetic field pressure is given by $P_{\rm mag}=B^2/(8\pi)$ and increases rapidly toward the NS.
In the case of $B$-field dominated by the dipole component (see eq. \ref{eq:dipole_field})
\be
P_{\rm mag}=\frac{1}{8\pi}\frac{\mu_{\rm d}^2}{r^6},
\ee
where $\mu_{\rm d}= B_0R^3/2$ is a dipole magnetic moment, $B_0$ is magnetic field strength at the NS magnetic poles, and $r$ is a distance from a NS centre.
Equating the magnetic field pressure and the ram pressure of accreting material, which is 
\be
P_{\rm ram}= \rho v^2\sim \frac{1}{4\pi}\frac{(2GM)^{1/2}\dot{M}}{r^{5/2}}
\ee
in the case of spherically symmetric accretion with free fall velocity, we estimate the radius from the NS where magnetic field disrupts the accretion flow
\beq\label{eq:R_A}
R_{\rm A}
&=&2.7\times 10^8\, \mu_{30}^{4/7}\dot{M}_{17}^{-2/7}m^{-1/7}\,\,{\rm cm}\\
&=&1.8\times 10^8\, B_{12}^{4/7}\dot{M}_{17}^{-2/7}m^{-1/7}R_6^{12/7}\,\,{\rm cm},\nonumber
\eeq 
the so-called Alfv\'en radius.
A similar estimate assuming the quadrupole magnetic field configuration (see eq. \ref{eq:quadrupole_field}) is given by 
\be\label{eq:R_A_quad}
R_{\rm A}^{\rm (q)}=3.44\times 10^{7}\,
B_{12}^{4/11}\dot{M}_{17}^{-2/11}m^{1/11}R_6^{16/11} \,{\rm cm}.
\ee 
The actual magnetospheric radius is of the order of the Alfv\'en radius but depends on the exact accretion flow geometry interacting with the NS magnetic field.
It is useful to introduce the coefficient of proportionality $\Lambda$ between the radius of the magnetosphere and Alfv\'en radius:
\be\label{eq:R_A_2_R_m} 
R_{\rm m}=\Lambda R_{\rm A}.
\ee 
In the particular case of disc accretion, the inner disc radius is typically assumed to be by a factor of $2$ smaller than the Alfv\'en radius ($\Lambda=0.5$), while in the case of spherical accretion $\Lambda\approx 1$ \citep{1978ApJ...223L..83G,1979ApJ...232..259G}.
However, this picture is oversimplified, and one has to keep in mind that the inner disc radius can be affected by the disc inclination with respect to the magnetic dipole of a NS, magnetic field structure \citep{1978SvA....22..702L,1978ApJ...219..617S,1980A&A....86..192A}, and by physical conditions in the accretion flow \citep{1999ApJ...521..332P,2017MNRAS.470.2799C}, which can be geometrically thin or thick, gas or radiation pressure dominated, advective or non-advective.

In wind-fed XRPs, accretion onto the NS is possible under the condition
\be 
R_{\rm m}<R_{\rm acc},
\ee 
i.e. in the case of
\be 
B\lesssim 
6\times 10^{15}\,\dot{M}_{17}^{1/2}m^2 R_6^{-3} v_{\rm rel,8}^{-7/2}\,\,\,{\rm G}
\ee
at the surface of a NS, or 
\be
\dot{M} > 2.8\times 10^9\,B_{12}^2 m^{-4} R_6^6\, v_{\rm rel,8}^7\,\,\,{\rm g\,s^{-1}}.
\ee
Otherwise, the magnetic barrier sets-in and the flow from the donor star cannot be captured properly and is deflected away.

For practical application, it is useful to have rough estimates of some physical conditions at $R_{\rm m}$.
Particularly, in the case magnetic field dominated by the dipole component, the field strength at the magnetospheric radius is
\be
B(R_{\rm m})\sim
10^5\,\Lambda^{-3} \mu_{\rm d,30}^{-5/7}
\dot{M}_{17}^{6/7}m^{3/7}\,\,{\rm G},
\ee 
and the Keplerian angular velocity is
\be\label{eq:Omega_K_Rm} 
\Omega_{\rm K}(R_{\rm m})=
\left(\frac{GM}{R_{\rm m}^3}\right)^{1/2}
\approx
2.6\,\Lambda^{-3/2} \mu_{\rm d,30}^{-6/7} \dot{M}_{17}^{3/7} m^{5/7}\,\,{\rm rad\,s^{-1}}.
\ee 
Note that the stronger the magnetic field is at the NS surface, the weaker it is at the magnetospheric boundary.

Because accreting material at $R_{\rm m}$ is expected to be hot and ionised, the penetration of the flow into the NS magnetosphere is problematic but possible due to the development of various instabilities, including magnetic Rayleigh-Taylor \citep{1976ApJ...207..914A,2008MNRAS.386..673K}, Kelvin-Helmholtz  instabilities \citep{1983ApJ...266..175B}, and reconnection.
Extremely hot plasma is known to be stable relative to the Rayleigh-Taylor instability, which is the main mechanism of plasma penetration into the magnetosphere of a NS \citep{1977ApJ...215..897E}.
Thus, for the effective matter penetration into the magnetosphere, the temperature of the accretion flow has to be below a certain critical value. 
This condition is typically satisfied in disc-fed XRPs, but can be violated in wind-fed sources. 
If so, the accretion flow onto the magnetised NS is below the value estimated by (\ref{eq:dotM_wind_approx}) and is determined by the ability of matter to cool below the critical temperature. 
If the luminosity of wind-fed XRPs is $\lesssim 4\times 10^{36}\,\ergs$, the material captured at $R_{\rm acc}$ cannot cool rapidly enough, which results in the formation of an extended quasi-static shell \citep{2012MNRAS.420..216S} around the NS magnetosphere.
The mass accretion rate through the shell is driven by the cooling processes and the ability of plasma to enter the magnetosphere.

\subsubsection{\red{4.2.2\,} Influence of the magnetospheric rotation}

\begin{figure}
\centering 
\includegraphics[width=7.cm, angle =0]{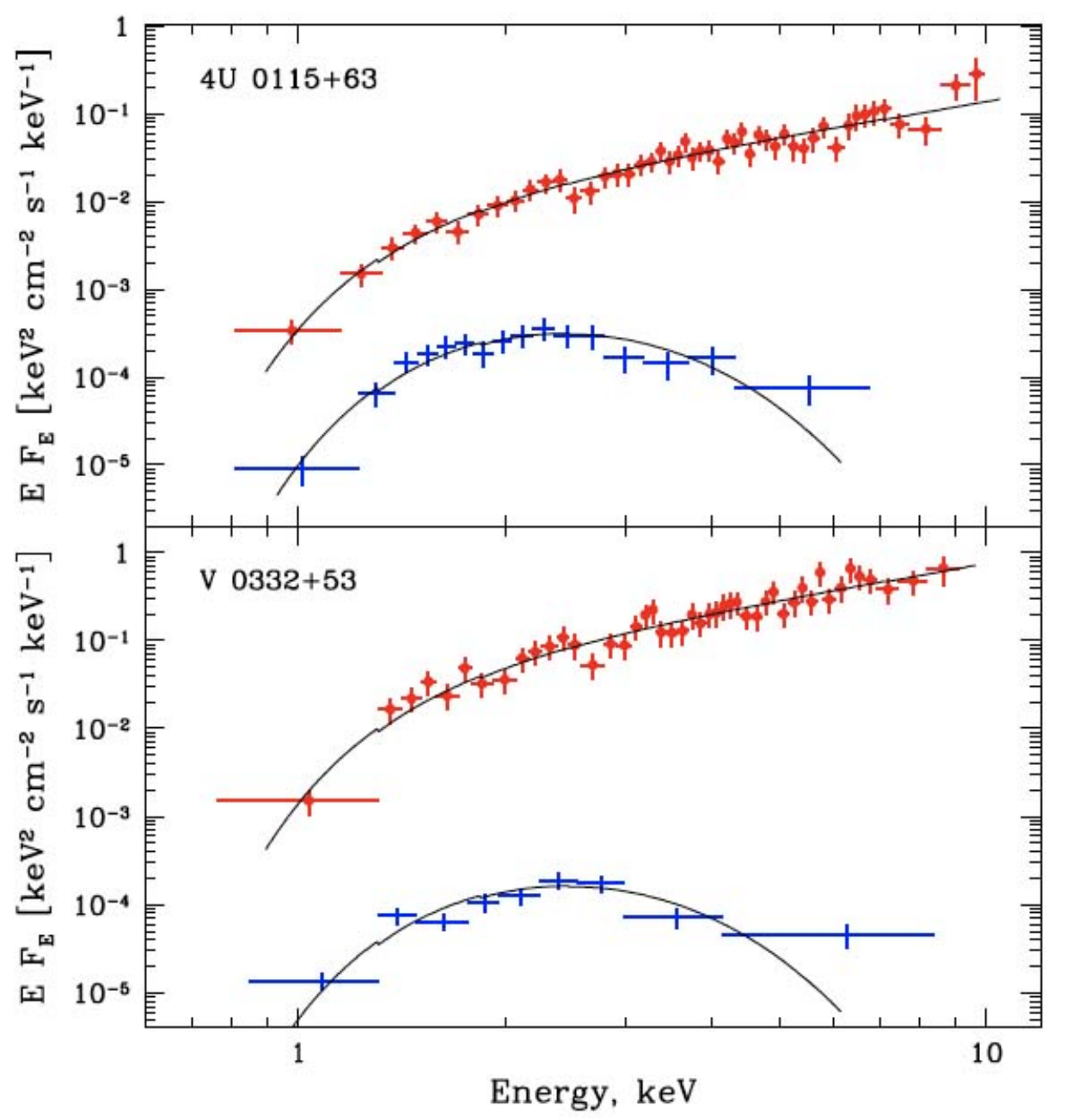} 
\caption{
Dramatic spectral changes observed during transition into the propeller state in XRPs 4U~0115+63 {\it(top)} and V~0332+53 {\it(bottom)}.
In both cases, the spectra in the propeller state (blue points) can be well fitted with the absorbed black-body model with the temperature $\sim 0.5\,\,{\rm keV}$ and radius of the emitting area $\sim 0.6-0.8\,\,{\rm km}$.
The observed soft blackbody-like spectra in the propeller state can be a sign of cooling of the NS polar regions after their heating during the outburst.
From \citep{2016A&A...593A..16T}.
}
\label{pic:propeller_sp}
\end{figure}

Another important linear scale for the accreting compact object is the corotation (or centrifugal) radius, where gravity is in balance with the centrifugal force acting on matter which is in corotation with the NS.
In another words, this is the radius where the magnetic field lines rotate with the same (Keplerian) velocity of matter in the accretion disc:
\be
R_{\rm cor}=\left(\frac{GMP^2}{4\pi ^2}\right)^{1/3} 
\simeq 1.5\times 10^{8}\,m^{1/3}P^{2/3}\,\,\,{\rm cm}.
\ee 
As follows from eq. (\ref{eq:R_A}) and (\ref{eq:R_A_quad}),  the Alfv\'en radius, i.e. inner radius of accretion flow unaffected by the field, 
depends on the mass accretion rate: the larger the mass accretion rate, the smaller the inner radius. 
If the accretion flow penetrates into the corotation radius ($R_{\rm m}<R_{\rm cor}$), the matter is not dynamically inhibited from falling onto the NS. 
In the opposite case, when $R_{\rm m}>R_{\rm cor}$, the centrifugal barrier will prevent direct accretion onto the central object, which results in appearance of the propeller effect \citep{1975A&A....39..185I,2006ApJ...646..304U}.
Equating the magnetospheric and corotation radii, one can estimate the limiting mass accretion rate, which separates regimes of accretion and propeller state:
\beq 
\dot{M}_{\rm prop} &\approx& 
7.4\times 10^{17}\, \Lambda^{7/2}\mu_{\rm d,30}^2 P^{-7/3}m^{-5/3}\,\,\,{\rm g\,s^{-1}} \\
&=& 2\times 10^{17}\, \Lambda^{7/2}B_{12}^2P^{-7/3}m^{-5/3}R_6^6\,\,\,{\rm g\,s^{-1}}.\nonumber 
\eeq 
The corresponding accretion luminosity is
\be\label{eq:L_prop} 
L_{\rm prop}=L_{\rm lim}(R)=\frac{GM\dot{M}_{\rm prop}}{R}\approx 2.7\times 10^{37} \Lambda^{7/2}B_{12}^2P^{-7/3}m^{-2/3}R_6^5\,\,\,\ergs.
\ee
As soon as XRP switches into the propeller state, accretion flow stops at $R_{\rm m}=R_{\rm cor}$ and does not reach the NS surface. 
Because $R_{\rm m}\gg R$ in XRPs, the accretion efficiency in propeller regime is much lower. 
As a result, the decrease of mass accretion rate below $\dot{M}_{\rm prop}$ leads to the drop of luminosity to the value expected for so-called ``magnetospheric accretion" case, when kinetic energy of matter is released at the magnetospheric radius 
\citep{1996ApJ...457L..31C,1998A&A...340...85R}:
\be 
L_{\rm lim}(R_{\rm cor})=\frac{GM\dot{M}_{\rm prop}}{R_{\rm cor}}=
L_{\rm lim}(R)\frac{R}{R_{\rm cor}}.
\ee 
In the absence of other sources of emission (such as, for example, NS surface heated up by accretion), the dramatic drop of accretion luminosity (about $10^{2}$ for a typical XRP) should be accompanied by significant spectral changes.
The effective temperature of accretion disc at radial coordinate $r$ can be estimated as
\be 
T_{\rm eff}\simeq 3.5\times 10^{-2}\,\dot{M}_{17}^{1/4}m^{1/4}r_8^{-3/4}\,\,{\rm keV},
\ee 
which gives a rough estimation of the effective temperature at the magnetospheric boundary
\be 
T_{\rm eff}(R_{\rm m})\sim 
0.02\,\Lambda^{-3/4} \mu_{\rm d,30}^{-3/7}\dot{M}_{17}^{13/28} m^{5/14}\,\,\,{\rm keV}.
\ee 
Thus, the accretion disc at the magnetospheric radius typical for XRPs hardly produces valuable amount of photons in the X-ray energy band.
However, leakage of matter through the centrifugal barrier or/and cooling of NS polar cups can provide noticeable X-ray emission \citep{2016A&A...593A..16T,2016MNRAS.463L..46W}.
This kind of emission with soft X-ray energy spectra was observed in two XRPs in the propeller state (see Fig.\,\ref{pic:propeller_sp}, \citep{2016A&A...593A..16T,2016MNRAS.463L..46W}).

The timescale of transition from the accretion to the propeller regime strongly depends on the structure of the disc-magnetosphere interaction region. 
Even in the case of pure disc accretion, when the sharpest onset of centrifugal barrier is expected, the material can only enter the magnetosphere due to instabilities arising in their interaction. 
These instabilities may drive the dramatic variability of X-ray flux at different timescales and make the transition essentially non-instant.

Unfortunately, the difficulty of predicting the exact time of the transition did not allow to observe such events “in real time” so far. 
The constraints on the transition timescale in the available data are limited mainly by the observations cadence. 
For instance, in the case of the pulsating ULX NGC 5907 such transition happened faster than in 6 days \citep{2017Sci...355..817I}, 
in the transient XRPs 4U~0115+63 and V~0332+53 faster than $\sim$1.5 day \citep{2016A&A...593A..16T}.  
However, the most detailed observation of the transition between the quiescent and accretion states was obtained fortuitously for 4U~0115+63 with the \textit{BeppoSAX} satellite \citep{2001ApJ...561..924C}. 
Namely, during the observation close to periastron, a huge luminosity increase by a factor of $\sim$250 in less than 15 hr was revealed. 
This was interpreted as the opening of the centrifugal barrier during  the  onset  of  accretion  state.

\begin{figure}
\centering 
\includegraphics[width=12.cm, angle =0]{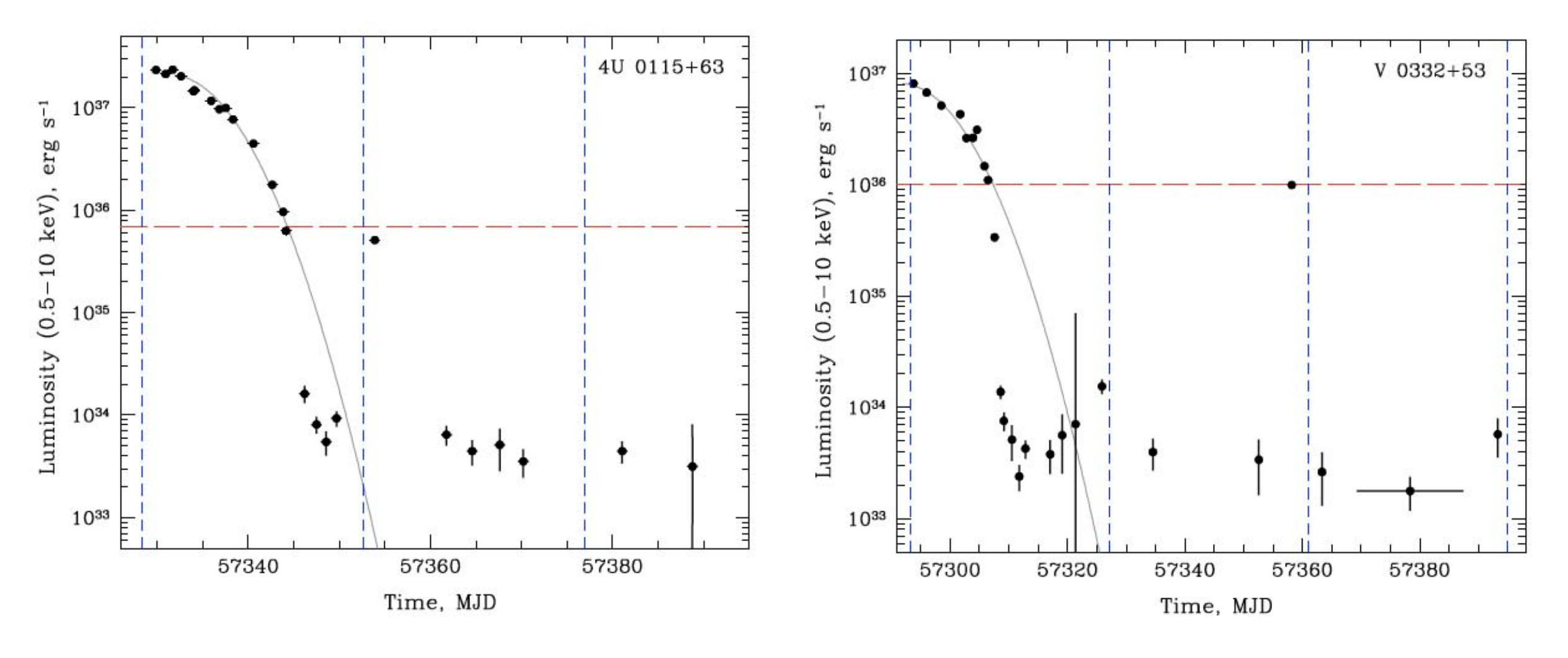} 
\caption{
Light curves obtained at the end of the outbursts observed from XRPs 4U~0115+63 {\it(left)} and V~0332+53 {\it(right)} and transition to the propeller state at the luminosity marked by red dashed lines.
Both sources were detected in the quiescent state and demonstrated brightening near the periastron passages (see vertical blue dashed lines).
From \citep{2016A&A...593A..16T}.
}
\label{pic:propeller_LC}
\end{figure}

\begin{figure}
\centering 
\includegraphics[width=12.cm, angle =0]{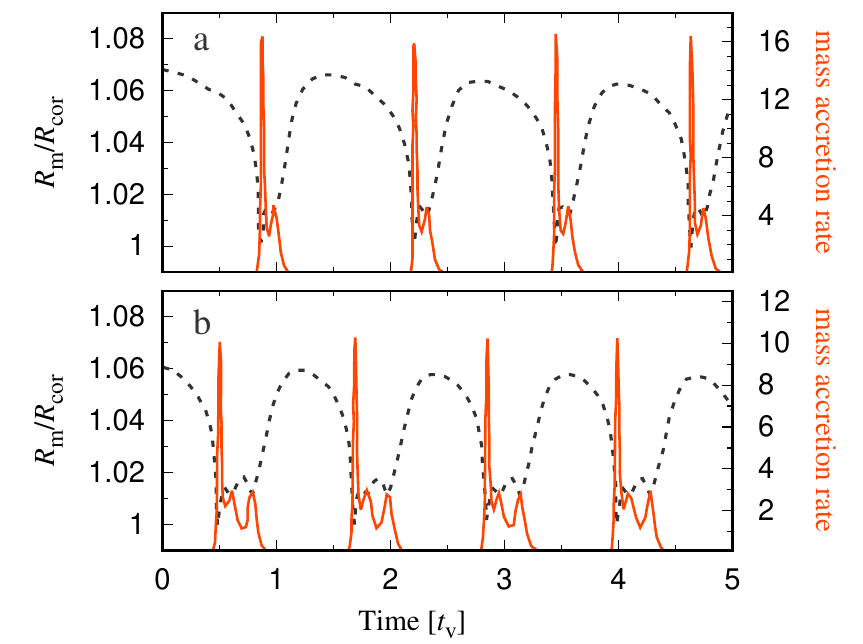}
\caption{
Accumulation of matter in the accretion disc during the propeller state can result in events of episodic accretion onto the NS surface. 
Variations of the inner disc radius (black dashed lines) and mass accretion rate (in relative units) onto the NS surface (red solid lines) are shown for different average mass accretion rates:
(a) $\dot{M}=4.6\times 10^{17}\,{\rm g\,s^{-1}}$
and
(b) $\dot{M}=7.6\times 10^{17}\,{\rm g\,s^{-1}}$.
The curves are reproduced from \citep{2010MNRAS.406.1208D}.
}
\label{pic:propeller_DAngelo}
\end{figure}

Accreting plasma in the state of magnetospheric accretion is accumulated at the magnetospheric boundary and forms a dead disc \citep{1977PAZh....3..262S}. 
If the magnetospheric radius remains to be close to the corotation radius, the accumulation of material should lead to the events of episodic accretion onto the NS surface (see Fig.\,\ref{pic:propeller_DAngelo}, \citep{2010MNRAS.406.1208D}). 
The episodes of accretion are expected to be quasi-periodic with periods determined by the viscous time scale in the disc.

\subsubsection{\red{4.2.3\,} Spin-ups and spin-downs of NS in XRPs}

\begin{figure}
\centering 
\includegraphics[width=0.49\columnwidth]{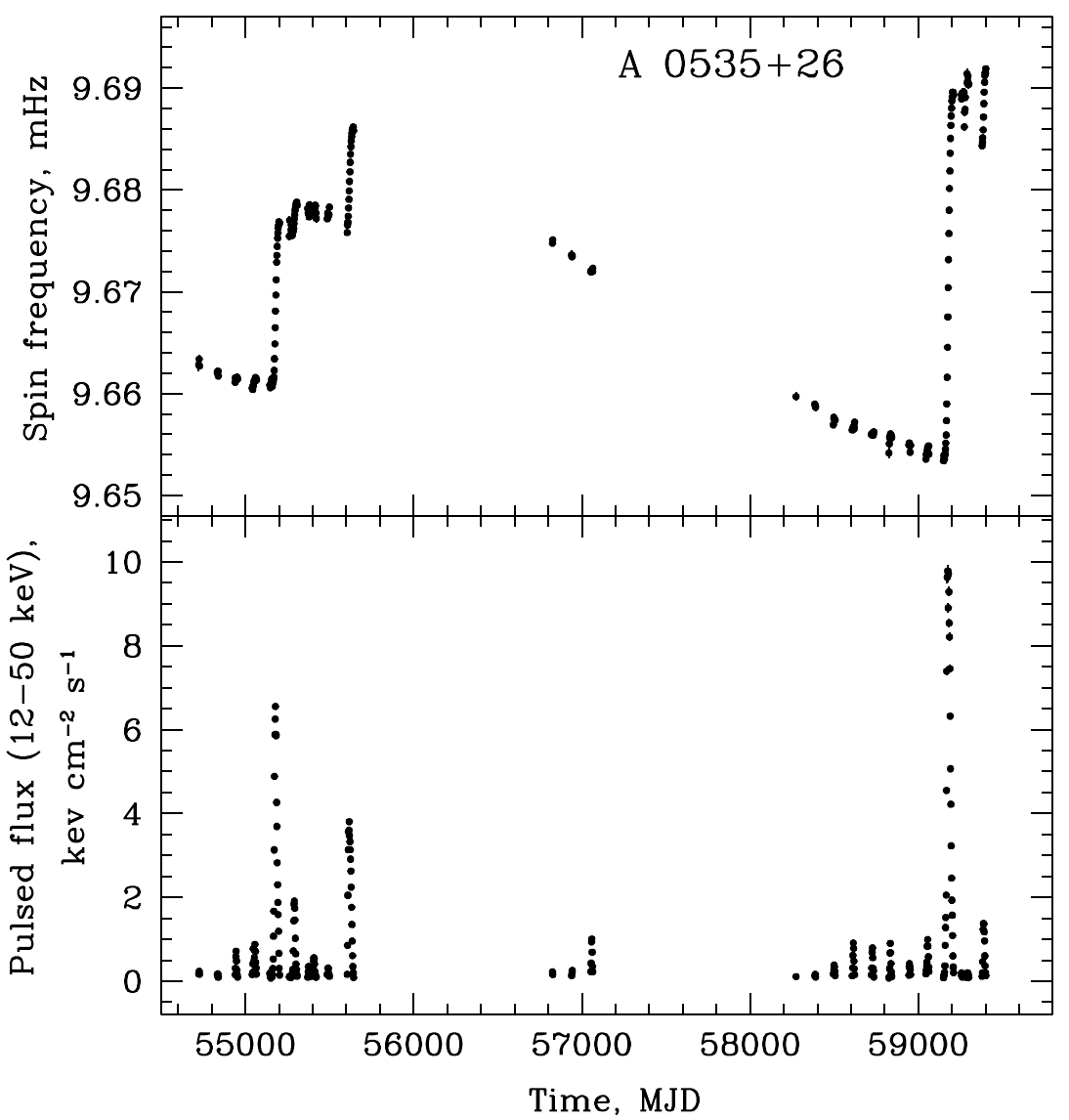} 
\includegraphics[width=0.49\columnwidth]{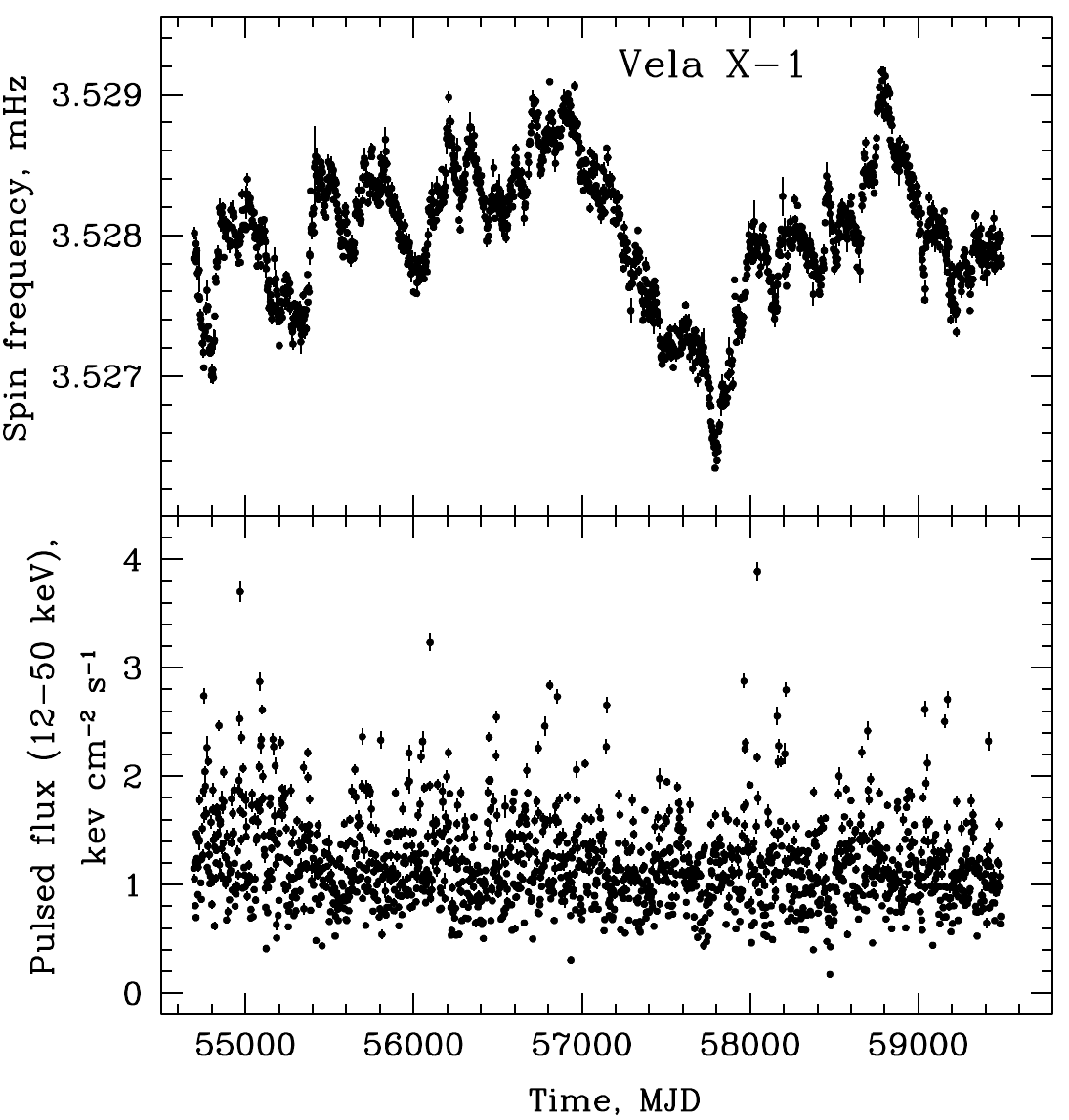} 
\vspace{0.4cm}

\includegraphics[width=0.56\columnwidth]{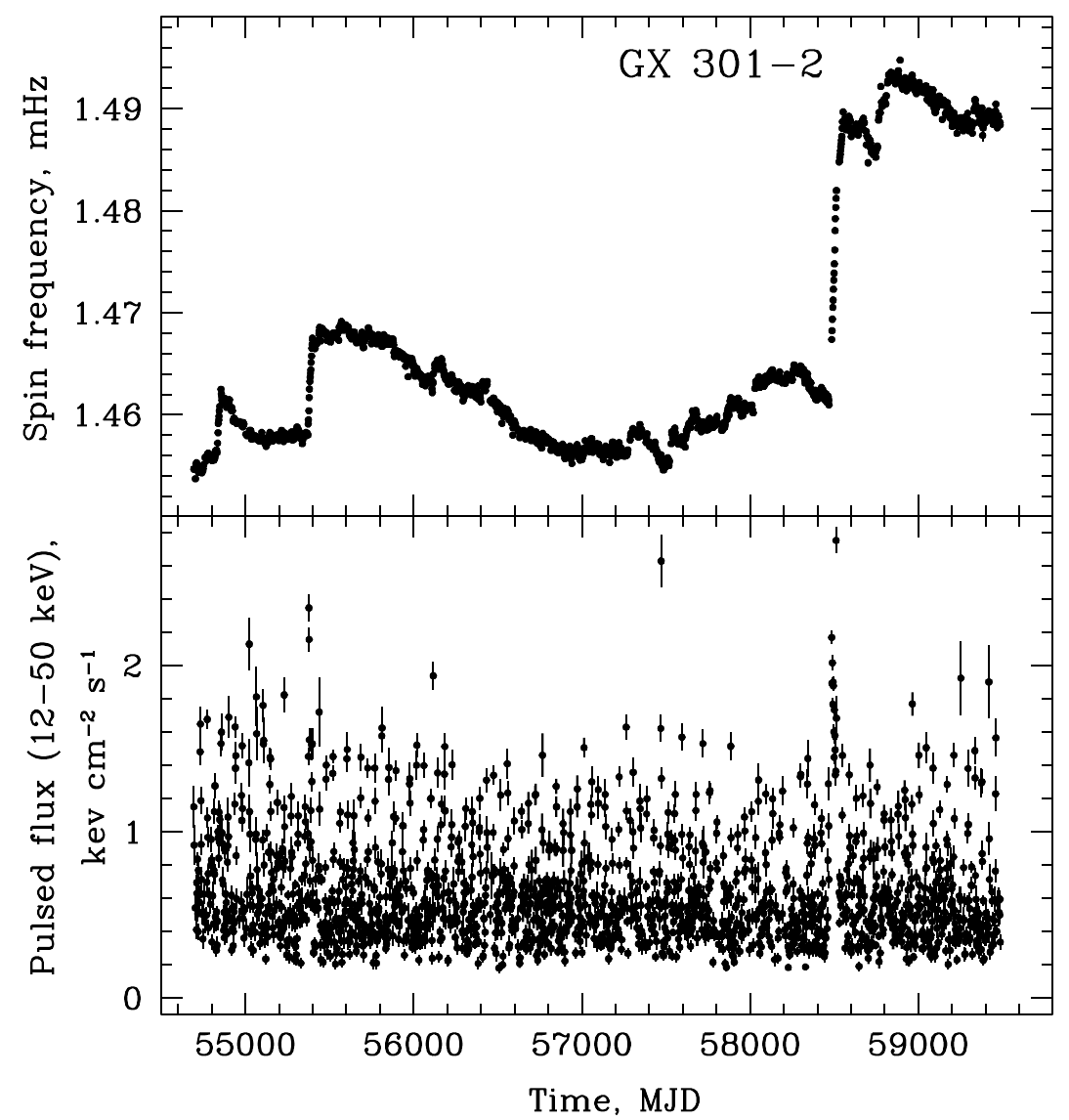} 
\caption{Binary-motion corrected spin frequency evolution of three XRPs with different accretion mechanisms: disc accretion for A~0535+26, wind accretion for Vela~X-1 and transient disc formation in the wind-fed GX~301$-$2  (top panels). The corresponding pulsed flux in the 12-50 keV energy band is shown in the bottom panels. From the \textit{Fermi}/GBM Accreting Pulsars Program (\url{http://gammaray.nsstc.nasa.gov/gbm/science/pulsars/}).
}
\label{pic:pdot_obs}
\end{figure}

The accretion process and interaction of accretion disc/wind with the magnetosphere of a NS results in angular momentum exchange and corresponding variations in the NS spin period \citep{2016ApJ...822...33P} and direction of its rotation axis \citep{2021MNRAS.505.1775B}.
The rate of change of the NS angular momentum $\vec{J}$ is determined by the total accretion torque $\vec{K}_{\rm tot}$
\be
\frac{{\rm d} \vec{J}}{{\rm d} t}=\vec{K}_{\rm tot}.
\ee  
In the case of accretion from stellar wind, the sign and magnitude of the torque are rather uncertain because of the stochastic fluctuations of the sign and magnitude of the specific angular momentum of the clumpy matter captured by the NS. 
It is nicely seen in the variation of NS spin frequency observed in XRP Vela~X-1 presented in Fig.\,\ref{pic:pdot_obs}.
In the case of accretion from the disc, the material is expected to move with nearly Keplerian velocity.
{In general, it is expected that XRPs, accreting at high mass accretion rates, spin up due to the interaction between the NS magnetosphere and the accretion flow. 
Indeed, in this case, the inner disc radius locates within the corotation radius ($R_{\rm m}<R_{\rm cor}$) and the angular velocity of accretion flow at the inner disc edge exceeds the angular velocity of the NS magnetosphere.
On the contrary, in the propeller mode, when the inner disc radius is located outside the corotation radius ($R_{\rm m}>R_{\rm cor}$), a gradual spin down is expected.
This kind of behaviour is seen in XRP A~0535+26 (see Fig.\,\ref{pic:pdot_obs}), where the accretion process occurs through the disc and the outbursts are correlated with spin-up phases, while the off-states are accompanied by a decrease of the NS spin frequency.
In some XRPs, like in GX~301$-$2, the accretion onto the central object may proceed both from the disc and stellar wind.
In this case, the spin frequency behaviour inherits traits from both accretion channels.}

The quantitative approach requires a bit more detailed analysis.
Let us consider the case of accretion from the disc.
Assuming alignment between accretion disc axis, NS spin axis and dipole magnetic axis, it is possible to make estimations of the torques.
Further, we will focus on this particular case. 
The torque applied to the NS has two contributions: the one associated with mass accretion \citep{1972A&A....21....1P}
\be 
K_0=\dot{M}R_{\rm m}^2 \Omega_{\rm K}(R_{\rm m})
\ee 
and the other related to the disc-star coupling $K_{\rm m}$.
As a result, the total torque is given by
\be
K_{\rm tot}=K_0+K_{\rm m}.
\ee 
Magnetic torque $K_{\rm m}$ can be either positive or negative depending on the location of the interaction zone relative to the corotation radius.

Several models have been proposed for estimating magnetic torque \citep{2016ApJ...822...33P}.
These are often expressed in the form
\be 
K_{\rm tot}=n(\omega_{\rm s})K_0,
\ee 
where $n$ is dimensionless function and
\be
\omega_{\rm s} \equiv \left(\frac{\Omega_{\rm s}}{\Omega_{\rm K,m}}\right)=\left(\frac{R_{\rm m}}{R_{\rm cor}}\right)^{3/2}
\ee 
is the ``fastness" parameter.
There are two simple analytical models proposed for the case of slowly rotating NSs:
\begin{itemize}
{\setlength{\itemsep}{0pt}
\item In \citep{1979ApJ...234..296G} was proposed
\be\label{eq:n_GhoshLamb}
n^{(GL)}(\omega_{\rm s})\simeq 1.39\frac{1-\omega_{\rm s}[4.03(1-\omega_{\rm s})^{0.173}-0.878]}{1-\omega_{\rm s}}.
\ee 

\item If the magnetic stress communicated by the magnetosphere is limited by its susceptibility to field line opening and reconnection,  the torque is given by \citep{1995ApJ...449L.153W}
\be\label{eq:n_Wang}
n^{(W)}(\omega_{\rm s})\simeq 
\frac{(7/6) - (4/3)\omega_{\rm s} + (1/9)\omega_{\rm s}^2}
{1-\omega_{\rm s}}.
\ee 

}
\end{itemize}
Note, that these models are only applicable when $R_{\rm m}<R_{\rm cor}$.
The corresponding changes of the NS spin period $P$ can be estimated as
\be\label{eq:Pdot}
\dot{P}\approx -2\times 10^{-12}\,n(\omega_{\rm s}) P^2 
\mu_{30}^{2/7} L_{37}^{6/7} I_{45}^{-1} m^{-3/7} R_6^{6/7}
\,\,\,
{\rm s\,s^{-1}}.
\ee 

The spin period derivative $\dot{P}$ turns to zero at some specific combination of the inner disc radius and the corotation radius, which means that the pulsar is in equilibrium.
The corresponding spin period is called the equilibrium spin period $P_{\rm eq}$. 
The equilibrium period is different in different torque models. 
In Ghosh\,\&\,Lamb model described by (\ref{eq:n_GhoshLamb}), it is reached at  $\omega_{\rm s}=0.35$ and equals to
\be 
P_{\rm eq}^{(GL)}=5.7\,\Lambda^{3/2}B_{12}^{6/7}L_{37}^{-3/7}m^{-2/7}R_6^{15/7}\,\,{\rm s}.
\ee 
In Wang's model described by (\ref{eq:n_Wang}), equilibrium corresponds to $\omega_{\rm s}=0.95$ and the equilibrium period is
\be 
P_{\rm eq}^{(W)}=2.1\,\Lambda^{3/2}B_{12}^{6/7}L_{37}^{-3/7}m^{-2/7}R_6^{15/7}\,\,{\rm s}.
\ee 

Propeller state arising under condition of $R_{\rm m}>R_{\rm cor}$ affects the torque applied to the NS.
When the accretion flow gets attached to the stellar field lines, the matter starts to corotate with the NS. 
If $2\pi R_{\rm m}/P$ exceeds the escape velocity at the magnetospheric radius $v_{\rm esc}=(2GM/R_{\rm m})^{1/2}$, i.e. if 
\be 
R_{\rm m}>2^{1/3}R_{\rm cor},
\ee
the accreting matter may be ejected due to the centrifugal force. 
Ejection of matter modifies the total torque:
\be 
K_{\rm tot}=K_0+K_{\rm m}
-\dot{M}_{\rm eject}R_{\rm m}^2\Omega,
\ee
where $\dot{M}_{\rm eject}$ is the mass ejection rate. 

Misalignment of the NS dipole and its rotation axis (that is required for the pulsations appearance) makes the problem of interaction between NS and accretion flow more complex, resulting in at least two additional effects:
(i) the plasma in the magnetosphere can flow to the polar cup more easily, and
(ii) misaligned stellar dipole can excite non-axisymmetric waves in the disc \citep{2008ApJ...683..949L}.
Note that the estimations of the inner disc radius are strongly model dependent for the case of accretion onto inclined magnetic dipoles \citep{1978SvA....22..702L,2018A&A...617A.126B}.
Because inclination is usually not known for particular XRPs, estimates of NS magnetic field strength based on measurements of $\dot{P}$ or spin equilibrium period remain quite uncertain.


\begin{figure}
\centering 
\includegraphics[width=10.cm, angle =0]{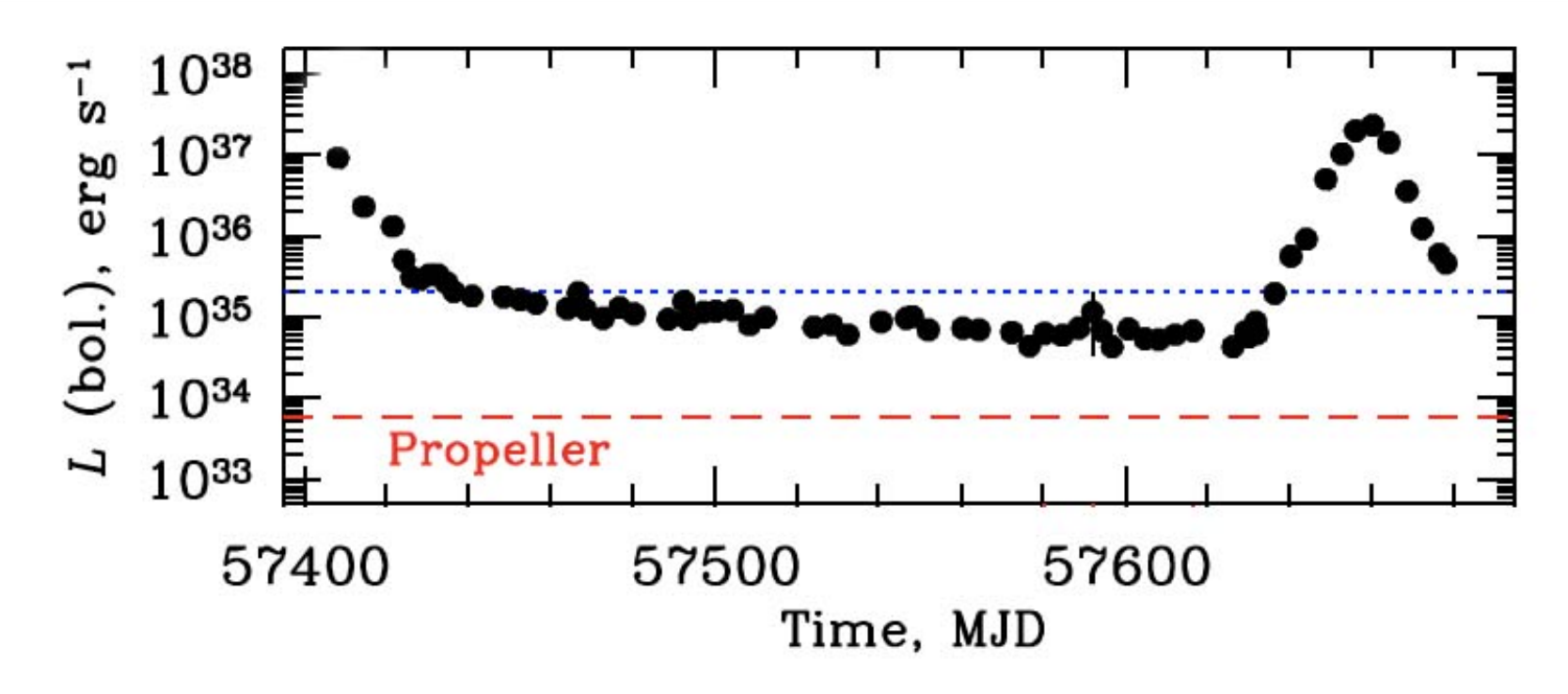} 
\caption{
Light curve observed from BeXRB GRO~J1008$-$57.
The source shows outbursts every periastron passage.
In the end of the outburst the source does not reach the luminosity level where one expect transition into the propeller state (red dashed line). 
Instead of that, at some luminosity, shown with the blue dotted line, the fast fading of the source is stopped and accretion process turns into a stable mode lasting until the next outburst. 
The stable mode in between the outbursts corresponds to the accretion from cold recombined disc of low viscosity.
From \citep{2017A&A...608A..17T}.
}
\label{pic:cold_disc_LC}
\end{figure}

\begin{figure}
\centering 
\includegraphics[width=11.5cm, angle=0]{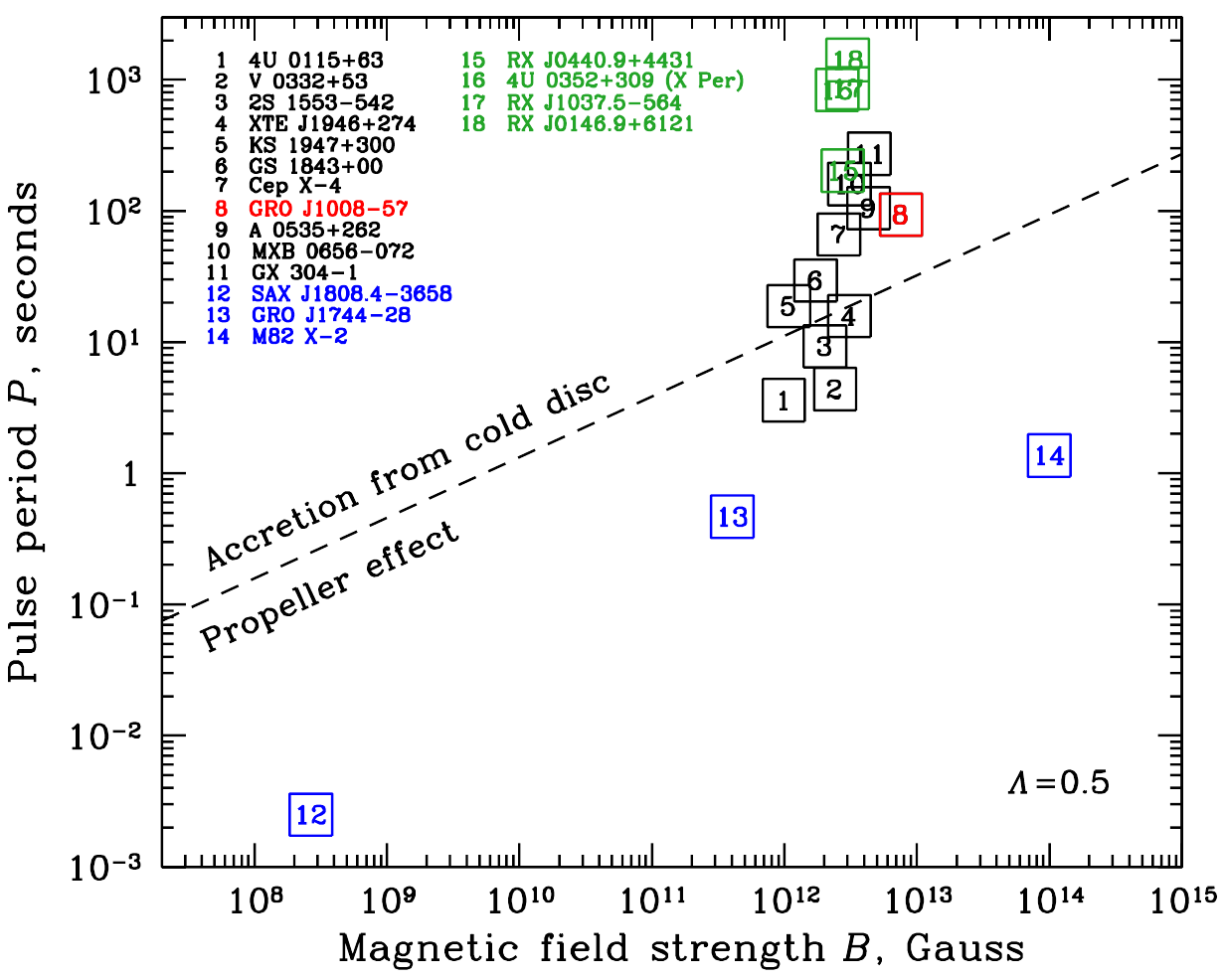} 
\caption{
Collection of some known BeXRBs (shown in black), as well as the accreting millisecond pulsar SAX~J1808.4$-$3658, the intermediate pulsar GRO~J1744$-$28, and the accreting magnetar M82~X-2 (all three shown in blue) on the $B-P$ plane. The dashed line corresponds to the prediction of $P^*(B)$ for parameter $\Lambda=0.5$. 
This line separates sources with different final states of an outburst: the propeller regime (below the line) and  stable accretion from the cold disc (above the line). 
Persistent low-luminous BeXRBs are shown in green.
}
\label{pic:cold_disc_vs_propeller}
\end{figure}

\begin{figure}
\centering 
\includegraphics[width=12.5cm, angle =0]{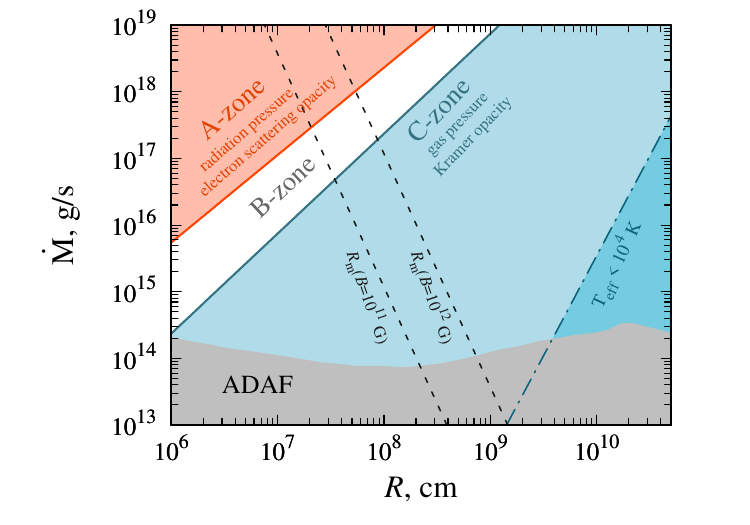} 
\caption{
Zones of different physical conditions in accretion disc at different mass accretion rates.
The disc in XRPs is extended down to the magnetospheric boundary at $R_{\rm m}$, which location is determined by the mass accretion rate and magnetic field of a NS. 
Dashed black lines show $R_{\rm m}$ calculated according to eq.\,\ref{eq:R_A} and \ref{eq:R_A_2_R_m} for the cases of the NS magnetic field $B=10^{11}\,{\rm G}$ and  $B=10^{12}\,{\rm G}$ (the field is assumed to be dominated by the dipole component).
Depending on the mass accretion rate, the zones of different physical conditions can present in the disc.
A-zone (red area) is radiation pressure dominated with the opacity determined by the electron scattering.
C-zone (blue area) corresponds to the gas pressure dominated part, where the opacity is given by the Kramers' law. 
The transitional B-zone (white area) is still gas pressure dominated but with the opacity dominated by the electron scattering. 
Dashed-dotted line corresponds to the radial coordinate, where the effective temperature drops below $\sim 10^4\,{\rm K}$
leading to the recombination of hydrogen in the accretion flow and development of the thermal instability. 
Because of that, the boundary given by dashed-dotted line is not stable.
At $\dot{M}\lesssim 10^{14}\,{\rm g\,s^{-1}}$, the accretion flow can turn into geometrically thick ADAF mode.
}
\label{pic:disc_zones}
\end{figure}

\subsubsection{\red{4.2.4\,} Different physical conditions in accretion discs around XRPs}

The picture of accretion disc interaction with the NS magnetosphere becomes even more complex if one takes into account physical conditions in the accretion disc \citep{2010arXiv1005.5279S}.
We would highlight a few states of accretion disc appearing at different mass accretion rates (see Fig.\,\ref{pic:disc_zones}):
\begin{itemize}[noitemsep,topsep=0pt]
{\setlength{\itemsep}{0pt}
\item At very low mass accretion rates ($\dot{M}\lesssim 10^{14}\,{\rm g\,s^{-1}}$), the mass density of accreting material is too low to cause sufficiently intensive cooling of accreting plasma. 
As a result, accretion flow reaches the virial temperature and turn into a state of geometrically thick advection dominated accretion flow (ADAF) \citep{1995ApJ...452..710N}.
This condition of accretion flow was not detected in XRPs so far.

\item Gas pressure dominated geometrically thin accretion disc composed of cold recombined material or relatively hot ionized gas (aka C-zone).

\item At sufficiently large mass accretion rates ($\dot{M}\gtrsim {\rm few}\times 10^{17}\,{\rm g\,s^{-1}}$, see red zone in Fig.\,\ref{pic:disc_zones}), the inner regions of accretion disc become radiation pressure dominated and geometrically thick (aka A-zone, see e.g. \citep{1973A&A....24..337S}).
Because of large optical thickness of accretion flow it can be advective \citep{1999AstL...25..508L}. 
Energy release in the disc leads to mass losses due to the radiation-driven outflows
\citep{2019MNRAS.484..687M,2019A&A...626A..18C}.
}
\end{itemize}

Gas pressure dominated accretion discs are most common under the observed conditions in the majority of XRPs (see Fig.\,\ref{pic:disc_zones}).
Under these conditions, the disc is geometrically thin.
Its relative scaleheight $H_{\rm m}$ at radial coordinate $r$ can be approximated by
\be\label{eq:H2R_Czone}
\left(\frac{H_{\rm m}}{r}\right)_{\rm C}\approx
0.03\,\alpha^{-1/10} \dot{M}_{17}^{3/20} m^{-3/8} r_8^{1/8},
\ee 
where $\alpha$ is the dimensionless viscosity parameter of order of $0.1$ \citep{2007ARep...51..549S}.
The phase transition of an accreting material between cold recombined state and hot ionised state causes development of thermal instability in the accretion flow.
Recombination of matter in the accretion flow leads to a decrease of the opacity, fast local cooling of accretion disc, reduction of its geometrical thickness and corresponding reduction of viscosity (similarly, ionisation of matter in the accretion flow leads to an increase of local viscosity).
Additionally, the phase transition between the ionised and recombined states can cause change of the dimensionless viscosity parameter $\alpha$ (see, e.g., \citep{2020AdSpR..66.1004H}), which is expected to be higher in ionised hot state than in recombined cold by a factor of $\alpha_{\rm hot}/\alpha_{\rm cold}\sim 5-10$.
Development of thermal instability leads to appearance of cooling and heating waves propagating inside-out or outside-in in the accretion disc.
Stable accretion onto the compact object is possible when the mass accretion rate is high enough to keep the entire accretion disc hot and fully ionised
\be\label{eq:dot_m_stable_up}
\dot{M}>\dot{M}_{\rm hot}\simeq 10^{16}\,\alpha_{\rm hot}^{0.01}m^{-0.89}
\left(\frac{r_{\rm out}}{10^{10}\,{\rm cm}}\right)^{2.68}\,\,{\rm g\,s^{-1}},
\ee 
where $r_{\rm out}$ is the effective outer disc radius,
or when the mass accretion rate is so low that the temperature is low enough to keep the accreting material recombined even at the inner radius
\be\label{eq:dot_m_stable_down}
\dot{M}<\dot{M}_{\rm cold}\simeq 4\times 10^{15}\,\alpha_{\rm cold}^{-0.004}m^{-0.88}
\left(\frac{r_{\rm in}}{10^{8}\,{\rm cm}}\right)^{2.65}\,\,{\rm g\,s^{-1}},
\ee 
where $r_{\rm in}=R_{\rm m}$ is the inner disc radius.
Note, that expressions (\ref{eq:dot_m_stable_up}) and (\ref{eq:dot_m_stable_down}) imply that accreting material has solar chemical composition and is dominated by hydrogen.
Different chemical composition results in different conditions for the critical mass accretion rates.

{The transition between the ionisation states can be significantly affected by accretion disc irradiation by the NS \citep{1998MNRAS.293L..42K,2022MNRAS.510.1837L}.
The irradiation can keep the inner regions of accretion disc in a hot state and prevent propagation of the cooling wave.
To clarify the influence of the irradiation, it is useful to estimate local irradiating flux
\be 
Q_{\rm irr}=\mathit{C}_{\rm irr}\frac{L}{4\pi r^2}, 
\ee 
where 
\be 
\mathit{C}_{\rm irr}=(1-A)\Psi(\theta)\frac{z_0}{r}q, \quad q=\left(\frac{{\rm d}\ln z_0}{{\rm d}\ln r}-1\right)
\ee 
is a dimensionless irradiation parameter \citep{2007ARep...51..549S,2022MNRAS.510.1837L}, $(1-A)\leq 1$ is the fraction of absorbed incident flux, $z_0$ is the semithickness of accretion disc, $\Psi(\theta)$ is the angular distribution of the irradiating flux (for the case of isotropic central source of radiation, $\Psi(\theta)=1$).
Taking $q\simeq 1/8$, which is valid for the C-zone of accretion disc \citep{1973A&A....24..337S}, assuming isotropic emission from the central source and using $(1-A)\simeq 0.1$ \citep{1999A&A...350...63S}, one gets $Q_{\rm irr}\approx 5\times 10^{16}\,L_{37} r_8^{-2}\,{\rm erg\,s^{-1}\,cm^{-2}}$.
Therefore, the irradiation can potentially keep the effective temperature in accretion disc above $10^4\,{\rm K}$ within the radial coordinate $r\sim 2\times 10^{10}\,L_{37}^{1/2}\,{\rm cm}$.
Note, however, that the internal temperature of accretion disc is affected by the illumination insignificantly (in the case of optically thick accretion disc, see, e.g., \citep{1999A&A...350...63S}) and therefore the influence of irradiation on the disc transition between hot and cold states is still questionable.
} 

Transitions between different ionisation states of an accretion disc are commonly considered to be responsible for bright outbursts in dwarf novae and soft X-ray transients  \citep{2001NewAR..45..449L}.
Recently, transitions were proposed to explain variations of the observed mass accretion rate in the end of X-ray outbursts in transient XRPs \citep{2017A&A...608A..17T}.
In this case, the outburst decay in transient XRP should be accompanied by two processes: 
(i) transition of accretion disc to the cold recombined state, when the cooling front moves outside-in, and
(ii) simultaneous expansion on the inner disc radius $R_{\rm m}$ due to reduction of mass accretion rate at the inner edge of the accretion disc (see expressions \ref{eq:R_A} and \ref{eq:R_A_2_R_m}).

At some point the cooling front reaches the inner disc radius, and, depending on the radial coordinate where it happens, one would expect different outcomes of the mass accretion rate decay: 
if the cooling front meets the inner disc radius outside the corotation radius, the outburst ends up with transition to the propeller state \citep{1975A&A....39..185I} (see Fig.\,\ref{pic:propeller_LC}); in the opposite case, accretion disc becomes entirely recombined and stable \citep{2017A&A...608A..17T}, i.e. further fast drop of the mass accretion rate stops (see Fig.\,\ref{pic:cold_disc_LC}).
Interestingly, the final state of the source after an outburst is determined by two fundamental parameters of the NS: its magnetic field and spin period. 
Equating the expressions for luminosities $L_{\rm cold}=GM\dot{M}_{\rm cold}/R$ and $L_{\rm prop}$ (\ref{eq:L_prop}), one can derive the critical value of the spin period, determining the source behaviour after an outburst, as a function of the NS magnetic field:
\be 
P^*\approx 38\,\Lambda^{6/7}B_{12}^{0.49}m^{-0.17}R_6^{1.22}\,\,{\rm sec}.
\ee 
If the spin period $P<P^*$, an XRP will end up in the propeller regime. 
Otherwise, the source will start to accrete steadly from the cold disc (see Fig.\,\ref{pic:cold_disc_vs_propeller}).
Some variations of the luminosity are possible even in the state of accretion from the cold disc, but because of low viscosity and correspondingly long viscose time scales the variations are expected to be slow.
It is still not known quite well, how cold and largely recombined accretion disc interacts with the NS magnetic field.
Larger mass density of a cold disc (due to the low viscosity) and peculiarities of its interaction with the $B$-field may lead to a dependence on the mass accretion rate different from that given by (\ref{eq:R_A}).

\begin{figure}
\centering 
\includegraphics[width=10.cm, angle =0]{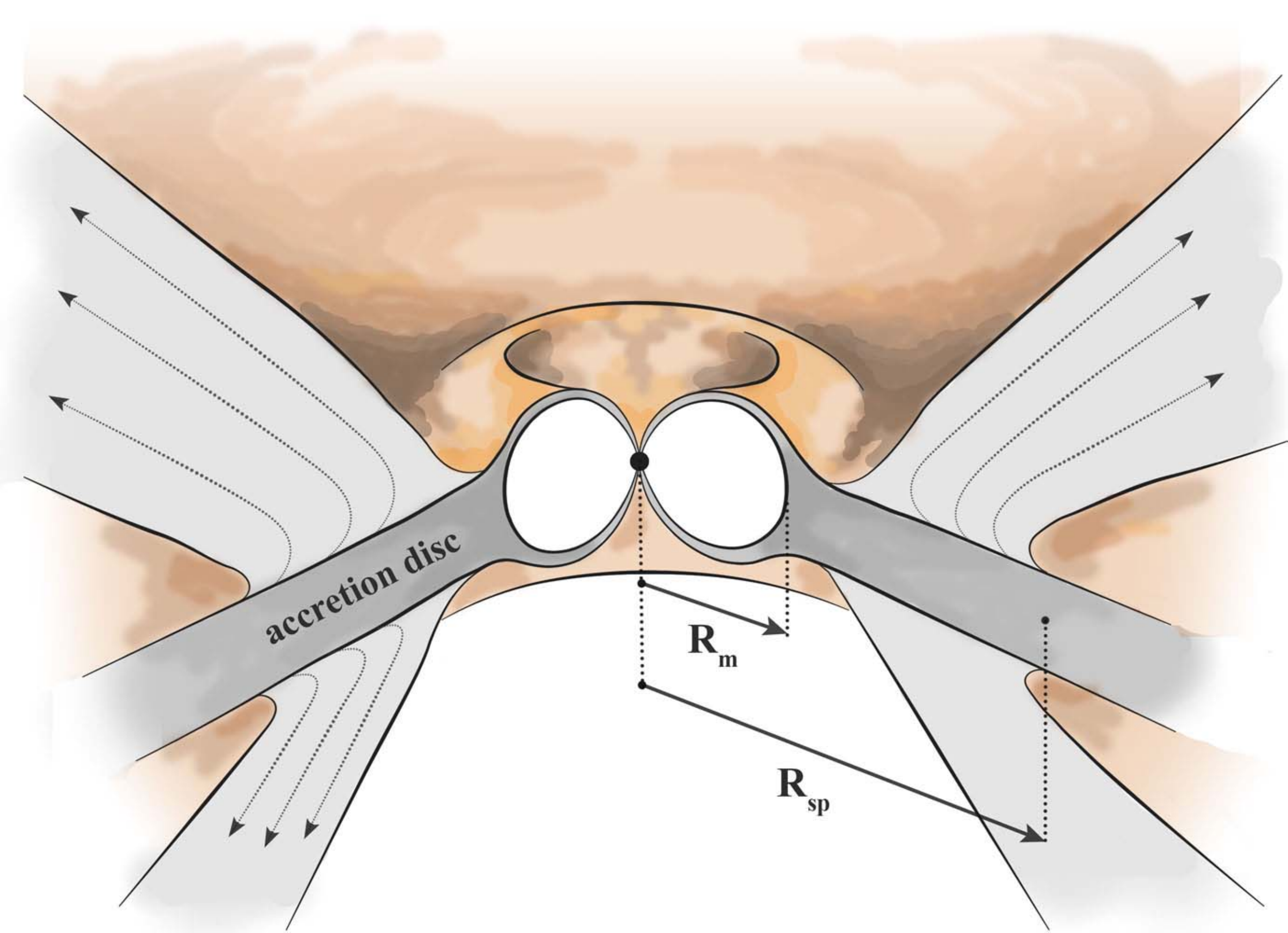} 
\caption{
Schematic structure of accretion flow at extremely high mass accretion rates onto strongly magnetised NS.
If the spherisation radius $R_{\rm sp}$ exceeds the magnetospheric radius $R_{\rm m}$, accretion disc starts to lose matter via wind. 
At sufficiently high mass accretion rates, the accretion flow covering magnetosphere becomes optically thick.
From \citep{2019MNRAS.484..687M}.
}
\label{pic:winds_ULX}
\end{figure}

Radiation pressure dominated accretion discs are typical for the brightest XRPs 
($L>3.4\times 10^{38}\,\Lambda^{21/22}B_{12}^{6/11}m^{6/11}\,\ergs$).
Accretion disc is expected to be radiation pressure dominated at radial coordinates \citep{2007ARep...51..549S}
\be 
r<R_{\rm rad}\approx 2.2\times 10^8\,
L_{39}^{16/21} m^{-3/7} R_6^{16/21}\,\,{\rm cm}.
\ee 
Geometrical scaleheight of such a disc is independent of radial coordinate:
\be 
H_{\rm d,rad}\approx 10^7\,L_{39}\,m^{-1} R_6\,\,\,{\rm cm}.
\ee 
At sufficiently high mass accretion rates, the radiation pressure gradient in the inner parts of accretion disc becomes high enough to compensate gravitational attraction in the direction perpendicular to the disc plane.
Then the accretion disc generates winds driven by radiation force, spending a fraction $\varepsilon_{\rm w}\in[0;1]$ of viscously dissipated energy to launch the outflows.
As a result, only a fraction of the mass accretion rate from the donor star reaches the boundary of NS magnetosphere and accretes onto the central object.

\begin{figure}
\centering 
\includegraphics[width=12.cm, angle =0]{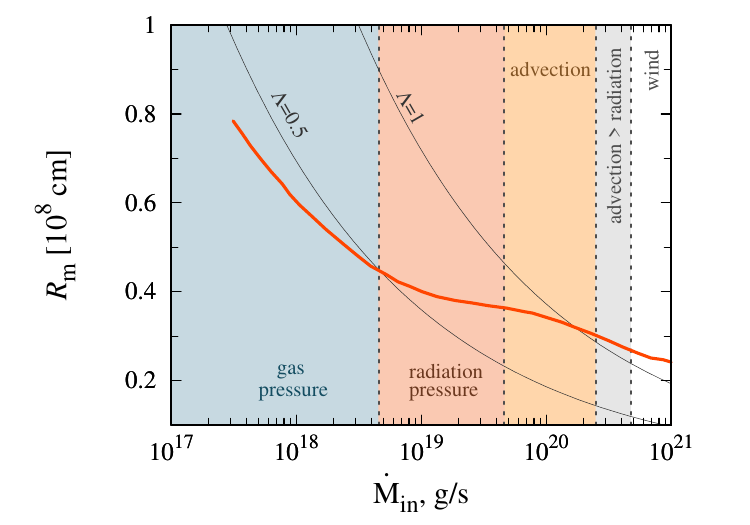} 
\caption{
Magnetospheric radius $R_{\rm m}$ as a function of the mass accretion rate at the inner disc radius for a NS with dipole magnetic moment $\mu=10^{30}\,\,{\rm G\,\,cm^3}$.
Segments of the red solid curve correspond to the different regimes of accretion near the magnetospheric boundary.
{One can see that $\Lambda\approx 0.5$ in the case of gas pressure dominated inner parts of the accretion disc.}
The inner disc radius is weakly dependent on the mass accretion rate in the case of radiation pressure dominated non-advective regime of accretion.
At higher mass accretion rates, when the inner disc parts are under the influence of advection, the dependence of the inner disc radius on the mass accretion rate becomes stronger, {but $\Lambda\approx 1$. }
The red curve is reproduced from \citep{2019A&A...626A..18C}.
}
\label{pic:Chashkina_R_m}
\end{figure}

Outflows launched by radiation pressure come into play within the spherization radius $R_{\rm sp}$ \red{(see Fig.\,\ref{pic:winds_ULX})}, inside of which the radiation force due to the energy release in the disc is no longer balanced by gravity. This radius can be roughly estimated as \citep{1999AstL...25..508L,2007MNRAS.377.1187P}:
\be\label{eq:R_sp}
R_{\rm sp}\approx 9\times 10^5\,\dot{m}_0\, 
\left[ 1.34-0.4\varepsilon_{\rm w}+0.1\varepsilon_{\rm w}^2 
- (1.1-0.7\varepsilon_{\rm w})\dot{m}_0^{-2/3} \right]\,\, {\rm cm},
\ee
where $\dot{m}_0=\dot{M}_0/\dot{M}_{\rm Edd}$ is the dimensionless mass accretion rate from the donor star in units of Eddington mass accretion rate at the NS surface:
\be 
\dot{M}_{\rm Edd}=\frac{L_{\rm Edd} R}{GM}\approx 1.9\times 10^{18}\,R_{6}\,\,\,{\rm g\,s^{-1}}.
\ee
In the case of accreting BHs with an accretion disc extends down to the innermost stable orbit (ISCO), and NS with low magnetic field, the outflow can carry out a significant amount of material from the accretion flow. 
In the case of XRPs, the inner disc radius is much larger than the radius of the ISCO and the fraction of material carried out by the outflow is dependent both on the mass accretion rate and magnetic field strength of the NS.
The mass accretion rate at a given radial coordinate $r$ can be estimated as 
\be\label{eq:dotM(R)}
\dot{M}(r)=\dot{M}_{\rm ISCO}+(\dot{M}_0 -\dot{M}_{\rm ISCO})\frac{r}{R_{\rm sp}}, 
\ee
where $\dot{M}_0$ is the mass accretion rate outside the spherisation raduis, and 
\be\label{eq:dotM_ISCO}
\dot{M}_{\rm ISCO}=\dot{M}_0\,\frac{1-A}{1-A\,(0.4\,\dot{m}_0)^{-1/2}}, 
\ee
is the expected mass accretion rate at ISCO, where $A\approx \varepsilon_{\rm w}(0.83-0.25\varepsilon_{\rm w})$.
Combining (\ref{eq:R_A}), (\ref{eq:R_A_2_R_m}) and (\ref{eq:dotM(R)}) one can get the dependence of the inner disc radius on the mass accretion rate from the companion star $R_{\rm m}(\dot{M}_0)$ and the mass accretion rate at the inner disc radius (i.e. onto the NS surface) for the case of advective accretion discs. 
The mass loss from the disc due to the wind requires that accretion flow penetrates deeper into the magnetosphere of a NS \citep{2019MNRAS.484..687M,2019A&A...626A..18C}. 
Note, however, that at high mass accretion rates, when the inner parts of the disc become geometrically thick, one additional ingredient starts to affect the displacement of the inner disc radius. 
Because thick accretion disc intercepts a large fraction of X-ray radiation from the central source, the inner disc radius is determined by the balance between ram pressure, magnetic field pressure and, additionally, radiative force acting on its inner edge \citep{2017MNRAS.470.2799C}.
Effectively, it can be considered as dependence of $\Lambda$ parameter in relation (\ref{eq:R_A_2_R_m}) on the mass accretion rate at the inner disc radius. 
Detailed analyses shows that the inner disc radius tend to decrease with the increase of the mass accretion rate till the disc is gas pressure dominated (as it is predicted by equations \ref{eq:R_A} and  \ref{eq:R_A_2_R_m}), then the inner disc radius is almost independent of the mass accretion rate, and starts to decrease with the mass accretion rate again as soon as advection comes into play (see, e.g., \citep{2019A&A...626A..18C} and Fig.\,\ref{pic:Chashkina_R_m}). 

Strong outflows from accretion discs, expected at high mass accretion rates, can cause beaming of X-ray radiation in the direction orthogonal to the disc plane \citep{2019MNRAS.485.3588K}.
The effectiveness of beaming due to the wind launch is still discussed.
At the same time, there is no doubts that the winds are launched at high mass accretion rates and their evidence were already found in observations \citep{2018MNRAS.479.3978K}.

\subsubsection{\red{4.2.5\,} Stochastic fluctuations of the mass accretion rate}

{
As we have already seen, the X-ray flux in XRPs shows strong aperiadic fluctuations observed in a wide range of Fourier frequencies (see examples of PDS in Fig.\,\ref{pic:PDS_2}).
The aperiodic variability is a natural feature of accretion process through the wind, where one would expect variability due to the clumpiness of accretion flow, or through the disc, where the variability is a result of propagating fluctuations of the mass accretion rate \citep{1997MNRAS.292..679L}.
}

\begin{figure}
\centering 
\includegraphics[width=8.cm, angle =0]{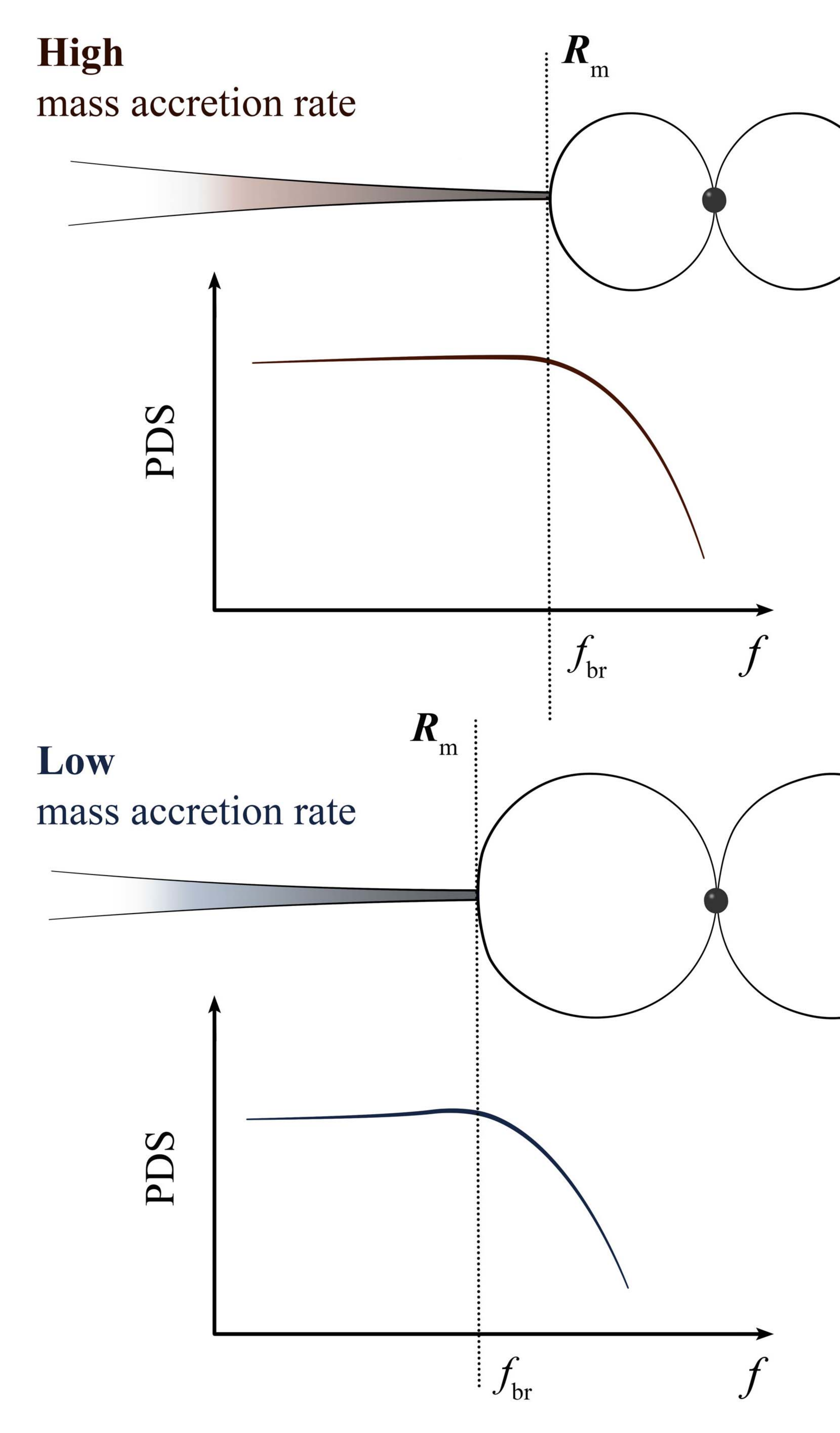} 
\caption{
Structure of the disc-magnetosphere interaction as a function of mass accretion rate in XRPs and the corresponding changes in the observed PDS.
The PDS shows break at Fourier frequency corresponding to the dynamo frequency at the inner disc radius.
The frequency of the magnetic dynamo is expected to be proportional to the local Keplerian frequency.
Thus, the increase of the mass accretion rate and corresponding decrease of the inner disc radius result in the break's shift towards higher frequencies.
Measurement of the break frequency in PDS of XRP can be used to estimate geometrical size of the NS magnetosphere.
}
\label{pic:PDS_theory}
\end{figure}

Let us focus on the accretion disc case as prevalent among known XRPs.
The inward mass transfer in the accretion disc is possible thanks to viscosity and friction between the adjacent rings in the disc, which helps to transfer the angular momentum of accreting material outwards (see, e.g., \citep{2010arXiv1005.5279S}).
The viscosity in the accretion disc is likely caused by the magnetic dynamo that generates a poloidal magnetic field component in a random fashion \citep{1991ApJ...376..214B,1995ApJ...440..742H}.
The initial fluctuations of the mass accretion rate arise all over the disc and then propagate inwards and outwards \citep{2018MNRAS.474.2259M}, modulating the fluctuations arising at other radial coordinates. 
The timescale of the magnetic dynamo is close to the local Keplerian time scale \citep{1992MNRAS.259..604T,1996ApJ...463..656S}:
\be\label{eq:TimeKepler}
t_{\rm dyn}\sim
t_{\rm K}=\frac{2\pi}{\Omega_{\rm K}}\propto r^{3/2}.
\ee 
Therefore, different time scales are introduced into the accretion flow at different distances from a NS, while the observed variability of X-ray flux reflects the variability of the mass accretion rate at the inner parts of accretion disc \citep{2001MNRAS.327..799K, 2016AN....337..385I,2018MNRAS.474.2259M}. 
The shortest timescale of the aperiodic variability corresponds to the inner disc radius and is expected to be close to the Keplerian frequency at $R_{\rm m}$ (\ref{eq:Omega_K_Rm}).
The detailed models of propagating fluctuations of the mass accretion rate predict broadband PDS with a break at the dynamo frequency corresponding to $R_{\rm m}$.
Because the inner disc radius depends on the NS magnetic field and mass accretion rate, the displacement of breaks in PDS is variable from one source to another.
In transient XRPs the break frequency depends on the accretion luminosity reflecting the variability of the magnetospheric radius (Fig.\,\ref{pic:PDS_theory}), as observed in some XRPs (for example A~0535+26, Fig.\,\ref{pic:PDS_2}\,{\it left} and \ref{pic:PDS_Mike}).

This schematic picture of the aperiodic variability in XRPs is not the ultimate truth. 
On top of variability caused by propagating fluctuations of the mass accretion rate, one would expect several other phenomena affecting broadband PDS.
Among them are 
(i) variability of the geometry of the emitting region at the NS surface, 
(ii) appearance of the photon bubbles at extremely high accretion luminosity (see next section, \citep{1996ApJ...457L..85K,1996ApJ...469L.119K}), 
(iii) partial reprocessing of X-ray emission by the accretion flow in between the inner disc radius and NS surface \red{(see Fig.\,\ref{pic:winds_ULX} and \citep{2019MNRAS.484..687M} for details)},
(iv) development of instabilities in radiation pressure dominated part of an accretion disc \citep{1996ApJ...466L..31C}.

\subsection{\red{4.3\,} Geometry and physics of the emitting region at the NS surface}

If the source is in accretion mode, plasma penetrates into the NS magnetosphere at $R_{\rm m}$ and following magnetic field lines reaches NS surface in a small regions located close the magnetic poles of a star. 
The size of the landing region could be roughly estimated as 
\be 
r_{\rm pol}\approx R \left(\frac{R}{R_{\rm m}}\right)^{1/2}
\approx 7\times 10^4\,\Lambda^{-1/2} B_{12}^{-2/7} \dot{M}_{17}^{1/7} m^{1/14} R_6^{9/14}\,\,{\rm cm},
\ee 
which gives $r_{\rm pol}\lesssim 1\,{\rm km}$ for typical XRPs parameters. 
The shape of the landing regions is determined by the geometry of accretion funnel above the NS surface. 
The accretion funnel is filled and the landing region can be close to the filled circle of area
\be \label{eq:S_spot_w}
S_{\rm spot}^{\rm (w)}=\pi r_{\rm pol}^2\approx
5\times 10^9\,\Lambda^{-1} B_{12}^{-4/7} \dot{M}_{17}^{2/7} m^{1/7} R_6^{9/7}\,\,{\rm cm^2}
\ee 
in the case of accretion  from the stellar wind.
In the case of disc accretion, the accretion funnel is hollow, and the landing region can form a ring-like structures around the NS magnetic poles. 
The geometrical thickness of the rings is determined by the inner disc radius and the width of the transition zone in it.
{Assuming that the transition zone at the magnetospheric boundary has a typical scale similar to the accretion disc scaleheight $H_{\rm m}$ at the inner disc radius, we can estimate the geometrical thickness of accretion channel in close proximity to the NS surface:
\be\label{eq:d_0}
d_0\approx R\left(\frac{R}{R_{\rm m}}\right)^{1/2}\frac{H_{\rm m}}{R_{\rm m}}
\ee 
and the area of accretion channel base
\be\label{eq:S_spot_d}
S_{\rm spot}^{\rm (d)}=2\pi r_{\rm pol} d_0 
\approx 2\pi\,\frac{R^3 H_{\rm m}}{R_{\rm m}^2}.
\ee 
This is a reasonable approximation in the case of accretion disc truncated in a C-zone (see Fig.\,\ref{pic:disc_zones}), where the disc is geometrically thick and gas pressure dominated.
Then using approximate disc scaleheight in a C-zone (\ref{eq:H2R_Czone}) we can rewrite (\ref{eq:d_0}) as 
\be 
d_{0,{\rm C}}\approx 2.3\times 10^3\,\alpha^{-1/10}\Lambda^{-3/8} B_{12}^{-3/14} \dot{M}_{17}^{9/35} m^{-9/28} R_6^{3/4}\,\,{\rm cm},
\ee 
and (\ref{eq:S_spot_d}) as 
\be \label{eq:S_spot_d_C}
S_{\rm spot,C}^{\rm (d)}\approx
10^9\,\alpha^{-1/10}\Lambda^{-7/8} B_{12}^{-1/2} \dot{M}_{17}^{2/5} m^{-1/4} R_6^{39/28}\,\,{\rm cm^2}.
\ee 
In the case of accretion disc truncated in the radiation pressure dominated zone (A-zone in Fig.\,\ref{pic:disc_zones}), the thickness of the transition region can be smaller than the accretion disc scaleheight, and estimates require additional analyses. 
}
Development of instabilities inside an accretion channel can result in filled funnel even in the case of accretion from the disc. 
Inclination of the dipole magnetic axis to the disc plane can result in asymmetric accretion funnel and fractionally filled ring-like structures at the NS surface (see numerical simulations performed in \citep{2013MNRAS.433.3048K}).
Moreover, possible exchange instability in the accretion flow above the stellar surface can result in a stochastic shape of the landing region \citep{1980A&A....90..359H}. 

Before the inevitable deceleration in the vicinity of the NS surface, the velocity of accreting material is expected to be close to the free-fall velocity (even if one accounts for matter motion along dipole magnetic filed lines).
The temperature of accreting plasma is determined by synchrotron emission and inverse Compton \citep{1976SvAL....2..111S}.
We have already seen that the effective temperature in the emitting region 
is expected to be $\gtrsim 1\,{\rm keV}$ (see equation \ref{eq:T_eff}).
Using the estimates for the hot spot area (\ref{eq:S_spot_w}) and  (\ref{eq:S_spot_d_C}), we can improve the estimate of the effective temperature and obtain
\be 
T_{\rm eff}^{\rm (w)}=\left(\frac{L}{2\sigma_{\rm SB}S_{\rm spot}^{\rm (w)}}\right)^{1/4}
\simeq 5.7\,\Lambda^{1/4} B_{12}^{1/7} L_{37}^{5/28} m^{1/28} R_6^{11/28}\,\,{\rm keV}
\ee 
for the case of accretion from the wind, and
\be 
T_{\rm eff}^{\rm (d)}=\left(\frac{L}{2\sigma_{\rm SB}S_{\rm spot}^{\rm (d)}}\right)^{1/4}
\simeq
8.6\,\alpha^{1/40}\Lambda^{7/32} B_{12}^{1/8} L_{37}^{3/20} m^{13/80} R_6^{-0.45}\,\,{\rm keV}
\ee 
for the case of accretion from the gas pressure dominated disc.

At the polar regions of the NS, accretion flow is decelerated via different mechanisms depending on the mass accretion rate.
At low mass accretion rates, when the radiation pressure is insignificant, the flow is stopped either by Coulomb collisions in the atmosphere of a NS \citep{1969SvA....13..175Z,1975A&A....42..311B,1976JETP...43..389P}, or in the collisionless shock above the NS surface \citep{1982ApJ...257..733L,2004AstL...30..309B}.  
At high mass accretion rates, radiation pressure affects dynamics of accretion flow and material can be stopped above the NS surface in a radiation-dominated shock \citep{1976MNRAS.175..395B}.
Let us consider each of these possibilities: 

\begin{itemize}
{\setlength{\itemsep}{0pt}
\item {\it Coulomb collisions and hot spots at the NS surface}

In this case, deceleration of the accretion flow happens on a short stopping length in the top layer of the NS atmosphere 
\citep{1981MNRAS.195P..45K,1984ApJ...278..369H,1989ApJ...346..405M}
and
accretion process results in hot spots at the stellar surface.
The most of the kinetic energy of the accretion flow is carried by ions, which loose their kinetic energy in ion-electron Coulomb collisions occurring with or without electron excitation to high Landau levels.
The stopping power of the plasma and the stopping length is also affected by the generation of the collective plasma oscillations.
Typical stopping length in the atmosphere of a NS can be estimated as a few tens of ${\rm g\,cm^{-2}}$ \citep{1984ApJ...278..369H,1993ApJ...418..874N}.
Accounting for the material spreading over the NS surface leads to the conclusion that the emitting region may have more complex geometry with accretion mound forming at the base of the accretion channel (see Fig.\,\ref{pic:accretion_mound}, \citep{2013MNRAS.430.1976M}). 

\begin{figure}
\centering 
\includegraphics[width=11.5cm, angle =0]{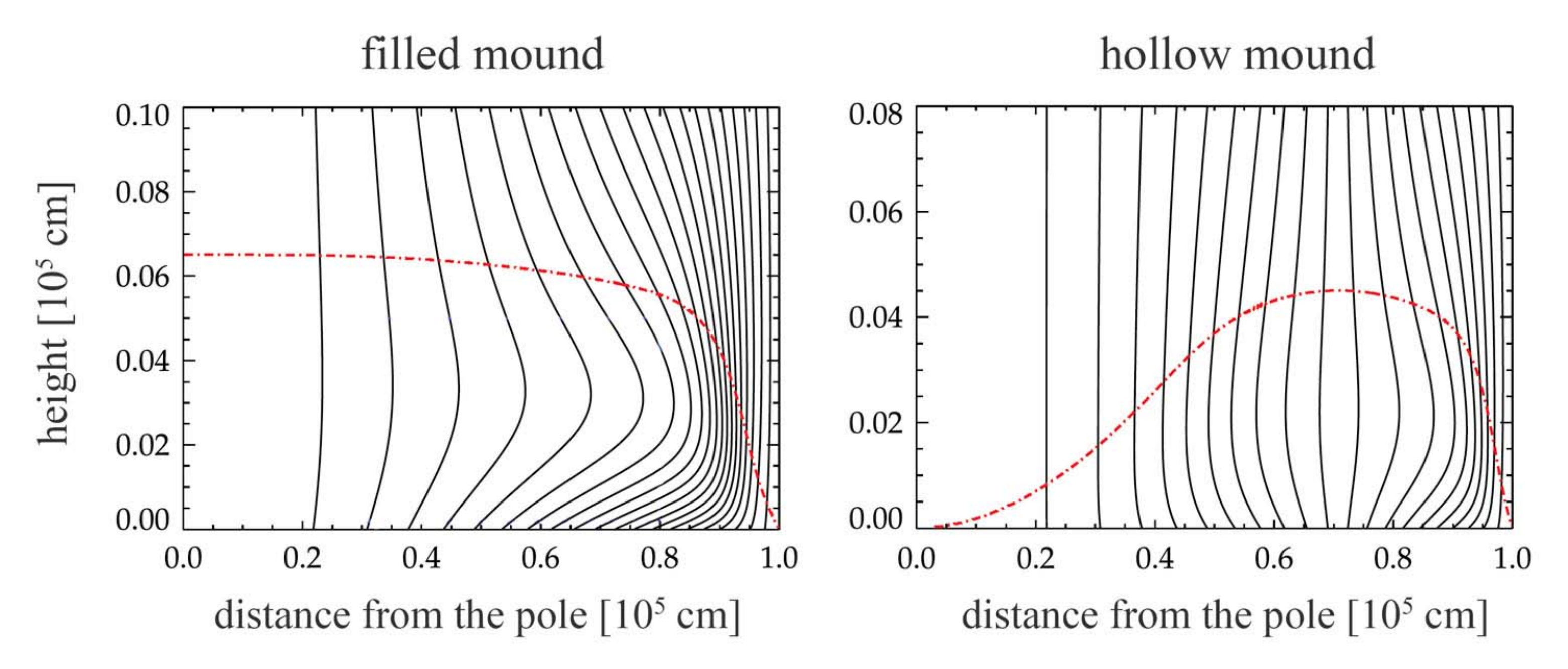} 
\caption{
The structure of the accretion mound in the case of filled {\it (left)} and hollow {\it (right)} accretion channel.
The horizontal axis corresponds to the distance from the NS magnetic pole, the vertical axis represents the height above the NS surface.
Magnetic field lines distorted by plasma spreading over the NS surface are shown as black solid lines.
Red dash–dotted line denotes the top of the mound.
Magnetic field at the NS surface is taken to be $\sim 10^{12}\,{\rm G}$.
The curves are reproduced from \citep{2013MNRAS.430.1976M}.
}
\label{pic:accretion_mound}
\end{figure}

\item{\it Collisionless shock}

Collisionless shock may arise due to the development of instabilities in plasma moving along magnetic field lines towards NS magnetic poles. 
To the best of our knowledge, no quantitative studies of such instabilities were performed up to date. 
However, some models of collisionless shocks were developed base of the assumption that such shock does develop \citep{1982ApJ...257..733L}.
Temperature of plasma right after the shock region can be estimated under the assumption that the process of plasma heating is adiabatic.
Then the electron gas is cooling rapidly due to the inverse Compton effect and synchrotron emission.
The cooling process of ions is much slower and the ion gas looses its energy mostly due to interaction with the cooler gas of electrons. 
As a result, one would expect two-temperature plasma below the shock region. 

\item {\it Radiation-dominated shock and accretion columns}

The increase of the mass accretion rate leads to the increase of accretion luminosity and strengthening of the radiative force applied to the accretion flow above the surface.
At certain accretion luminosity the radiation pressure becomes strong enough to stop accretion flow above the NS surface \citep{1976MNRAS.175..395B}:
\be\label{eq:L_crit}
L_{\rm crit}\approx 4\times 10^{36}\,
\left(\frac{\sigma_{\rm T}}{\sigma_{\rm eff}}\right)
\left(\frac{l_0}{2\times 10^{5}\,{\rm [cm]}}\right)
\frac{m}{R_6}\,\,\ergs,
\ee 
where $\sigma_{\rm T}$ is non-magnetic Thomson scattering cross section, $\sigma_{\rm eff}$ is the effective cross section in accretion channel above NS surface, and $l_0$ is the length of accretion channel base.
This luminosity is called  ``critical".
Note, that it is not enough to compensate the gravitational force by the radiative one in order to stop accretion flow above the NS surface (which moves with free-fall velocity). 
Because the scattering cross section in a strong magnetic field depends on the photon energy and $B$-field strength, effective scattering cross section $\sigma_{\rm eff}$ and the critical luminosity (\ref{eq:L_crit}) depends on the field strength at the NS surface \citep{2015MNRAS.447.1847M}.
The critical luminosity was determined observationally in two transient XRPs up to now: 
V~0332+53 \citep{2017MNRAS.466.2143D} 
and 
A~0535+262 \citep{2021ApJ...917L..38K}. 

\begin{figure}
\centering 
\includegraphics[width=8.cm, angle =0]{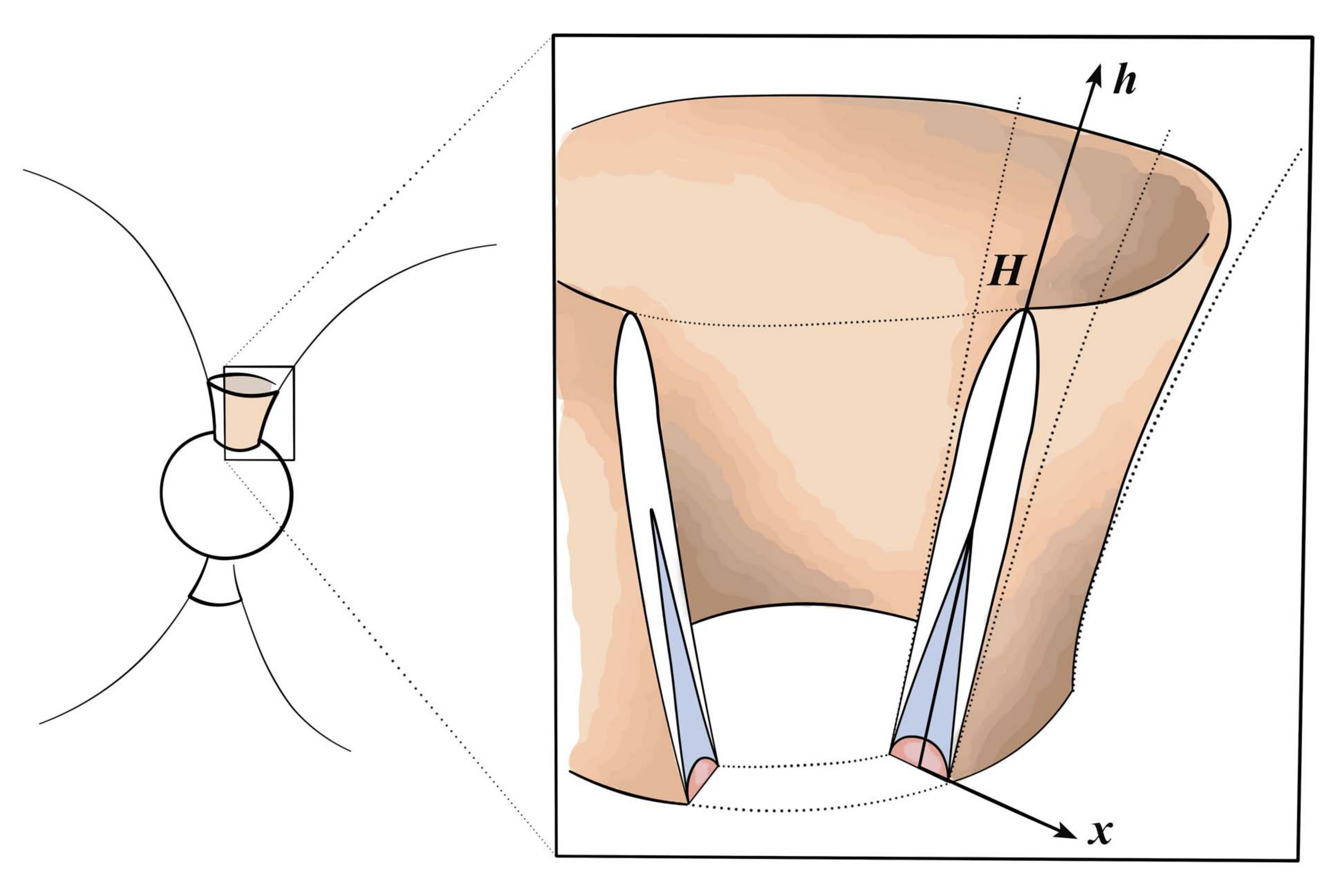} 
\caption{
Schematic structure of hollow accretion column confined by a strong magnetic field and supported by radiation pressure above the surface of a NS.
The height of the column can be comparable to the NS radius. 
At extreme mass accretion rates, the column can potentially turn in the advective regime, when X-ray photons are confined within the accretion flow.
The photons originated from the region marked in blue on the plot cannot leave accretion channel because the diffusion time scale for them are larger than the dynamic time scale.
There is possible a small region on the bottom of the column (marked in red), where the mass density is so high that the gas pressure dominates over the radiation pressure. 
From \citep{2018MNRAS.476.2867M}.
}
\label{pic:accretion_column}
\end{figure}

Luminosity higher than the critical one results in appearance of radiation-dominated shock and accretion columns above the polar regions of accreting NS \citep{1975PASJ...27..311I,1976MNRAS.175..395B,1981A&A....93..255W,2015MNRAS.454.2539M,2021MNRAS.501..564G}.
Accretion columns are confined by a strong magnetic field and supported by radiation pressure. 
The theory of the accretion column structure is one of the key unsolved problems in the physics of XRPs. 
The problem is highly complex, especially at high accretion rates when there is strong coupling between matter and radiation in the accretion channel.
Even accounting for non-magnetic radiative transfer requires  usage of advanced numerical calculations and huge computational resources \citep{2016PASJ...68...83K}.
Order of magnitude estimations show that accretion columns can provide luminosity as high as $10^{40}\,\ergs$ and, therefore, can explain the observed luminosity of recently discovered ULX pulsars \citep{2014Natur.514..202B,2017Sci...355..817I}.
At the same time, it is known that the accretion column structure and maximal luminosity depend on the geometry of the accretion channel (in particular on its geometrical thickness, determined by accretion flow interaction with the NS magnetosphere at $R_{\rm m}$, see \citep{2015MNRAS.454.2539M}) and magnetic field structure in the vicinity of the NS surface \citep{2021MNRAS.504..701B}.

Additionally, at very high mass accretion rates, photon diffusion from accretion column becomes inefficient. 
It leads to development of the photon bubble instability \citep{1989ESASP.296...89K,1992ApJ...388..561A,1996ApJ...457L..85K}, which causes strong time-dependent density fluctuations inside the accretion column.
The presence of these fluctuations may alter the time-averaged structure of the column. 
Because photons will tend to preferentially escape along low-density channels, photon bubbles may ultimately determine the angular distribution of X-ray photon emission, reduce the efficacy of radiation pressure, and affect the estimates of the mass accretion rates above which photons can be trapped by advection process.
It was predicted that photon bubble instability can be observed as quasi-periodic oscillations at Fourier frequences from $\sim 10^2$ to $10^4\,{\rm Hz}$ \citep{1996ApJ...457L..85K}.
The signs of photon bubble instability were not detected solidly so far \citep{2015MNRAS.451.4253R}, but excess of aperiodic variability detected in some sources at high mass accretion rates (see, for example, Fig.\,\ref{pic:PDS_2} {\it right}) were interpreted by some authors as an evidence of photon bubbles \citep{1996ApJ...469L.119K}.

Recent studies have shown that accretion columns can be strongly influenced by advection \citep{2018MNRAS.476.2867M}, when photons are locked inside the accretion column, at extremely high mass accretion rates ($>10^{20} {\rm g\,s^{-1}}$) typical for the brightest XRPs. 
In this case the opacity in accretion channel and, therefore, the dynamics of accretion flow can be affected by the production of electron-positron pairs \citep{2019MNRAS.485L.131M} and possibly strong neutrino emission \citep{2018MNRAS.476.2867M,1992PhRvD..46.4133K}.

}
\end{itemize}

\begin{figure}
\centering 
\includegraphics[width=10.cm, angle =0]{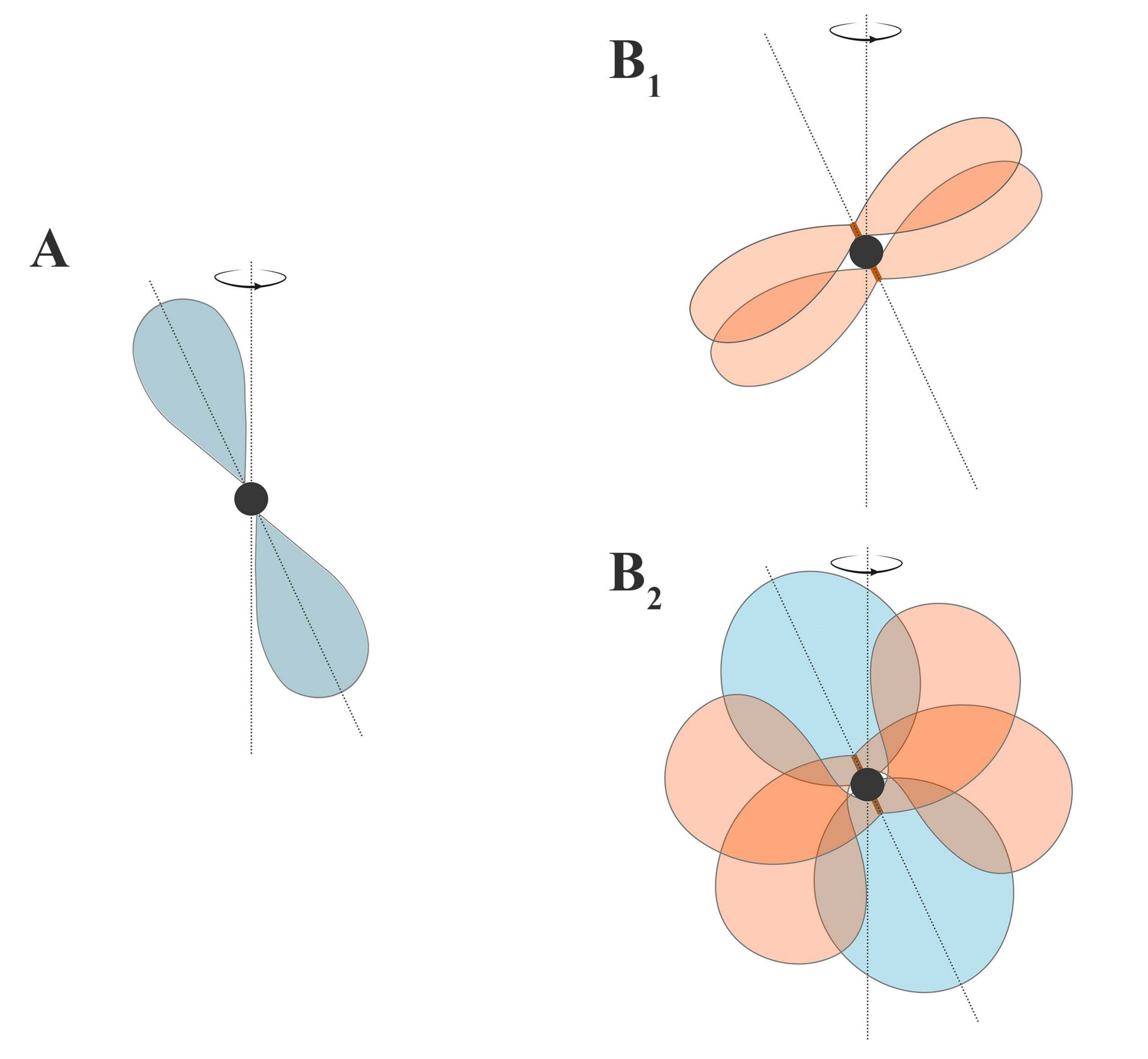} 
\caption{
Beaming of X-ray radiation from the NS in XRP at sub-critical mass accretion rates, when the accretion process results in hot spots at the NS surface (A), and at super-critical mass accretion rates, when the accretion columns arise above the stellar surface (B$_1$ and B$_2$).
The B$_2$ case implies beaming of accretion column radiation towards the NS surface. 
In this case, the X-ray flux detected by a distant observer is composed of the direct radiation from the columns (red pieces of the beam pattern) and the radiation intercepted and reflected by the NS atmosphere (blue pieces of the beam pattern).
}
\label{pic:beam_paterns}
\end{figure}

Different geometries of emitting regions above the NS surface imply different scenarios of spectra formation and different beam patterns \citep{1973A&A....25..233G}. 
In particular, it is natural to expect pencil beam diagram from the accretion with hot spots at the NS surface (see case A in Fig.\,\ref{pic:beam_paterns}), while collisionless shocks and radiation-dominated accretion columns would cause fan beam diagram (see case B$_1$ in Fig.\,\ref{pic:beam_paterns}). 
The beam pattern naturally determines the pulse profiles in XRPs.
Therefore, observations of XRPs and variations of their pulse profiles over a wide range of accretion luminosity can provide an evidence of changes in geometry of emitting region caused by changes of the mass accretion rate.
In particular, switch from the hot spot geometry to accretion column and corresponding change of a beam pattern can cause phase shift in the observed pulse profile. 
In fact, the process of pulse profile formation is far more complicated than it might seem on the base of simplified models. 
There are at least two features to be taken into account in the models. 
One of them is influence of gravitational light bending on the pulse profiles formation 
\citep{1988ApJ...325..207R,2001ApJ...563..289K,2006MNRAS.373..836P}
, which, in the case of bright XRPs with accretion columns, can cause significant increase of the observed pulsed fraction and might affect the relation between the apparent and actual luminosity \citep{2018MNRAS.474.5425M,2020PASJ...72...34I}.
The second feature which complicates the pulse profile modelling is the inclusion of additional sources of X-ray emission.
In the case of accretion column formation in bright XRPs, a fraction of X-rays emitted by the column is intercepted by the NS surface and reprocessed by its atmosphere  (see case B$_2$ Fig.\,\ref{pic:beam_paterns}, \citep{2013ApJ...777..115P}).
This process influences both spectra \citep{2015MNRAS.452.1601P} and pulse profile formation \citep{2015MNRAS.448.2175L,2018MNRAS.474.5425M}.

\subsection{\red{4.4\,} Spectra formation}

\subsubsection{\red{4.4.1\,} Challenges and complications}

We have to confess that there is still no self-consistent model of spectra formation in XRPs that would cover a wide range of accretion luminosities. 
Observational data tell us that over a wide luminosity range the broadband energy spectra of XRPs can be described by an absorbed power-law model modified with an exponential cut-off at high energies. 
Some sources show significant deviations at low luminosity, but this is not a general trend.

The basic ingredients of the spectra formation modelling in XRPs are (i) geometry of the emitting region (which can be dependent on the mass accretion rate), (ii) details of energy release mechanism and production of seed photons, and (iii) all major processes influencing radiative transfer. 
The set of possible geometries of the emitting region can be reduced to two general configurations: the case of a plane-parallel atmosphere (implying sub-critical mass accretion rates) and the case of a cylindrical emitting region (implying super-critical mass accretion rates or formation of collisionless shock in the low luminosity state).
The major processes determining spectral formation in XRPs are magnetic Compton scattering \citep{1986ApJ...309..362D}, cyclotron emission/absorption \citep{1991ApJ...374..687H}, and magnetic bremsstrahlung (see \citep{1992herm.book.....M} for review).
At extreme mass accretion rates, electron-positron pair creation and annihilation can come into play.
In strongly magnetised plasma, the cross section of all of these processes show dramatic dependence on the photon polarization state, photon energy and momentum direction. 
All these make calculation of radiative transfer very complicated task, especially in the case of its strong coupling with the dynamical structure of the emitting region.

Because cross sections of the major processes responsible for the radiative transfer are polarization dependent \citep{1992herm.book.....M,2006RPPh...69.2631H}, the radiation leaving the emitting regions in XRPs is expected to be highly polarised.
Models predicting polarisation in XRPs should account both for radiative transfer in the emitting regions and influence of an extended NS magnetosphere on polarisation of X-ray photons.
The latter is due to the QED vacuum polarisation \citep{1978SvAL....4..117G,1984ASPRv...3..197P}, which effectively results in rotation of photon polarisation plane while the photons propagate in the NS magnetosphere. 
Rotation of the polarisation plane due to the photon interaction with NS magnetic field is noticeable within the so-called adiabatic radius $R_{\rm ad}$.
For the case of the dipole magnetic field configuration, the adiabatic radius can be estimated as
\be 
R_{\rm ad}\simeq 7.6\times 10^6\,B_{12}^{0.4}E_{\rm keV}^{0.2}R_6\,\,{\rm cm},
\ee 
where $E_{\rm keV}$ is the photon energy in keV \citep{2016MNRAS.459.3585G}.
Outside the adiabatic radius, the effects of vacuum polarization can be neglected.

\subsubsection{\red{4.4.2\,} Broadband energy spectra}

The first attempts to model spectral formation in XRPs were based on the Feautrier numerical scheme \citep{1978stat.book.....M} and performed by Nagel \citep{1981ApJ...251..278N,1981ApJ...251..288N}, who studied radiative transfer in  two specific geometries: the slab and the cylinder. 
The calculations were performed either taking into account angular redistribution of photons but  assuming coherent scattering \citep{1981ApJ...251..288N}, or allowing the energy exchange in scattering events but neglecting angular redistribution \citep{1981ApJ...251..278N}.
These calculations were improved later by taking into account both energy exchange and angular redistribution of X-ray photons due to the scattering, as well as the vacuum polarization effects \citep{1985ApJ...298..147M,1985ApJ...299..138M}. 
These works, however, have solved the radiative transfer problem under the assumptions of constant temperature, constant mass density in the atmosphere, non-relativistic Maxwellian distribution of electrons, and the electron gas being at rest.
It has been shown that the classical power-law spectra with the cut-off at high energies can be produced in the accretion shock due to bulk and thermal Comptonisation of soft seed photons \citep{1982SvAL....8..330L,2007ApJ...654..435B}.
In \citep{2007ApJ...654..435B} it was possible to get analytical solutions for cylindrical geometry in the form of Green functions.
The analytical solutions, however, do not account for coupling between hydrodynamical and radiative transfer parts of the problem, {which can result} in solution violating energy conservation \citep{2021A&A...656A.105T}.
Recently, the model by \citep{2007ApJ...654..435B} was generalised by accounting for an X-ray polarisation in the radiative transfer part \citep{2021MNRAS.501..109C}. 

\red{
Under the condition of high mass accretion rate, additional ingredients can influence the process of boadband spectra formation: \begin{itemize}
{\setlength{\itemsep}{0pt}
\item{\it X-ray reflection from the atmosphere of a NS}\\
Super-critical mass accretion rates result in appearance of accretion column above magnetic poles of a NS. 
Because the radiation from the walls of accretion column can be beamed towards the NS surface \citep{1976SvA....20..436K,1988SvAL...14..390L}, a large fraction of X-ray luminosity can be intercepted by the atmosphere of a NS \citep{2013ApJ...777..115P}. 
Reprocessing and reflection of X-rays can influence spectra making it harder \citep{2015MNRAS.452.1601P}. 
\item{\it Reprocessing of X-rays by accretion flow within the magnetospheric radius}\\
A fraction of X-ray photons can be reprocessed by accretion flow in between the inner disc radius and NS surface. 
Such a reprocessing becomes especially important at accretion luminosity $>{\rm few}\times 10^{38}\,\ergs$, when the flow can turn in optically thick state due to the Compton scattering \citep{2017MNRAS.467.1202M}.  
\item{\it Reprocessing of X-rays by the outflows launched from the disc}\\
Extreme mass accretion rates onto the NS surface lead to radiation driven outflows from accretion disc. 
Strong outflows cause geometrical collimation of X-ray photons.
Multiple reflections and reprocessing of the photons due to their interaction with the outflow can influence X-ray spectra making it softer. 
}
\end{itemize}
}

\begin{figure}
\centering 
\includegraphics[width=10.cm, angle =0]{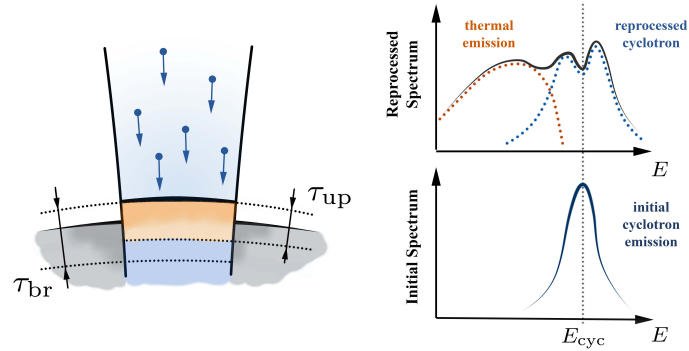} 
\caption{
Schematic picture of the theoretical model explaining spectra formation at extremely low mass accretion rates.
The X-ray energy spectrum is originated from the atmosphere of a NS with the upper layer overheated by low-level accretion.
The accretion flow is stopped in the atmosphere due to collisions.
Collisions result in excitation of electrons to upper Landau levels.
The following radiative de-excitation of electrons produces cyclotron photons.
The cyclotron photons are partly reprocessed by magnetic Compton scattering and partially absorbed in the atmosphere.
The reprocessed photons form a high-energy component of the spectrum, while the absorbed energy is released in thermal emission and forms a low-energy part of the spectrum.
From \citep{2021MNRAS.503.5193M}.}
\label{pic:low_state_sp_model}
\end{figure}

\begin{figure}
\centering 
\includegraphics[width=11.5cm, angle =0]{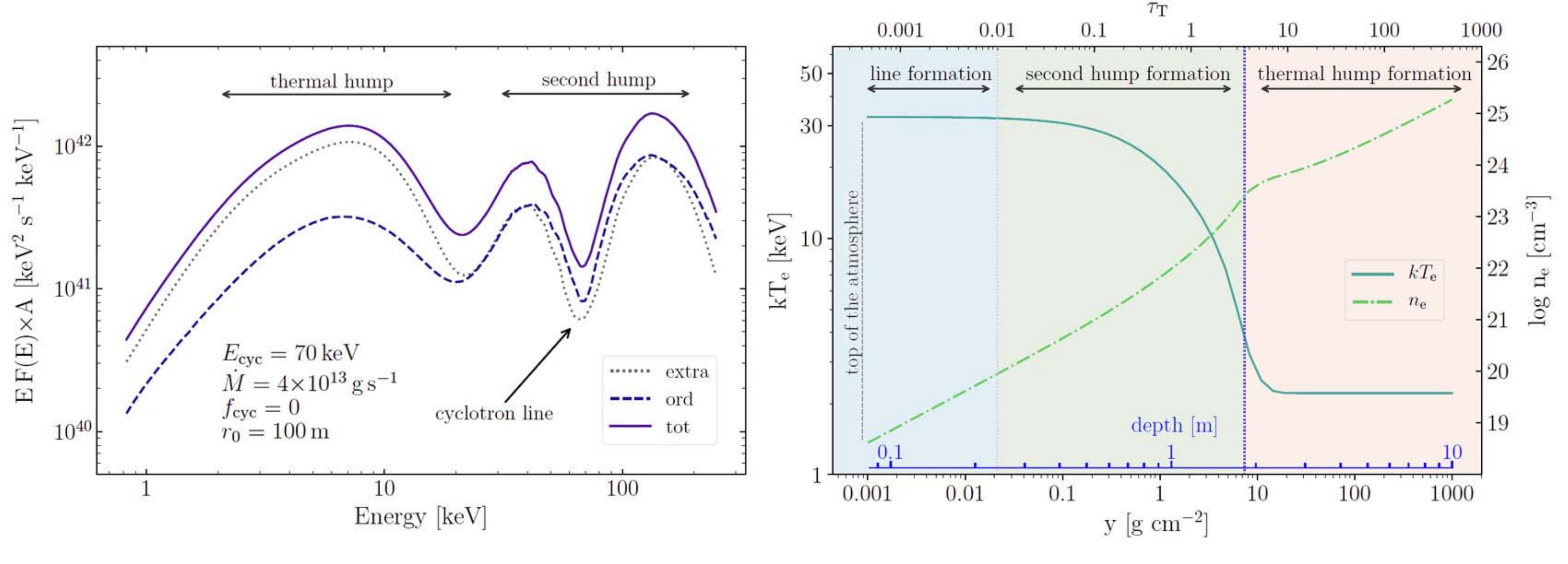} 
\caption{
{\it(Left:)} XRP energy spectra in two polarisation modes at low mass accretion rates.
Solid line describes the total flux, while the dotted and the dashed lines correspond to extraordinary and ordinary modes respectively.
{\it(Right:)} Atmospheric structure: electron temperature (solid) and electron number density (dashed-dotted).
The major spectral components and associated formation regions are indicated in both panels.
The plots are reproduced on the base of \citep{2021A&A...651A..12S}.
}
\label{pic:low_state_sp_SL}
\end{figure}

In contrast to the standard pulsar spectra observed at high luminosities, the transition of some XRP to the state with low mass accretion rate results in very different spectral shape, consisting of two broad components (see Fig.\,\ref{pic:low_state_sp}, \citep{2019MNRAS.483L.144T,2019MNRAS.487L..30T,2021ApJ...912...17L}).
Such spectral distribution can be explained by models where the upper layers of NS atmosphere are overheated by the accretion process, and seed photons are produced by magnetic free-free emission and cyclotron de-excitation of electrons from high Landau levels (see Fig.\,\ref{pic:low_state_sp_model}, \citep{2021MNRAS.503.5193M,2021A&A...651A..12S}).
According to these models, accretion flow is stopped in the atmosphere by Coulomb collisions with some fraction of kinetic energy going into heat, while anothergoing into the excitation of electrons to the upper Landau levels.
Excited electrons almost immediately transfer to the ground Landau level emitting one or several cyclotron photons.
The low-energy component in X-ray spectra is formed due to the thermal emission of strongly magnetised plasma, while the high-energy component is a result of cyclotron emission. 
Both components experience magnetic absorption in the atmosphere and Comptonisation. 
Resonant Compton scattering at the cyclotron energy forces cyclotron photons to leave the atmosphere in the wings of a cyclotron line.
It results in the broadening of high-energy component and the appearance of a cyclotron scattering feature on top of it (see Fig.\,\ref{pic:low_state_sp_model},\,\ref{pic:low_state_sp_SL}).
Current models of spectra formation at a low luminosity state require the upper layers of the NS atmosphere to be overheated up to temperatures $T\gtrsim 40\,{\rm keV}$. 
This result is aligned with earlier calculations of the atmospheric structure under conditions of its heating by the accretion process \citep{1969SvA....13..175Z}.
Under the condition of high temperatures, a large fraction (if not the major) of seed cyclotron photons are produced due to the magnetic bremsstrahlung in the atmosphere.
One of the significant parameters in the models is the typical depth of accretion flow brake in the atmosphere.
In that sense, accurate theoretical models of spectra formation at a low mass accretion rate can shed light on the aspects of plasma physics in a strong field regime. 
Note also that the models developed here \citep{2021MNRAS.503.5193M,2021A&A...651A..12S} do not require the formation of collisionless shock (at least, at the luminosity range and at magnetic field strength, where two-component spectra are observed).
The problems of the models describing spectra formation at low luminosity are related to the lack of their self-consistency. 
In particular, there is still no self-consistent calculations of the temperature structure in the atmosphere and it is still not explained, how the spectra composed of two component at low state transforms to ``classical" spectra observed at higher mass accretion rates. 

\smallskip\smallskip

In a couple of transient XRPs in the propeller state \citep{2016A&A...593A..16T}, soft spectra composed of one black-body-like component were detected (see Fig.\,\ref{pic:propeller_sp}). 
In both cases, the spectra in the propeller state were well fitted by the absorbed black-body model with the temperature $\sim 0.5\,{\rm keV}$ and radius of the emitting area $\sim 0.6-0.8\,{\rm km}$.
Because the centrifugal barrier largely prevents matter penetration to the NS surface in the propeller state, the observed black-body-like emission can be due to cooling of accretion heated NS crust.
The accretion-induced heating of the NS crust and its subsequent cooling has been studied extensively for NSs with low surface magnetic field ($B\lesssim 10^8\,{\rm G}$), but not for strongly magnetised sources ($B\gtrsim 10^{12}\,{\rm G}$). 
The strong magnetic field could alter both heating and cooling processes due to magnetic field influence on the  heat transfer and details of energy release during the accretion state \citep{2015SSRv..191..171P,2016MNRAS.463L..46W}.
The models involving cooling of the NS crust predict slow reduction of X-ray flux from the NS surface and gradual decrease of black-body temperature.
The observational data available up to date is largely consistent with the cooling hypothesis (see, however, \citep{2017MNRAS.472.1802R}).

\subsubsection{\red{4.4.3\,} Cyclotron lines: the fingerprints of a strong magnetic field}

In addition to the continuum emission, in the spectra of some XRPs, a narrow absorption feature, a cyclotron resonant scattering feature (aka cyclotron line), may be observed.
Photons experience resonant Compton scattering by the electron in rest at energies
\be 
\label{eq:resonanceSimp}
E^{(n)}_{\rm res}(b) \! =\! 511\,{\rm keV} 
 \left\{  \begin{array}{ll}
 \strut\displaystyle \! \! 
 \frac{\sqrt{1+2nb\sin^2\theta}-1}{\sin^2\theta}, &  \mbox{for}\ \theta\neq0, \  n=1,2,..., \\
\! \!  b ,                                             & \mbox{for}\ \theta=0.  
\end{array} \right.
\ee
where Landau level number $n\in \{0,1,2,3,...\}$,
and $\theta$ is the angle between the photon momentum and local direction of magnetic field. 
Expression (\ref{eq:resonanceSimp}) naturally results in the approximate relation between the magnetic field strength and the energy of the fundamental cyclotron line (\ref{eq:Ecyc_vs_B}).
Currently, cyclotron lines are known in a few dozens of XRPs \citep{2019A&A...622A..61S} providing an important (and sometimes the only) tool to measure magnetic field strength in accreting strongly magnetised NSs (see expression \ref{eq:Ecyc_vs_B} relating cyclotron energy and local magnetic field strength).
It is great luck if the cyclotron line is observed in X-ray energy spectra with its harmonics because, in this case, one can be more confident that the feature in the spectra is nothing else but the cyclotron line. 

Photons can be scattered resonantly by any charged particles including protons and ions.
However, the scattering cross section is significantly smaller due to the larger mass of ions.
The energy of the fundamental line in the case of scattering by ion is relatively small and given by
\be 
E_{\rm res, ion}\simeq 6.3\times 10^{-3}\,B_{12}\left(\frac{Z}{A}\right)\,{\rm keV},
\ee 
where $Z$ is the atomic number and $A$ is the mass number of the ion.
The proton resonant scattering feature was discussed in interpretation of X-ray spectra in ULX NGC~M51~X-8 \citep{2018NatAs...2..312B}.

Variations of mass accretion rate and luminosity in XRPs result in changes of dynamics and geometry of the emitting regions in the vicinity of the NS surface. 
These changes can cause variations in the cyclotron line properties. 
The most investigated phenomenon here is variations of the cyclotron line centroid energy with the accretion luminosity.
As we have already mentioned, there are sources showing a positive correlation between the line centroid energy and luminosity 
\citep{2007A&A...465L..25S,2012A&A...542L..28K,2017MNRAS.466.2752R,2021ApJ...919...33C}
and sources, where the correlation is negative \citep{2004ApJ...610..390M,2006MNRAS.371...19T}. 
The first ones show systematically lower accretion luminosity (see Fig.\,\ref{pic:cyc_lines_variations}). 
The explanation of observed cyclotron line behaviour is based on expected variations of geometry of the line forming regions with changes in a mass accretion rate.
In particular, the positive correlation is considered as evidence for the sub-critical accretion regime, when the accretion process results in hot spots at the magnetic poles of a NS or in collisionless shock above them.
The negative correlation, on the contrary, can be considered as evidence for super-critical accretion when radiation-dominated shock appears above the stellar surface. 
Let us consider some theoretical models explaining the variation of cyclotron line centroid energy: 

\begin{itemize}
{\setlength{\itemsep}{0pt}

\item{\it Positive correlation}

There are two models explaining the positive correlation between observed cyclotron line energy and accretion luminosity: one is based on the assumption that collisionless shock is formed above the NS surface, and another is assuming that the energy of accreting material is released in hot spots at the surface. 

\begin{figure}
\centering 
\includegraphics[width=11.cm, angle =0]{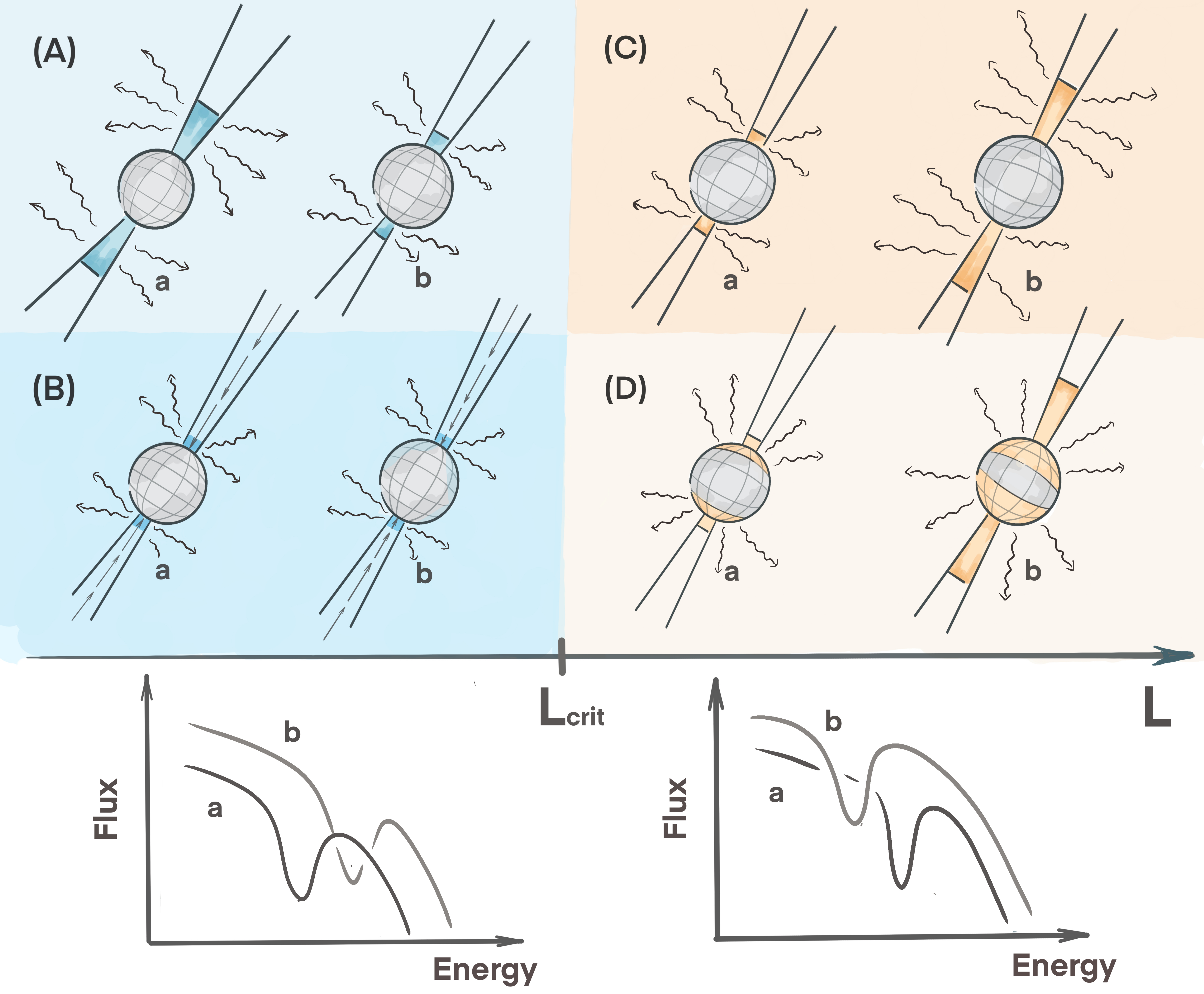} 
\caption{
\red{Schematic illustration of the models explaining positive (A,B) and negative (C,D) correlations between accretion luminosity and cyclotron line centroid energy in XRPs:\\
(A) According to the models accounting for the collisionless shock, the cyclotron centroid energy increase with the luminosity is associated with the decrease of collisionless shock height. \\
(B) In the model accounting for the Doppler effect in the accretion channel, radiation pressure affects the velocity of the accreting matter: 
the higher the luminosity, the lower the velocity in the vicinity of the NS surface, the lower the redshift, and the higher the cyclotron line centroid energy. \\
(C) The most straightforward model explaining the negative correlation associates the decay of the cyclotron line centroid energy with the growth of accretion column and a corresponding shift of a line forming region to a larger height, where the B-fields is weaker. \\
(D) According to the reflection model, the cyclotron features form due to the reflection/reprocessing of X-rays by the atmosphere of a NS. 
Magnetic field strength decreases towards the equator of a NS. 
Thus, the larger is the luminosity, the higher is the accretion column, the larger the illuminated fraction of a NS surface, 
the weaker the average magnetic field, and the lower the cyclotron line energy.}
}
\label{pic:cyc_lines_all}
\end{figure}

\begin{enumerate}

\item[(A)] 
The models based on the assumption of {\it collisionless shock} formation \citep{2007A&A...465L..25S,2017MNRAS.466.2752R} use the fact that the shock height above NS surface is anticorrelated with the mass accretion rate: the higher the mass accretion rate, the closer collisionless shock locates to the stellar surface \citep{1982ApJ...257..733L,2004AstL...30..309B}.
Thus, at higher mass accretion rates, the typical field strength in a line forming region is higher, that results in a positive correlation between the line centroid energy and luminosity (see Fig.\,\ref{pic:cyc_lines_all}A).

\item[(B)] 
The alternative model explaining the positive correlation assumes that the broadband X-ray energy spectrum forms in the {\it hot spots} at the NS surface (see Fig.\,\ref{pic:cyc_lines_all}B, \citep{2015MNRAS.454.2714M}).
Because of the relatively low mass accretion rate, the accretion regime is sub-critical and the accretion flow above NS surface is optically thin for continuum X-ray radiation. 
The flow, however, remains optically thick for X-ray photons at energies close to the cyclotron one. 
Variations in mass accretion rate and luminosity cause variations in the dynamics of accreting material: 
the higher the luminosity, the larger the radiative force acting on the accretion flow, and the smaller the velocity of the flow just above the surface of a NS. 
The resonant scattering of X-ray photons by the accretion flow is affected by the Doppler effect in accretion channel: the resonant scattering by slower accretion flow results in the appearance of cyclotron feature at higher energies. 
Thus, the model naturally predicts the positive correlation.
Note, that the displacement of the cyclotron centriod energy is affected both by bulk velocity of accreting material and direction of photon propagation. 
Therefore, correlation properties are influenced by the beam pattern as well \citep{2014ApJ...781...30N}. 

\end{enumerate}

\red{We would speculate that precise accounting for the MHD structure of a line forming region can provide additional hypothesis on the nature of the positive correlation. 
In particular, hot spots have more complex structure than we usually assume in the models.
Finite timescale of matter spreading over the NS surface from the polar regions results in appearance of accretion mounds at the polar cups of a NS \citep{2013MNRAS.430.1976M}. 
The spreading process results in magnetic field distortion in the line forming regions (unless there is some mechanism that allows mass to slip through the magnetic field lines, see e.g. \citep{2020JPlPh..86f9002K}). 
Such distortions can cause variations of typical magnetic field strength in a line forming region and affect variations of the line centroid energy with the luminosity.}

\item{\it Negative correlation}

The negative correlation is expected in super-critical XRPs and all models explaining the correlation involve radiation dominated accretion columns. 

\begin{enumerate}

\item[\red{(C)}] 
The most straightforward model explaining the negative correlation assumes that the cyclotron line is formed in radiation dominated shock, which height above the NS surface depends on the mass accretion rate: the higher the mass accretion rate, the higher the shock above the NS surface (see Fig.\,\ref{pic:cyc_lines_all}D).
Because the field strength $B\propto r^{-3}$, the higher mass accretion rate places the line forming region in the area of lower magnetic field strength, that naturally leads to the negative correlation. 
This model, however, has a couple of problematic points: 
(i) in the case of high accretion columns, the range of magnetic field strength represented in the line forming region is so broad that the cyclotron lines are expected to be very wide or even disappear from the spectra \citep{2008ApJ...672.1127N}, and 
(ii) because $B\propto r^{-3}$, even small variations of accretion column height have to result in a significant variations of cyclotron line centroid energy, while observations demonstrate $\Delta E_{\rm cyc}/E_{\rm cyc}\lesssim 0.1$ (see {\it top right} panel in Fig.\,\ref{pic:cyc_lines_variations}, \citep{2006MNRAS.371...19T,2016MNRAS.460L..99C,2017MNRAS.466.2143D}). 

\item[\red{(D)}] 
Another possible explanation of the negative correlation is related to the specific features of accretion column. 
Radiation leaving accretion column from its sides is expected to be beamed towards the NS surface due to the relativistic effects \citep{1976SvA....20..436K,1988SvAL...14..390L}.
Because of that, a large fraction of X-ray luminosity produced by accretion columns is intercepted by the atmosphere of a NS. 
Reprocessing of X-ray radiation by the atmosphere results in the appearance of an absorption-like feature at the cyclotron energy corresponding to the local magnetic field strength \citep{1973trsl.book.....I,2019Ap.....62..129G}. 
In the case of magnetic field dominated by dipole component, the field strength at the NS surface is given by $B=0.5\,B_0\sqrt{1+3\cos^2\theta}$, where $B_0$ is the field strength at the magnetic pole, and $\theta$ is co-latitude at the stellar surface.
Then the variations of the accretion column height with luminosity naturally results in appearance of the negative correlation: the higher the mass accretion rate, the larger the accretion column height, the larger the area illuminated by accretion column around the magnetic poles of a NS, the smaller the average magnetic field strength over the illuminated part of a star, and the smaller the cyclotron line centroid energy in X-ray spectrum (see  Fig.\,\ref{pic:cyc_lines_all}E and \citep{2013ApJ...777..115P} for details). 
The reflection model naturally explains small amplitudes of observed variations of the cyclotron line energy: magnetic field strength varies by a factor of 2 only over the NS surface and possibly large changes of accretion column height result in a small variations of line displacement.
Recent Monte-Carlo simulations \citep{2021A&A...655A..39K} show that observed variations of cyclotron line energy might be to large to be explained by the model.
However, the final conclusions on the model applicability can be done only after detailed simulations of the reflection process.

\end{enumerate}

Note, that the discussed theoretical models explaining the negative correlation do not exclude each other, and the cyclotron lines can appear in X-ray energy spectra due to combination of all these mechanisms of cyclotron line formation.

}
\end{itemize}

\section{Open issues}

In this section we highlight key open issues in physics and astrophysics of XRPs.

\begin{enumerate}

  \item Advanced spin-up/-down theory in XRPs in the view of super-fluidity of NS inner crust and core.

  \item Detailed theory and numerical model of accretion from a cold disc. Quantitative description of XRP transition into cold disc mode. Inner disc radius in the state of accretion from the cold disc.  

  \item First principal simulations of plasma (with different physical properties) penetration  into the magnetosphere of a NS. 

  \item {Detailed theoretical model of disc accretion onto inclined magnetic dipoles and NS with a more complicated geometry of NS magnetic field.}

  \item Detailed theory and numerical simulations of plasma deceleration in the atmosphere of magnetised NS.
 
  \item Theoretical models of NS crustal heating induced by accretion and the crustal cooling under conditions of a strong magnetic fields ($B\gtrsim 10^{12}\,{\rm G}$).
 
  \item Existence of collisionless shocks above NS surface at low mass accretion rates. Accurate numerical simulations of collisionless shock formation.
  
  \item Self-consistent theoretical and numerical model of accretion column accounting for radiative transfer in a strongly magnetised accreting plasma, appearance of the photon bubble instability, nuclear reactions in accretion channel, neutrino emission and possible deviations of accreting material from the condition of local thermodynamic equilibrium.
  
  \item {Unified theory explaining spectra, polarisation and pulse profile formation in XRPs over a wide range of accretion luminosity ($10^{33}-10^{41}\,\ergs$). }

  \item { \red{Explanation of unexpectedly low polarisation degree observed in sub-critical X-ray pulsars by {\it IXPE}.} }
  
  \item Numerical model reproducing formation, shape and variability (i.e. phase-resolved variability and variability with accretion luminosity) of cyclotron scattering features in spectra of XRPs over a wide range of accretion luminosity.  
  
  \item {Theory of magnetic field evolution under the influence of accretion.}
  
  \item Fraction of accreting NSs and BHs in ULXs. Magnetic field strength and structure in ULXs powered by accretion onto NSs. 
  
  \item Influence of ULXs powered by accretion onto NSs on global evolution of high-mass X-ray binaries and production of gravitational wave sources. 
  
  
\end{enumerate}

\vspace{0.3cm}

{\bf Key-points to have in mind}:
\begin{enumerate}
\item 
{\bf XRPs are} accreting strongly magnetised NSs in close binary systems.
The strong magnetic field of the order of $10^{12}$~G in XRPs affects both geometry of accretion flow on the spatial scales of $\sim 10^8\,{\rm cm}$ and physical processes in close proximity to the NS surface.
The magnetic field at the NS surface in XRPs is orders of magnitude stronger than the magnetic field achievable in labs.
\item 
{\bf Luminosity of XRPs is powered by accretion process.}
Nuclear burning in XRPs is going in a stable regime, and its contribution to the luminosity is insignificant.
\item 
Strong {\bf magnetic field in XRPs disrupts accretion flow towards a NS at $R_{\rm m}\sim 10^8\,{\rm cm}$ and forces accreting material to land NS surface in small regions} located close to magnetic poles of a star.
The energy release occurs predominantly in these regions. 
Observed pulsations of the X-ray flux, thus, is a result of misalignment between the rotational and magnetic axes of a NS. 
\item 
{\bf XRPs provide a rich phenomenology.}
Spin periods of NSs in XRPs cover a few orders of magnitude from a fraction of a second up to thousands of seconds.
The apparent X-ray luminosity of some sources covers more than 7 orders of magnitude on the time scales from minutes to months.
Many XRPs properties, including their X-ray energy spectra, power density spectra and pulse profiles, are known to be variable with the luminosity. 
\item 
There are {\bf three primary mechanisms of the mass transfer} between NS and its companion in XRPs: 
(i) Roche lobe overflow,
(ii) capture of matter from the decretion disc in Be-system, and
(iii) companion mass loss due to the stellar wind.
\item 
{\bf Rotation of a NS sets up a centrifugal barrier} for the accreting material.
The accretion flow can penetrate through the barrier in the case of a sufficiently high mass accretion rate.
At low mass accretion rates, the centrifugal barrier stops accretion flow at the magnetospheric boundary and largely prevents accretion into a NS surface (propeller effect).
\item 
On the other hand, {\bf interaction of accretion flow with the NS magnetosphere affects rotation of a compact object} and spin periods are known to be variable in XRPs.
Analyses of spin period variability allow estimating magnetic field strength.
However, the interaction of accretion flow with NS magnetosphere is affected by many parameters (inclination of a magnetic dipole in respect to the accretion flow, the exact structure of NS magnetic field, etc.), which influence is not investigated sufficiently.  
\item 
Because of a wide range of mass accretion rates represented in XRPs, {\bf accretion discs show a great variety of possible physical conditions there}. 
The physical conditions in the discs largely affect the observational manifestation of XRPs. 
\item 
{\bf Depending on the mass accretion rate onto the NS surface, one would expect different geometry of the emitting regions} at the poles of a star.
Relatively low mass accretion rates ($\lesssim 10^{17}\,{\rm g\,s^{-1}}$) lead to the hot spot geometry. 
Large mass accretion rates ($\gtrsim 10^{17}\,{\rm g\,s^{-1}}$) result in the appearance of radiation-dominated shock above the surface and a sinking region below it - the structure called "accretion column".
Accretion columns, supported by radiation pressure and confined by a strong magnetic field, allow luminosity well above the Eddington limit.
\item 
At extreme mass accretion rates, {\bf accretion columns might be advective}. 
In this case, the internal condition can be sufficient to provide {\bf strong neutrino emission} with luminosity comparable to the photon luminosity of XRPs.
On the other hand, large mass accretion rates can lead to the development of {\bf the photon bubble instability}.
\item 
{\bf Geometry of the emitting region affects spectra and beam pattern formation} in XRPs.
\item 
{\bf Cyclotron absorption features} often observed in the XRPs spectra arise due to the resonant Compton scattering of X-ray photons in a strong magnetic field. 
The cyclotron features serve as the only direct method of measuring the NS magnetic field strength. 
The dependence of the emission region structure at the NS surface on the mass accretion rate is reflected in the variations of the cyclotron energy with the source luminosity.

\end{enumerate}

\vspace{0.3cm}

\section{\red{Cross-References}}

\red{
Below we list a selection of other chapters in this book that provide more details on various aspects mentioned in this chapter.}
\begin{itemize}[leftmargin=15pt]
\item \red{``Fundamental physics with neutron stars"
{\it by Joonas Nättilä  and Jari Kajava}}
\item \red{``Ultra-luminous X-ray sources: extreme accretion and feedback" 
{\it by Ciro Pinto and Dominic Walton}}
\item \red{``Low-Mass X-ray Binaries"
{\it by Arash Bahramian and Nathalie Degenaar}}
\end{itemize}


{

}

\end{document}